\documentclass[12pt]{article}

\usepackage{amsfonts,amssymb,amstext}

\usepackage[normal]{subfigure}
\usepackage{epsfig}
\usepackage{graphicx}
\usepackage{enumerate}
\usepackage{mathtools}
\usepackage{color}
\usepackage[normalem]{ulem}
\usepackage{float}
\usepackage{amsmath}
\usepackage{amsthm}
\allowdisplaybreaks

\input epsf.tex

\newcommand{\leftlim}{(}
\newcommand{\rightlim}{)}

\newcommand{\inta}{\int \limits_{z_{\alpha-1/2}}^{z_{\alpha+1/2}}}

\newcommand{\abs}[1]{\left\vert #1 \right\vert}

\newtheorem{theorem}{Theorem}

\newtheorem{remark}{Remark}

\def\Q{\vec{\bold Q}}

\def\R{\mathbb{R}}

\def\vu{\boldsymbol{\vec{u}}}

\def\vv{\boldsymbol{\vec{v}}}

\def\vn{\boldsymbol{\vec{n}}}

\def\a{\alpha}
\def\b{\beta}
\def\r{\rho}
\def\O{\Omega}
\def\S{\Sigma}
\def\G{\Gamma}

\def\g{\nabla}
\def\p{\partial}

\def\bS{\boldsymbol{\Sigma}}

\def\dint{\displaystyle\int}
\def\dsum{\displaystyle\sum}

\def\dfrac{\displaystyle\frac}

\def\bd{\boldsymbol}
\def\td{\tilde}

\def\mb{\mbox}

\title{A multilayer shallow model for dry granular flows with the $\mu(I)$ rheology: Application to granular collapse on erodible beds}

\author{E.D. Fern\'andez-Nieto \thanks{Dpto. Matem\'atica Aplicada I. ETS Arquitectura - Universidad de Sevilla.
  Avda. Reina Mercedes S/N, 41012-Sevilla, Spain, (edofer@us.es)}\ , J. Garres-D{\'i}az \thanks{IMUS {\&} Dpto. Matem{\'a}tica Aplicada I. ETS Arquitectura - Universidad de Sevilla.
  Avda. Reina Mercedes S/N, 41012-Sevilla, Spain, (jgarres@us.es)}\ , A. Mangeney
 \thanks{Institut de Physique du Globe de Paris, Equipe Sismologie, University Paris-Diderot, Sorbonne Paris Cit\'e, Paris, France, (mangeney@ipgp.fr)}\ \thanks{ANGE team, CEREMA, INRIA, Lab. J. Louis Lions, Paris, France}\ , G. Narbona-Reina\thanks{Dpto. Matem\'atica Aplicada I. ETS Arquitectura - Universidad de Sevilla.
  Avda. Reina Mercedes S/N, 41012-Sevilla, Spain, (gnarbona@us.es)}}

\begin{document}
\date{}
\maketitle


\abstract{
In this work we present a multilayer shallow model to approximate the Navier-Stokes equations with hydrostatic pressure and the $\mu(I)$-rheology. The main advantages of this approximation are (i) the low cost associated with the numerical treatment of the free surface of the modelled flows, (ii) exact conservation of mass and (iii) the ability to compute 3D profiles of the velocities in the directions along and normal to the slope. The derivation of the model follows \cite{EnriqueMultilayer} and introduces a dimensional analysis based on the shallow flow hypothesis. The proposed first order multilayer model fully satisfies a dissipative energy equation. A comparison with an analytical solution with a non-constant normal profile of the downslope velocity demonstrates the accuracy of the numerical model. Finally, by comparing the numerical results with experimental data, we show that the proposed multilayer model with the $\mu(I)$-rheology reproduces qualitatively the effect of the erodible bed on granular flow dynamics and deposits, such as the increase of runout distance with increasing thickness of the erodible bed. We show that the use of a constant friction coefficient in the multilayer model leads to the opposite behaviour. This multilayer model captures the different normal profiles of the downslope velocity during the different phases of the flow (acceleration, stopping, etc.) including the presence of static and flowing zones within the granular column.
}

\bigskip

\tableofcontents


\frenchspacing   

\section{Introduction}
Granular flows have been widely studied in recent years because of their importance in industrial processes and geophysical flows such as avalanches, debris or rock avalanches, landslides, etc. In particular, numerical modelling of geophysical granular flows provides a unique tool for hazard assessment. \\

The behaviour of real geophysical flows is very complex due to topography effects, heterogeneity of the material involved, presence of fluid phases, fragmentation, etc. \cite{Delannay2015}. One of the major issues is to quantify erosion/deposition processes that play a key role in geophysical flow dynamics but are very difficult to measure in the field. Laboratory experiments of granular flows are very useful to test flow models on simple configurations where detailed measurements can be performed, even if some physical processes may differ between the large and small scale. These experiments may help in defining appropriate rheological laws to describe the behaviour of granular materials. Recent experiments by Mangeney et al. \cite{MangeneyErosion} and Farin et al. \cite{FarinErosion} on granular column collapse have quantified how the dynamics and deposits of dry granular flows change in the presence of an erodible bed. They showed a significant increase of the runout distance (i.e. maximum distance reached by the deposit) and flow duration with increasing thickness of the erodible bed. This strong effect of bed entrainment was observed only for flows on slopes higher than a critical angle of about $16^\circ$ for glass beads. The question remains as to whether this behaviour can be reproduced by granular flow models.\\

Understanding the rheological behaviour of granular material is a major challenge. In particular, a key issue is to describe the transition between flow (fluid-like) and no-flow (solid-like) behaviour. Granular flows have been described by viscoplastic laws and especially by the so-called $\mu(I)$ rheology, introduced by Jop et al. \cite{JopForterrePouliquen}. It specifies that the friction coefficient $\mu$ is variable and depends on the inertial number $I$ that is related to the pressure and strain rate. Lagr{\'e}e et al \cite{LagreeStaronPopinet} implemented it in a full Navier-Stokes solver (Gerris) by defining a viscosity from the $\mu(I)$ rheology. They validated the model with a 2D analytical solution and compared it to 2D discrete element simulations of granular collapses over horizontal rigid beds and with other rheologies.  Staron et al. \cite{StaronLagreePopinet2012} and \cite{StaronLagreePopinet2014} applied this model to granular flows in a silo. Using an Augmented Lagrangian method combined with finite element discretisation to solve the 2D full Navier-Stokes equations, Ionescu et al. \cite{IonescuMangeney} showed that this rheology reproduces quantitatively laboratory experiments of granular collapses over horizontal and inclined planes. By interpreting the $\mu(I)$ rheology as a viscoplastic flow with a Drucker-Prager yield stress criterion and a viscosity depending on the pressure and strain rate, they showed that using a constant or variable viscosity only slightly changes the results when simulating granular column collapses of small aspect ratio.  In \cite{ChauchatMedale2}, Chauchat and M\'edale implemented the $\mu(I)$ rheology in a three-dimensional numerical model with a finite element method combined with the Newton-Raphson algorithm with a regularisation technique. The numerical model was validated by an analytical solution for a dry granular vertical-chute flow and a dry granular flow over an inclined plane and by laboratory experiments. Previously, Chauchat and M\'edale \cite{ChauchatMedale1} simulated the bed-load transport problem in 2D and 3D with a two-phase model that considers a Drucker-Prager rheology for the granular phase. Lusso et al. \cite{LussoBouchut} used a finite element method to simulate a 2D viscoplastic flow considering a Drucker-Prager yield stress criterion and a constant viscosity. They obtained similar results taking into account either a regularisation method or the Augmented Lagrangian algorithm. By comparing the simulated normal velocity profiles and the time change of the position of the static/flowing interface with laboratory experiments of \cite{FarinErosion}, they concluded that a pressure and rate-dependent viscosity can be important to study flows over an erodible bed. Similar conclusion is presented in \cite{LussoBouchutSimplified} after comparing the normal velocity profiles and the position of the static/flowing interface during the stopping phase of granular flows over erodible beds calculated with a simplified thin-layer but not depth-averaged viscoplastic model with those measured in laboratory experiments.\\

Because of the high computational cost of solving the full 3D Navier-Stokes equations, in particular in a geophysical context, granular flows have often been simulated using depth-averaged shallow models. The shallow or thin-layer approximation (the thickness of the flow is assumed to be small compared to its downslope extension) associated with depth-averaging leads to conservation laws like the Saint-Venant equations. These approximations have been applied to granular flows by Savage and Hutter \cite{SavageHutter} by assuming a Coulomb friction law where the shear stress at the bottom is proportional to the normal stress, with a constant friction coefficient $\mu$. However, this  model does not reproduce the increase in runout distance observed with increasing thickness of the erodible bed. The analytical solution deduced in  \cite{FaccanoniMangeney} proves that this system leads to the opposite effect. The question is as to whether this opposite behaviour between the experiments and simulations is due to the thin-layer approximation and/or depth-averaging process or to the rheological law implemented in the model (i.e. constant friction coefficient).\\

Gray and Edwards \cite{GrayEdwards} introduced the $\mu(I)$ rheology in a depth-averaged model by adding a viscous term. However, in depth-averaged models, only the mean velocity over the whole thickness of the flow is calculated (i.e. the whole granular column is either flowing or at rest). Granular collapse experiments and simulations have shown on the contrary that the velocity of the grains near the free surface is higher than that of the grains located near the bottom. During the stopping phase and when erosion/deposition processes occur, static zones may develop near the bottom and propagate upwards. The resulting normal gradient of the downslope velocity is a significant term in the strain rate and therefore strongly influences the $\mu(I)$ coefficient. \\

To take into account the change of the velocity field in the direction normal to the topography, we present here a multilayer shallow model that we have developed with the $\mu(I)$ rheology. This model consists of subdividing the domain into several layers in the normal direction and applying the thin-layer approximation within each layer. As a result, a velocity is calculated for each layer, providing a normal velocity profile. Multilayer models were introduced by Audusse \cite{Audusse2005} and extended by Audusse et al. \cite{Audusse2008}. A different multilayer model, which takes into account the exchange of mass and momentum between the layers, has since been derived by Audusse et al \cite{Audusse2011}, \cite{Audusse2011bis} and Sainte-Marie \cite{SainteMarie}.

A new procedure to obtain a multilayer model has been introduced by Fern{\'a}ndez-Nieto et al. \cite{EnriqueMultilayer}. Several differences appear between this multilayer model and the ones deduced by Audusse et al. First, in \cite{EnriqueMultilayer}, the multilayer model is derived from the variational formulation of  Navier-Stokes equations with hydrostatic pressure by considering a discontinuous profile of the solution at the interfaces of a vertical partition of the domain. This procedure proves that the solution of this multilayer model is a particular weak solution of the Navier-Stokes system. Moreover, the mass and momentum transfer terms  at the interfaces of the normal partition are deduced from the jump conditions  verified by the weak solutions of the Navier-Stokes system. In addition, the definition of the vertical velocity profile is easily obtained using the mass jump condition combined with the incompressibility condition.\\

By comparing this model with granular flow experiments on erodible beds (\cite{MangeneyErosion}, \cite{FarinErosion}), we evaluate (1) if the model with the $\mu(I)$ rheology gives a reasonable approximation of the flow dynamics and deposits of real granular flows, (2) if it reproduces the increase in runout distance observed for increasing thickness of the erodible bed above a critical slope angle $\theta_c \in [12^o, 16^o]$ and (3) how the multilayer approach improves the results compared to the classical depth-averaged Saint-Venant model (i.e. monolayer model).\\

The paper is organised as follows. In Section \ref{se:initialSystem} we introduce the $\mu(I)$ rheology and the associated viscosity as well as a dimensional analysis of the 3D Navier-Stokes equations. In Section \ref{se:multilayer_approach} we present the multilayer approach following \cite{EnriqueMultilayer} to derive a 3D multilayer model for dry granular flows up to first order when considering the thin-layer or shallow approximation. The final $\mu(I)$ rheology Multilayer Shallow Model (MSM) is deduced in Section \ref{se:weakalfa}. In Section \ref{se:numericalTest} we validate our model using the 2D analytical solution presented in \cite{LagreeStaronPopinet} and compare our results with those of laboratory experiments done by Mangeney et al. \cite{MangeneyErosion}. We show that the $\mu(I)$ rheology can reproduce qualitatively the increase in runout distance of granular flows over erodible beds as opposed to the constant friction model and that the multilayer approach significantly improves results compared to the monolayer (i.e. Saint-Venant) model.

\section{The 3D initial system}
\label{se:initialSystem}
We consider the space variables $(x,z)\in\O \times \mathbb{R}^+ \subset \R^{3}$, where $x = (x_{1},x_{2})\in\O\subset\R^{2}$ corresponds to the horizontal and $z\in\mathbb{R}^+$ to the vertical variable, the velocity $\vu\in\mathbb{R}^3$ with horizontal and vertical components $(\vu_{H},w)$, the density $\r\in\mathbb{R}$ that is assumed to be known and gravity $g\in\mathbb{R}$. We set $\smash{\g = (\p_{x_{1}}, \p_{x_{2}}, \p_{z})}$, the usual differential operator in the space variables, and $\;\smash{\g\!_{x} := (\p_{x_{1}}, \p_{x_{2}})}$, the reduced operator to the horizontal variable. \\

The 3-dimensional Navier-Stokes equations are written as\\

\begin{equation}
\label{eq:NS_3D_comp}
\left\{
\begin{array}{l}
\g\cdot \vu\ = \ 0,\\
\\
\r\p_{t}\vu \;+\; \r\g \cdot \left(\vu\otimes\vu \right) - \g\cdot\Sigma\  = \r\vec{\bd{g}},\\
\end{array}
\right.
\end{equation}

\noindent where $\vec{\bd{g}} = \left(0,-g\right)\in\mathbb{R}^3$ and

\begin{equation}
\label{not_sigma}
\bS= -p\bd{I} + \bd{T}
\end{equation}
is the stress tensor, with $p\in\mathbb{R}$ the pressure, $\bd{I}$ is the identity tensor and $\bd{T}$ the deviatoric tensor given by
\[\bd{T} =\eta D(\vu);\quad
\bd{T} = \left( \begin{matrix}
\bd{T}_{H} & T_{xz}\\
T_{xz} &  T_{zz}
\end{matrix}\right),\]

\noindent where $\eta\in\mathbb{R}$ denotes the viscosity and $D(\vu)$ is the strain rate tensor

\begin{equation}\label{eq:defDu}
D(\vu) = \g\vu+\left(\g\vu\right)' = \left( \begin{matrix}
D_{H}\left(\vu_{H}\right) & \p_{z}\vu_{H}+\left(\g_{x}w\right)'\\
\\
\left(\p_{z}\vu_{H}\right)'+\g_{x}w &  2\p_{z}w
\end{matrix}\right),
\end{equation}

\noindent where $D_{H}(\vu_{H}) = \g_x\vu_{H} + \left(\g_x\vu_{H}\right)'$. With these definitions, system \eqref{eq:NS_3D_comp} can be developed as

\begin{equation}
\label{eq:NS_3D}
\left\{
\begin{array}{l}
\g_{x}\cdot \vu_{H}+\p_{z}w= 0,\\
\\
\r\p_{t}\vu_{H}+\r\g_{x}\cdot (\vu_{H}\otimes\vu_{H})+\r\p_{z}\left(\vu_{H}w\right)+\g_{x}p = \g_{x}\cdot \Big(\eta D_{H}(\vu_{H})\Big)+\p_{z}\Big(\eta \Big(\p_{z}\vu_{H}+\left(\g_{x}w\right)'\Big)\Big),\\
\\
\r\p_{t}w+\r\vu_{H}\g_{x}w+\r w\p_{z}w +\p_{z}p+\r g=\g_{x}\cdot \Big(\eta\Big(\left(\p_{z}\vu_{H}\right)'+\g_{x}w\Big)\Big)+2\p_{z}\Big(\eta\p_{z}w\Big).
\end{array}
\right.
\end{equation}

In the following subsection, the rheology and boundary conditions are presented. In subsection \ref{se:dimensional_analysis}, a dimensional analysis of the system is performed.
\subsection{Closures}
\subsubsection{Rheology}
\label{se:muI}
We consider the so-called $\mu(I)$ rheology (see \cite{JopForterrePouliquen}), which is defined by
\begin{equation}
\label{eq:viscosity_prev}
\eta = \dfrac{\mu(I)p}{\|D(\vu)\|},
\end{equation}
where $\|D\| = \sqrt{(D:D)/2}$, the usual second invariant of a tensor $D$. The friction coefficient $\mu(I)$ depends on the inertial number
 \begin{equation} \label{eq:def_I}
 I = \frac{d_s\|D(\vu)\|}{\sqrt{p/\r_{s}}},
 \end{equation}
 where $d_s$ is the particle diameter and $\r_{s}$ the particle density. The solid volume fraction, denoted by $\varphi_s$, is assumed to be constant, leading to an apparent flow density
 \begin{equation} \label{eq:def_rho}
\r = \varphi_s \r_s.
\end{equation}

\noindent The variable friction coefficient is written
\[\mu(I) = \mu_{s} + \dfrac{\mu_2-\mu_s}{I_{0}+I}I ,\]
where $I_{0}$ is a constant value and $\mu_{2}>\mu_s$ are constant parameters. Note that when the shear rate is equal to zero, $\mu(I)$ is reduced to $\mu_s$ and, for high values of $I$, converges to $\mu_2$.\\

The $\mu(I)$ rheology includes a Drucker-Prager plasticity criterion, that is, the material flows when
\[\|\bd{T}\| > \mu(I)p.\]
\\
Note that the $\mu(I)$ rheology can equivalently be written as a decomposition of the deviatoric stress in a sum of a plastic term and a rate-dependent viscous term (see \cite{IonescuMangeney}):
\[\left\{\begin{array}{ll}
\bd{T} = \dfrac{\mu_s p}{\|\bd{D}\|}\bd{D} + \tilde \eta \bd{D} & \mbox{if } \bd{D} \neq 0,\\
\\
\|\bd{T}\| \leq \mu_s p & \mbox{if }\bd{D} = 0;
\end{array}\right.\]
 \\
with a viscosity defined as $\tilde{\eta}=\dfrac{(\mu_2 - \mu_s)p}{\frac{I_0}{d_s}\sqrt{p/\r_s}+\|\bd{D}\|}$.
Here we investigate the rheology defined by a variable friction $\mu(I)$ and a constant friction $\mu_s$. Note that assuming $\mu=\mu_s$, i.e. $\tilde{\eta}=0$, is different than taking $\tilde{\eta}=K$, with $K$ a non-zero constant as in \cite{IonescuMangeney}.\\

The model that considers a viscosity defined by (\ref{eq:viscosity_prev}) presents a discontinuity when $\|D(\vu)\| = 0$. To avoid this singularity there are several ways to proceed. One of them is to apply a duality method, such as Augmented Lagrangian methods \cite{GLT89} or Berm\'udez-Moreno algorithm \cite{BermudezMoreno}. Another option is to use a regularisation of $D(\vu)$, which is cheaper computationally, however it does not give an exact solution, contrary to duality methods.\\

In this work, we take into consideration two kinds of regularisations of $D(\vu)$. First, we use the regularisation proposed in \cite{LagreeStaronPopinet}, which consist in bounding the viscosity by $\eta_{M} = 250\r\sqrt{g h^{3}}$ Pa$\cdot$s, considering instead of \eqref{eq:viscosity_prev},
\begin{equation}\label{eq:regularisation}
\eta = \frac{\mu(I)p}{\text{max}\left(\|D(\vu)\|,\frac{\mu(I)p}{\eta_{M}}\right)}.
\end{equation}
\\
In this way, we obtain $\eta = \eta_{M}$ if $\|D(\vu)\|$ is close to zero. We used this regularisation in the simulation of the granular flow experiments. However, as explained in Section \ref{se_sub:solAnal}, we cannot consider this regularisation in the simulation of the analytical solution, for which we take into account the regularisation introduced in \cite{EngelmanReg},
\[\eta = \frac{\mu(I)p}{\sqrt{\|D(\vu)\|^2 + \delta^2}},\]
where $\delta > 0$ is a small parameter.
\subsubsection{Boundary and kinematic conditions}
$\bullet$ At the free surface $z_b+h$, we consider the usual kinematic condition
\begin{equation}\label{eq:KinCond}
\p_{t}h + \vu\cdot\vn^{h}  = 0,
\end{equation}
with $\vn^{h} = (\g_{x}(z_b+h),-1)/\sqrt{1+\abs{\g_{x}(z_b+h)}^{2}}$ the downward unit normal vector to the free surface.
We also assume a normal stress balance
\begin{equation}\label{eq:surfpres}
p = p_S,
\end{equation}
with $p_S$ the surface pressure.

$\bullet$ At the bottom $z = z_b$ we consider the no penetration condition
\begin{equation}\label{eq:nopenet}
\vu \cdot \vn^b = 0,
\end{equation}
where $\vn^{b} = (\g_x z_b,-1)/\sqrt{1+\abs{\g_x z_b}^2}$ is the downward unit normal vector to the bottom.\\

\noindent We also consider a Coulomb type fiction law involving the variable friction coefficient $\mu(I)$:

\begin{equation}\label{eq:BC_fondo}
\bS\ \vn^{b} - \left(\left(\bS\ \vn^{b}\right)\cdot\vn^{b}\right)\vn^{b} = \left(\begin{matrix}
\mu(I)p\dfrac{\vu_{H}}{\abs{\vu_{H}}}\\
\\
0
\end{matrix}\right).
\end{equation}


\subsection{Dimensional analysis}\label{se:dimensional_analysis}
In this subsection we carry out a dimensional analysis of the system \eqref{eq:defDu}-\eqref{eq:BC_fondo}. We consider a shallow domain by assuming that the ratio $\varepsilon = H/L$ between the characteristic height $H$ and the characteristic length $L$ is small. We define the dimensionless variables, denoted with the tilde symbol ($\tilde{.}$), as follows:
\begin{equation*}
\label{nondim_var}
\begin{array}{c}
(x,z,t) = (L\td{x},H\td{z},(L/U)\td{t}),
\quad
(\vu_{H},w) =  (U\td{\bd{u}}_{H},\varepsilon U\td{w}),\\
\\
h = H\td{h},
\quad
\r = \r_{0}\td{\r},
\quad
p = \r_{0}U^{2}\td{p},
\quad
p_S = \r_0 U^2 \td{p_S}, \\
\\
\eta= \r_{0}LU\td{\eta},
\quad
\eta_M = \r_{0}LU\td{\eta_M}.
\end{array}
\end{equation*}

\noindent Let us also denote
\begin{equation}\label{eq:DuEps}
D_{\varepsilon}(\vu) = \left( \begin{matrix}
D_{H}(\td{\bd{u}}_{H}) & \frac{1}{\varepsilon}\p_{z}\td{\bd{u}}_{H}+\varepsilon\left(\g_{x}\td{w}\right)'\\
\\
\frac{1}{\varepsilon}\left(\p_{z}\td{\bd{u}}_{H}\right)'+\varepsilon\g_{x}\td{w} &  2\p_{z}\td{w}
\end{matrix}\right),
\end{equation}

\noindent and the Froude number
\[Fr = \frac{U}{\sqrt{gH}}.\]

Then, the system of equations (\ref{eq:NS_3D}) can be re-written using this change of variables as (tildes have been dropped for simplicity):\\

\begin{equation}
\label{eq:nondim_NS_3D}
\left\{
\begin{array}{l}
\g_{x}\cdot \vu_{H}+\p_{z}w\ = \ 0, \\
\\
\r\p_{t}\vu_{H}+\r\g_{x}\cdot (\vu_{H}\otimes\vu_{H})+\r\p_{z}\left(\vu_{H}w\right)+\g_{x}p =\g_{x}\cdot \Big(\eta D_{H}(\vu_{H})\Big)+\p_{z}\Big(\eta \Big(\dfrac{1}{\varepsilon^2}\p_{z}\vu_{H}+\left(\g_{x}w\right)'\Big)\Big), \\
\\
\varepsilon^2\Big(\r\p_{t}w+\r\vu_{H}\g_{x}w+\r w\p_{z}w\Big)+\p_{z}p+\r \dfrac{1}{Fr^2}=\g_{x}\cdot \Big(\eta\Big(\left(\p_{z}\vu_{H}\right)'+\varepsilon^2\g_{x}w\Big)\Big)+2\p_{z}\Big(\eta\p_{z}w\Big).
\end{array}
\right.
\end{equation}

\noindent We also write the boundary and kinematic conditions using dimensionless variables

$\bullet$ At the free surface
\[\p_{t}\left(z_b+h\right) + \vu_{H}|_{z=z_b+h}\cdot\g_{x}\left(z_b+h\right) - w|_{z=z_b+h} = 0;\hspace{1cm}p = p_S.\]

$\bullet$ At the bottom
\begin{equation}
\label{eq:boundCond_bot}
\begin{array}{c}
\vu_H\ \cdot\ \g_x z_b = w\ ;\\
\\
\dfrac{\eta}{\varepsilon}\p_{z}\vu_{H} = \mu(I)p\dfrac{\vu_{H}}{\abs{\vu_{H}}} + O\left(\varepsilon^2\right).
\end{array}
\end{equation}

In addition, we assume an asymptotic regime in the rheology for the friction coefficient $\mu(I)$, namely:
\[\mu(I) = \varepsilon\mu^{0}(I).\]
Consequently,
\begin{equation}\label{eq:etaepsilon}
\eta  =
\varepsilon\ \frac{\mu^{0}(I)p}{\text{max}\left(\|D_{\varepsilon}(\vu_{\a})\|,\frac{\mu(I)p}{\eta_{M}}\right)}  = \varepsilon\eta^{0}.
\end{equation}

\section{A multilayer approach}\label{se:multilayer_approach}
\begin{figure}[!h]
\begin{center}
\includegraphics[width=0.6\textwidth]{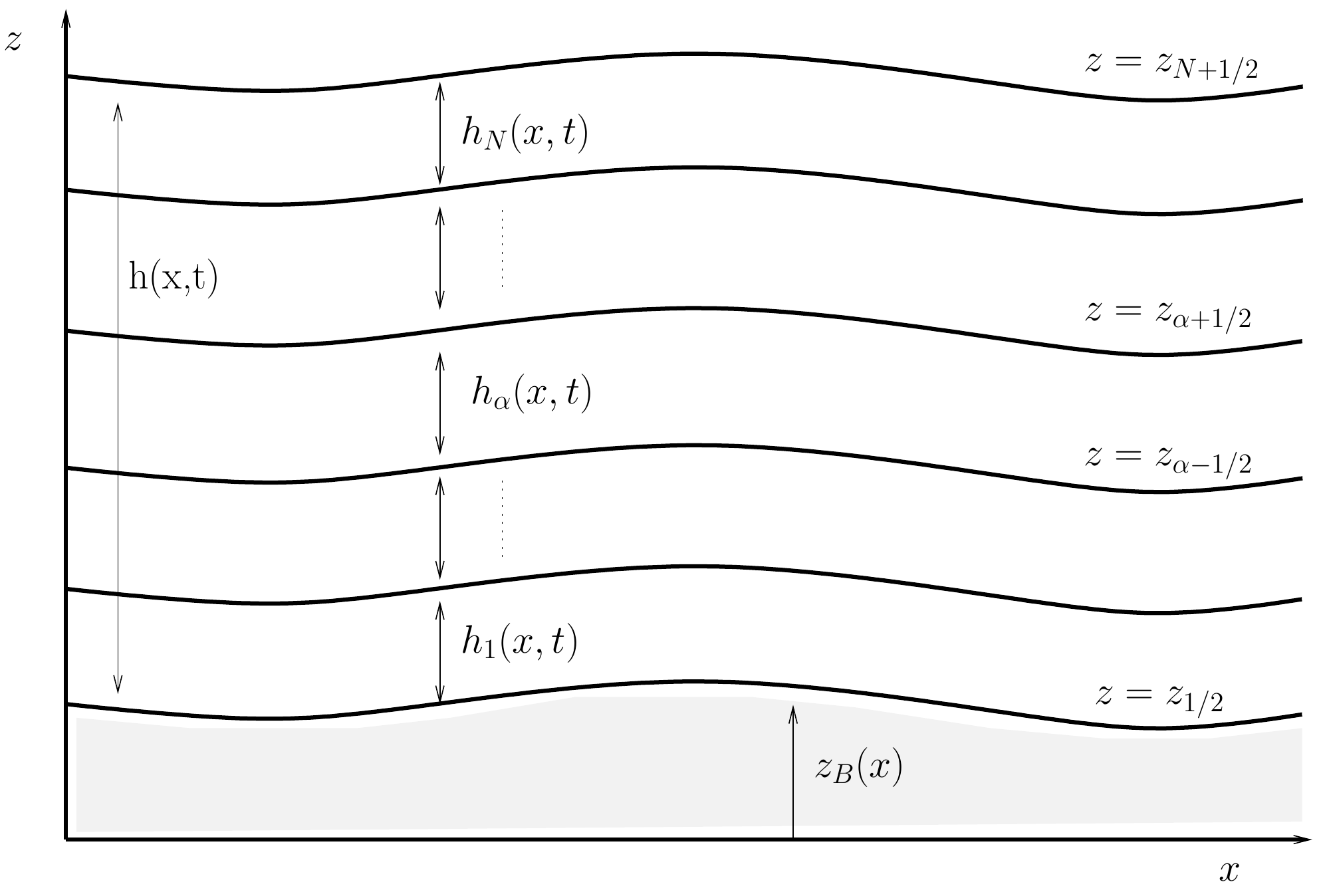}
\end{center}
 \caption{\label{fig:Multilayers} \it{Sketch of the multilayer division of the fluid domain.}}
 \end{figure}

We apply the multilayer approach proposed in \cite{EnriqueMultilayer}. Using the same notation, we denote the fluid domain $\Omega_{F}(t)$ and its projection $I_{F}(t)$ on the horizontal plane, for a positive $t\in [0,T]$, i.e.

 \[I_{F}(t) = \Big\{x\in \mathbb{R}^{2} ; (x,z) \in \O_F(t)\Big\}.\]

This approach considers a vertical partition of the domain in $N\in\mathbb{N}^*$ layers with preset thicknesses $h_\a(t,x)$ (see figure \ref{fig:Multilayers}). Note that $\sum^N_{\a=1}h_\a = h$. These layers are separated by $N+1$ interfaces $\G_{\a+1/2}(t)$, which are described by the equations $z = z_{\a+1/2}(t,x)$ for $\a = 0,1,..,N$, $x\in I_{F}(t)$, where $z_b = z_{1/2}$ and $z_{s} = z_{N+1/2}$ are the bottom $\G_{b}$ and free surface $\G_{s}$ respectively. We assume that these interfaces are smooth enough. Note that $z_{\a+1/2} = z_{b} + \sum_{\b=1}^{\a}h_{\b}$, for $\a = 1,...,N$ and $h_\a = z_{\a+\frac{1}{2}} - z_{\a-\frac{1}{2}}$.\\

  Denoting $\O_\a(t)$ the subdomain between $\G_{\a-1/2}$ and $\G_{\a+1/2}$ and $\Theta_{\a}(t)$ the lateral vertical boundary, for a positive $t\in [0,T]$, we obtain
 \[
\begin{array}{l}

\O_{\a}(t) =  \Big\{(x,z); \; x\in I_{F}(t) \mb{ and } z_{\a-\frac{1}{2}}< z <z_{\a+\frac{1}{2}}\Big\},\\
\p\O_{\a}(t) = \G_{\a-\frac{1}{2}}(t) \cup\G_{\a+\frac{1}{2}}(t)\cup\Theta_{\a}(t), \mb{ with}\\
\Theta_{\a}(t) =  \Big\{(x,z); \; x\in \p I_{F}(t) \mb{ and } z_{\a-\frac{1}{2}}< z <z_{\a+\frac{1}{2}}\Big\}.\\
\\

\end{array}\]

\begin{remark}
We need to introduce a specific notation:
\begin{enumerate}
\item For two tensors $\smash{\bd a}$ and $\smash{\bd b}$ of sizes $\smash(n,m)$ and $\smash(n,p)$, we denote by $\smash {\leftlim \bd a; \;\bd b \rightlim}$ the concatenation of $\smash{\bd a}$ and $\smash{\bd b}$, which is a tensor of size $\smash(n,m+p)$.
\item For a function $\smash f$ and for $\smash\a=0,1,...,N$, we set
  $$ f_{\a+\frac{1}{2}}^{-} := ( f_{|_{\O_{\a}(t)}})_{|_{\G_{\a+\frac{1}{2}}(t)}} \mb{\;\;and\;\;}  f_{\a+\frac{1}{2}}^{+} := (f_{|_{\O_{\a+1}(t)}})_{|_{\G_{\a+\frac{1}{2}}(t)}}.$$
  Note that if the function $\smash f$ is continuous,
  $$ f_{\a+\frac{1}{2}} := f_{|_{\G_{\a+\frac{1}{2}}(t)}} = f_{\a+\frac{1}{2}}^+ = f_{\a+\frac{1}{2}}^-.$$

  \item For a given time $t$, we denote
\begin{equation*}
\label{eq:unit_norm}
\vn_{T,\a+1/2}=\dfrac{\left(\p_t z_{\a+\frac{1}{2}}, \g\!_{x}z_{\a+\frac{1}{2}}, -1\right)^{'}}{\sqrt{1+\abs{\g\!_{x}z_{\a+\frac{1}{2}}}^2 + \left(\p_t z_{\a+\frac{1}{2}}\right)^2}} \hspace{0,3cm}\mbox{and}\hspace{0,3cm} \vn_{\a+1/2}=\dfrac{\left(\g\!_{x}z_{\a+\frac{1}{2}}, -1\right)^{'}}{\sqrt{1+\abs{\g\!_{x}z_{\a+\frac{1}{2}}}^2}}.
\end{equation*}
the space-time unit normal vector and the space unit normal vector to the interface $\smash \Gamma_{\a+1/2}(t)$ outward  to the layer $\smash \Omega_{\a}(t)$  for $\smash\a=0,...,N$.
\\
\end{enumerate}
\end{remark}

For convenience, we write the set of equations (\ref{eq:nondim_NS_3D}) in matricial notation before applying the multilayer approach. First, we focus on the equations of momentum. We multiply the horizontal momentum equation by $\varepsilon$, which gives
\[\begin{array}{l}
\varepsilon\r\p_{t}\vu_{H} +\varepsilon\r\g_{x}\cdot (\vu_{H}\otimes\vu_{H}) +\varepsilon\r\p_{z}\left(\vu_{H}w\right) +\varepsilon\g_{x}p =\varepsilon\g_{x}\cdot \Big(\eta D_{H}(\vu_{H})\Big)+\p_{z}\Big(\eta \Big(\dfrac{1}{\varepsilon}\p_{z}\vu_{H}+\varepsilon\left(\g_{x}w\right)'\Big)\Big),\\
\\
\varepsilon^2\Big(\r\p_{t}w+\r\vu_{H}\g_{x}w+\r w\p_{z}w\Big)+\p_{z}p +\r \dfrac{1}{Fr^2}=\g_{x}\cdot \Big(\eta\Big(\left(\p_{z}\vu_{H}\right)'+\varepsilon^2\g_{x}w\Big)\Big)+2\p_{z}\Big(\eta\p_{z}w\Big).
\end{array}\]

\noindent Note that the terms involving the stress tensor (without divergence operator) are:

\[\eta\left( \begin{matrix}
\varepsilon D_{H}(\vu_{H}) & \frac{1}{\varepsilon}\p_{z}\vu_{H}+\varepsilon\left(\g_{x}w\right)'\\
\\
\left(\p_{z}\vu_{H}\right)'+\varepsilon^{2}\g_{x}w &  2\p_{z}w
\end{matrix}\right)= \eta\,D_{\varepsilon}(\vu)\,\left(\begin{matrix}
\varepsilon \bd{I}_2 & 0\\
\\
0 & 1
\end{matrix}\right),\]
where $D_{\varepsilon}(\vu)$ is defined by \eqref{eq:DuEps} and $I_2$ is the two-dimensional identity matrix. We introduce the notation:
\[\begin{array}{l}
\vu_{\varepsilon} = (\vu_{H},\varepsilon w)',\\
\\
\vec{\bd{f}} = \left(0,\dfrac{1}{Fr^{2}}\right)' \text{ with }\dfrac{1}{Fr^{2}} = \dfrac{gH}{U^{2}},\\
\\
\mathcal{E} = \left(\begin{matrix}
\varepsilon \bd{I}_2 & 0\\
\\
0 & 1
\end{matrix}\right).\\
\end{array}\]

\noindent With this notation, we can write the momentum equation as follows
\[\varepsilon\r\p_{t}\vu_{\varepsilon} + \varepsilon\r\g\cdot (\vu_{\varepsilon}\otimes\vu) + \g\cdot(p\mathcal{E}) + \r\vec{\bd{f}} = \g\cdot (\eta D_{\varepsilon}(\vu)\mathcal{E}),\]
and we obtain the set of equations (\ref{eq:nondim_NS_3D}) in matricial notation:
\begin{equation}
\label{eq:NS_matricial}
\left\{
\begin{array}{l}
\g \cdot \vu = \ 0,\\
\\
\r\p_{t}\vu_{\varepsilon}\;+\;\r\g\cdot (\vu_{\varepsilon}\otimes\vu) - \dfrac{1}{\varepsilon}\g\cdot(\S\mathcal{E})\; = \;-\dfrac{1}{\varepsilon}\r\vec{\bd{f}},
\end{array}
\right.
\end{equation}
where now $\S = -p\bd{I} + \eta D_{\varepsilon}(\vu).$\\

In subsections \ref{se:weak_sol} and \ref{se:Stress_tens} we define the weak solutions for our system and the process to calculate the vertical velocities is presented in subsection \ref{se:vert_vel}.\\

\subsection{Weak solution with discontinuities}\label{se:weak_sol}

\medskip

\par Following \cite{EnriqueMultilayer}, we look for a weak solution $(\vu,p,\r)$ of \eqref{eq:NS_3D}. We assume that the velocity $\vu$, the pressure $p$ and the density $\r$ are smooth in each $\Omega_\a(t)$ but may be discontinuous across the interfaces $\G_{\a+1/2}$ for $\a = 1,...,N-1$. Then the following conditions must hold:
\begin{itemize}
  \item[(i)] $(\vu,p,\r)$ is a standard weak solution of \eqref{eq:NS_matricial} in each layer $\O_{\a}(t)$.
  \item[]
  \item[(ii)] $(\vu,p,\r)$ satisfies the normal flux jump conditions at  $\G_{\a+\frac{1}{2}}(t)$, for $\alpha=0,\dots, N$ for the mass and momentum laws:
  \begin{itemize}
  \item[$\bullet$] Mass conservation law,
\color{black}
    \begin{equation}
\label{eq:mass_interfCond_gen}
\left[ \leftlim\r; \;\r\vu \rightlim\right]_{|_{\G_{\a+\frac{1}{2}}(t)}}\!\!\!\cdot\;\;\;\vn_{T,\a+1/2}  = 0,
\end{equation}

  \item[]
  \item[$\bullet$] Momentum conservation law,

 \color{black}
  \begin{equation}
\label{eq:mom_interfCond_gen}
\Big[\leftlim \r\vu_{\varepsilon}; \;\r\vu_{\varepsilon}\otimes\vu - \frac{1}{\varepsilon}\bS\mathcal{E}\rightlim\Big]_{|_{\G_{\a+\frac{1}{2}}(t)}}\!\!\!\!\!\cdot\;\;\;\vn_{T,\a+1/2}  = 0,
\end{equation}

where $\left[ \leftlim a; \;b \rightlim\right]_{|_{\G_{\a+\frac{1}{2}}(t)}}$ denotes the jump of the pair $\leftlim a;\;b \rightlim$ across $\G_{\a+\frac{1}{2}}(t)$,
$$
\left[ \leftlim a; \;b \rightlim\right]_{|_{\G_{\a+\frac{1}{2}}(t)}}=\left( \leftlim a; \;b\rightlim _{|_{\O_{\a+1}(t)}}  - \leftlim a; \;b \rightlim _{|_{\O_{\a}(t)}}\right)_{|_{\G_{\a+\frac{1}{2}}(t)}}.
$$
\end{itemize}
\end{itemize}

\par We consider a particular family of velocity functions by assuming that the thickness of each layer is small enough to make the horizontal velocities independent of the vertical variable $z$. From this and the incompressibility condition in each layer, we obtain that vertical velocities are linear in $z$ and may be discontinuous, that is
 $$ \vu_{|_{\O_{\a}(t)}} := \vu_{\a} := (\vu_{H,\a}, w_{\a})^{'}
$$
where $\vu_{H,\a}$ and $w_{\a}$ are the horizontal and vertical velocities, respectively, on layer $\alpha$, and the particular family satisfies

\begin{equation}
\label{eq:strucvp}
\partial_z \vu_{H,\a}=0;  \quad \partial_z w_{\a}=d_\a(t,x)
\end{equation}
for some smooth function $d_\a(t,x)$. 
Note that
\begin{equation*}
\label{eq:horiz_vel}
\vu_{H,\a-\frac{1}{2}}^+(t,x) \;=\; \vu_{H,\a+\frac{1}{2}}^-(t,x) \;=\; \vu_{H,\a}(t,x).
\end{equation*}

\noindent Let us denote

\begin{equation}\label{eq:defG}
  \left\{ \begin{array}{l}
\displaystyle   G_{\a+\frac{1}{2}}^{\, +}=\p_{t}z_{\a+\frac{1}{2}} + \vu_{H,\a+1}\cdot\g\!_{x}z_{\a+\frac{1}{2}} - w_{\a+\frac{1}{2}}^{+}, \\ \\
\displaystyle  G_{\a+\frac{1}{2}}^{\, -}=\p_{t}z_{\a+\frac{1}{2}} + \vu_{H,\a}\cdot\g\!_{x}z_{\a+\frac{1}{2}} - w_{\a+\frac{1}{2}}^{-}.
\end{array} \right.
  \end{equation}
Then $\vu$ satisfies the jump conditions for the mass conservation law (\ref{eq:mass_interfCond_gen}) if both coincide. In this case we set\\
\begin{equation}
\label{eq:mass_interfCond}
G_{\a+\frac{1}{2}}:=  \, G_{\a+\frac{1}{2}}^{\, -}=G_{\a+\frac{1}{2}}^{\, +},
\end{equation}

\noindent Note that  $G_{\a+\frac{1}{2}}$ is the normal mass flux at the interface $\smash\G_{\a+\frac{1}{2}}(t)$.\\

Moreover, using (\ref{eq:mass_interfCond}), the momentum conservation jump conditions (\ref{eq:mom_interfCond_gen}) can be written in terms of the normal mass flux as

\begin{equation}\label{eq:mom_cond_salto}
 \begin{array}{l}
\displaystyle \Big[\dfrac{1}{\varepsilon}\bS\mathcal{E}\Big]_{|_{\G_{\a+\frac{1}{2}}(t)}}   \left( \g\!_{x}z_{\a+\frac{1}{2}}, - 1\right)   =
\displaystyle \Big[\leftlim \r\vu_{\varepsilon}; \;\r\vu_{\varepsilon}\otimes\underline{\vu}  \rightlim \Big]_{|_{\G_{\a+\frac{1}{2}}(t)}}\!\!\!\!\!\;  \left(\p_{t}z_{\a+\frac{1}{2}}, \g\!_{x}z_{\a+\frac{1}{2}}, -1\right).
\end{array}
\end{equation}

\noindent Also, by   (\ref{eq:mass_interfCond}),

\[\begin{array}{l}
\displaystyle \Big[\leftlim \r\vu_{\varepsilon}; \;\r\vu_{\varepsilon}\otimes\vu  \rightlim \Big]_{|_{\G_{\a+\frac{1}{2}}(t)}} = \displaystyle \left[\left( \begin{matrix}
\r \vu_{H} & \r \vu_{H}\otimes\vu_{H} & \r \vu_{H}w\\
\varepsilon\r w & \varepsilon\r \vu_{H}w & \varepsilon\r w^{2}
\end{matrix} \right) \right]_{|_{\G_{\a+\frac{1}{2}}(t)}} \, =\\
\\
=\left(\begin{matrix}
\bd{I}_2 &0\\
0 & \varepsilon
\end{matrix}\right) \left[\left( \begin{matrix}
\r \vu_{H} & \r \vu_{H}\otimes\vu_{H} & \r \vu_{H}w\\
\r w & \r \vu_{H}w & \r w^{2}
\end{matrix} \right) \right]_{|_{\G_{\a+\frac{1}{2}}(t)}}\,
= \mathbb{B}\ \Big[\leftlim \r\vu; \;\r\vu\otimes\vu\rightlim \Big]_{|_{\G_{\a+\frac{1}{2}}(t)}},
\end{array}\]

\noindent with $\mathbb{B} = \varepsilon\mathcal{E}^{-1}$. Using (\ref{eq:mass_interfCond}) we deduce
$$
\begin{array}{l}
\displaystyle \Big[\leftlim \r\vu; \;\r\vu\otimes\vu \rightlim \Big]_{|_{\G_{\a+\frac{1}{2}}(t)}}\!\!\!\!\!\;  \left(\p_{t}z_{\a+\frac{1}{2}}, \g\!_{x}z_{\a+\frac{1}{2}}, -1\right)
\displaystyle = {\r\,G_{\a+\frac{1}{2}}}\,\left[\vu\right]_{|_{\G_{\a+\frac{1}{2}}(t)}}.
\end{array}
$$

\noindent Therefore,
\[ \begin{array}{l}
\displaystyle \Big[\leftlim \r\vu_{\varepsilon}; \;\r\vu_{\varepsilon}\otimes\vu \rightlim \Big]_{|_{\G_{\a+\frac{1}{2}}(t)}}\!\!\!\!\!\;  \left(\p_{t}z_{\a+\frac{1}{2}}, \g\!_{x}z_{\a+\frac{1}{2}}, -1\right)
\displaystyle = \displaystyle \mathbb{B}\Big[\leftlim \r\vu; \;\r\vu\otimes\vu \rightlim \Big]_{|_{\G_{\a+\frac{1}{2}}(t)}}\!\!\!\!\!\;  \left(\p_{t}z_{\a+\frac{1}{2}}, \g\!_{x}z_{\a+\frac{1}{2}}, -1\right)
\displaystyle\\
\\
=\r\,G_{\a+\frac{1}{2}}\,\mathbb{B}\left[\vu\right]_{|_{\G_{\a+\frac{1}{2}}(t)}}.
\end{array}\]

\noindent Finally, we obtain, from the previous equality and from \eqref{eq:mom_cond_salto}, the momentum jump condition

\begin{equation}
\label{eq:mom_interfCond}
\dfrac{1}{\varepsilon}\Big[\bS\mathcal{E}\Big]_{|_{\G_{\a+\frac{1}{2}}(t)}}\vn_{\a+\frac{1}{2}} = \dfrac{\r\,G_{\a+\frac{1}{2}}}{\sqrt{1+\abs{\g\!_{x}z_{\a+\frac{1}{2}}}^2 }}\,\mathbb{B}\left[\vu\right]_{|_{\G_{\a+\frac{1}{2}}(t)}}.
\end{equation}

\subsection{Stress tensor approximation}\label{se:Stress_tens}

\noindent For $\a =1, ..., N-1$, the total stress is written
\begin{equation*} \label{newsigma}
\bS_{ \a+\frac{1}{2}}^\pm \;=\; - {p_{\a+\frac{1}{2}}}\,\bd{I} + \bd{T}_{\a+\frac{1}{2}}^\pm,
\end{equation*}
where $p_{\a+\frac{1}{2}} = p_{\a+\frac{1}{2}}^{+} = p_{\a+\frac{1}{2}}^{-}$ is the pressure and $\bd{T}_{\a+\frac{1}{2}}^\pm$ are approximations of $\eta D_{\varepsilon}(\vu_{\a})$ at $\Gamma_{\a+\frac{1}{2}}$. By the momentum jump condition \eqref{eq:mom_interfCond_gen}, rewritten as \eqref{eq:mom_interfCond}, $\bd{T}_{\a+\frac{1}{2}}^\pm$ must satisfy

\begin{equation} \label{apa4_defdifsigma_b}
\frac{1}{\varepsilon}\Big(  \bd{\S}_{ \a+\frac{1}{2}}^+ - \bd{\S}_{\a+\frac{1}{2}}^-\Big)\ \mathcal{E}\ \vn_{\a+\frac{1}{2}} = \frac{1}{\varepsilon}\Big(  \bd{T}_{\a+\frac{1}{2}}^+ - \bd{T}_{\a+\frac{1}{2}}^-\Big)\ \mathcal{E}\ \vn_{\a+\frac{1}{2}} = \dfrac{\r\,G_{\a+\frac{1}{2}}}{\sqrt{1+\abs{\g\!_{x}z_{\a+\frac{1}{2}}}^2 }}\,\ \mathbb{B}\ \left[\vu\right]_{|_{\G_{\a+\frac{1}{2}}(t)}}.
\end{equation}

Moreover, by consistency, we consider the following condition
\begin{equation} \label{apa4_defsumsigma_b}
\frac{1}{2} \Big(  \bd{T}_{ \a+\frac{1}{2}}^+ + \bd{T}_{\a+\frac{1}{2}}^-\Big) =\widetilde{\bd{T}}_{\a+\frac{1}{2}} ,
\end{equation}
where
\begin{equation*}\label{apa4_defsumsigma_b_2}
\tag{\ref{apa4_defsumsigma_b}$'$} \widetilde{\bd{T}}_{\a+\frac{1}{2}} = \eta \widetilde{D}_{\varepsilon,\a +\frac{1}{2}}
\end{equation*}
is an approximation of $\eta D_{\varepsilon}(\vu_{\a})_{|\Gamma_{\a+\frac{1}{2}}}$. Concretely, we set
\begin{equation*} \label{apa4_defsumsigma_b_3}
\tag{\ref{apa4_defsumsigma_b}$''$}\widetilde{D}_{\varepsilon,\a+\frac{1}{2}}=
\left(
\begin{array}{ccc}
\displaystyle D_{H} \left( \frac{\vu_{H,\a+\frac{1}{2}}^+ + \vu_{H,\a+\frac{1}{2}}^-}{2} \right) &\quad &
\widetilde{D}_{\varepsilon,\a+\frac{1}{2},xz}
 \\
&&\\
\left( \widetilde{D}_{\varepsilon,\a+\frac{1}{2},xz} \right)' & \quad & 2 \, Q_{v,\a+\frac{1}{2}} \\
\end{array}
\right),
\end{equation*}
where,
$$
\widetilde{D}_{\varepsilon,\a+\frac{1}{2},xz}=
\varepsilon\left( \g\!_{x} \left( \dfrac{w_{\a+\frac{1}{2}}^+ + w_{\a+\frac{1}{2}}^-}{2} \right)  \right)'+ \displaystyle \dfrac{1}{\varepsilon}\Q_{H,\a+\frac{1}{2}},
$$
and $(\vec{\textbf{Q}}_{H,\a+\frac{1}{2}},Q_{v,\a+\frac{1}{2}})$ is defined as follows.\\

\noindent We approximate the second order derivatives in $z$ using a mixed formulation because of the possible vertical discontinuous profile. We set an additional auxiliary unknown $\Q$ that satisfies
\begin{equation*} \label{def_q}
\Q-\partial_z \vu=0, \quad \mbox{with} \quad \Q=(\Q_H,Q_v).
\end{equation*}
And to approximate $\Q$, we approximate $\vu$ by $\widetilde{\vu}$, a $\mathbb P_1(z)$ interpolation such that $\widetilde{\vu}_{|z=\frac{1}{2}(z_{\a-\frac{1}{2}}+z_{\a+\frac{1}{2}})}={\vu_{\alpha}}$. Then $\Q_{\a+\frac{1}{2}} = \left(\Q_{H,\a+\frac{1}{2}},Q_{v,\a+\frac{1}{2}}\right)$ is an approximation of  $\Q(\widetilde{\vu})$ at $\Gamma_{\a+\frac{1}{2}}$.\\

\noindent Finally, we multiply equation (\ref{apa4_defdifsigma_b}) by $\varepsilon/2$ and multiply scalarly (\ref{apa4_defsumsigma_b}) by vector $\mathcal{E}\ \vn_{\a+\frac{1}{2}}$. Then, we get a linear system whose unknowns are $\bd{T}_{ \a+\frac{1}{2}}^{\pm}\  \mathcal{E}\ \vn_{\a+\frac{1}{2}}$. As a result we obtain
\begin{equation} \label{rk_momemtum_prop}
\begin{array}{ll}
  \bd{T}_{ \a+\frac{1}{2}}^{\pm}\  \mathcal{E}\ \vn_{\a+\frac{1}{2}}= & \displaystyle \widetilde{\bd{T}}_{\a+\frac{1}{2}}\mathcal{E} \ \vn_{\a+\frac{1}{2}}
   \displaystyle \pm \frac{1}{2} \dfrac{\varepsilon\r G_{\a+\frac{1}{2}}}{\sqrt{1+\abs{\g\!_{x}z_{\a+\frac{1}{2}}}^2 }}\mathbb{B}\ \left[\vu\right]_{|_{\G_{\a+\frac{1}{2}}(t)}},
\end{array}
\end{equation}
where $\widetilde{\bd{T}}_{\a+\frac{1}{2}}$ is defined by \eqref{apa4_defsumsigma_b_2}-\eqref{apa4_defsumsigma_b_3}.

\subsection{Vertical velocity}\label{se:vert_vel}
Let us recall the velocity structure requirements set in equations \eqref{eq:strucvp}. This makes the vertical velocity linear in $z$ in each layer. Concretely, if $\vu_\a$ is a solution of system \eqref{eq:NS_matricial} in $\O_{\a}(t)$, the vertical velocity can be recovered by integrating the continuity equation between $z_{\a-\frac{1}{2}}$ and $z \in (z_{\a-\frac{1}{2}}, z_{\a+\frac{1}{2}})$,

$$
w_{\a}(t,x,z) \;=\; w_{\a-\frac{1}{2}}^+(t,x) \;-\; (z-z_{\a-\frac{1}{2}})\g\!_{x}\cdot\vu_{H,\a}(t,x),
\quad \mbox{for } \smash\a=1,...,N.
$$

\noindent Moreover, from conditions \eqref{eq:mass_interfCond} at the interfaces, we obtain the relation
\begin{equation} \label{wdiscon}
w_{\a+\frac{1}{2}}^{+} \;=\; (\vu_{H,\a+1} - \vu_{H,\a})\cdot\g\!_{x}z_{\a+\frac{1}{2}} \;+\; w_{\a+\frac{1}{2}}^{-}.
\end{equation}

\noindent We therefore use the horizontal velocities deduced from the model to compute the vertical velocities in the layers following the algorithm:
\\

\begin{itemize}
\item From the mass transfer $G_{1/2}$, which is given as data, we obtain $\smash w_{\frac{1}{2}}^{+}$ using condition \eqref{eq:mass_interfCond} at the bottom,
 $$
 w_{\frac{1}{2}}^{+} = \vu_{H,1}\cdot\g\!_{x}z_{B} + \p_{t}z_{B} - G_{\frac{1}{2}}.
 $$
\item Then, for $\smash\a=1,...,N$ and $z \in (z_{\a-\frac{1}{2}}, z_{\a+\frac{1}{2}})$, we set
\begin{equation}
\label{eq:vert_vel}
\left\{
\begin{array}{l}
 w_{\a}(t,x,z) \;=\; w_{\a-\frac{1}{2}}^+(t,x) \;-\; (z-z_{\a-\frac{1}{2}})\g\!_{x}\cdot\vu_{H,\a}(t,x),\\
    \\
     \\
  w_{\a+\frac{1}{2}}^{+} \;=\; (\vu_{H,\a+1} - \vu_{H,\a})\cdot\g\!_{x}z_{\a+\frac{1}{2}} \;+\; w_{\a+\frac{1}{2}}^{-};\\
  \\
\end{array}
\right.
\end{equation}
where
$$
 w_{\a+\frac{1}{2}}^{-} =   {w_{\alpha}}_{|_{\Gamma_{\a+1/2}(t)}}=w_{\a-\frac{1}{2}}^{+} - h_{\a}\g\!_{x}\cdot\vu_{H,\a}.\\
$$

\end{itemize}

\noindent In this way, the velocity vector $\vu$ is the piecewise smooth function such that $\vu(t,x,z)|_{\O_{\a}(t)}=\vu_{\a}(t,x,z)$ for $\smash\a=1,...,N$, where
\begin{equation}
\label{eq:vel_vec}
\vu_{\a}(t,x,z) = \left(\vu_{H,\a}(t,x), \;w_{\a-\frac{1}{2}}^+(t,x) \;-\; (z-z_{\a-\frac{1}{2}})\g\!_{x}\cdot\vu_{H,\a}(t,x)\right)^{'},
\end{equation}
and $w_{\a-\frac{1}{2}}^+(t,x)$ is computed using (\ref{eq:vert_vel}). \\

\section{Weak solution of the first order model}
\label{se:weakalfa}
In this section we derive the model of order $\varepsilon$.


\subsection{Pressure}\label{se_sub:pressure}
Using the vertical momentum equation in (\ref{eq:nondim_NS_3D}) we can deduce an expression for the pressure. We write this equation up to order $\varepsilon^2$ for each layer:
\[\p_{z}p_{\a} = -\r\dfrac{1}{Fr^{2}} + \g_{x}\cdot\left(\eta\p_{z}\vu_{H,\a}\right)+ \p_{z}\left(2\eta\p_{z}w_{\a}\right).\]

\noindent Taking into account the requirements $\p_{z}\vu_{H,\a} = 0$ (see \eqref{eq:strucvp}) and $\eta = \varepsilon\eta^{0}$ (see \eqref{eq:etaepsilon}), we get:
\[\p_{z}p_{\a} = -\r\dfrac{1}{Fr^{2}} + \varepsilon\p_{z}\left(2\eta^{0}\p_{z}w_{\a}\right).\]
Then, we obtain the hydrostatic pressure framework (up to order $\varepsilon$) in each layer:

\[\p_{z} p_{\a} = -\r \dfrac{1}{Fr^{2}}.\]

\noindent Now, by  the continuity of the dynamic pressure (see \cite{EnriqueMultilayer}), we can deduce that

\begin{equation}
\label{eq:pressure}
p_{\a}(z) = p_{S}+ \dfrac{\r}{Fr^{2}}(z_{b}+h-z),
\end{equation}
where $p_S$ is the pressure at the free surface.\\

\subsection{A particular weak solution}
Noting that $\vu_{\a}$ is a weak solution of the system \eqref{eq:NS_matricial} in $\O_{\a}(t)$, let us consider the weak formulation of \eqref{eq:NS_matricial} in $\O_{\a}(t)$ for  $\smash\a=1,...,N$. Assuming $\vu_{\a} \in L^2(0,T;H^1(\O_{\a}(t))^3)$, $\partial_t \vu_{\a} \in L^2(0,T;L^2(\O_{\a}(t))^3)$ and $p_{\a} \in L^2(0,T;L^2(\O_{\a}(t)))$, then a weak solution  in $\O_{\a}(t)$ should satisfy

\color{black}
\begin{equation} \label{eq:weak_mom}
\left\{
\begin{array}{rl}
0 = &  \dint_{\O_{\a}(t)} (\g\cdot {\vu_{\a}})\,\varphi \,d\O, \\
&\\
\displaystyle -\frac{1}{\varepsilon}\dint_{\O_{\a}(t)}\r\vec{\bd{f}}\cdot\vv\,d\O = & \dint_{\O_{\a}(t)}\r\p_{t}\vu_{\varepsilon,\a}\cdot\vv\,d\O + \dint_{\O_{\a}(t)}\r\Big(\vu_{\varepsilon,\a}\cdot\g\vu_{\a}\Big)\cdot\vv\,d\O \, +\\
& \\
&  + \displaystyle \frac{1}{\varepsilon}\dint_{\O_{\a}(t)} \left( \g\cdot \left(p_{\a}\mathcal{E}\right) \right)\cdot \vv\ d\O  - \displaystyle \frac{1}{\varepsilon}\dint_{\O_{\a}(t)} \left( \g\cdot \left(\bd{T}_{\a}\ \mathcal{E}\right) \right)\cdot \vv\ d\O ,\end{array}
\right.
\end{equation}
for all $ \varphi \in L^2(\O_{\a}(t))$ and for all $ \vv \in H^1(\O_{\a}(t))^3$.\\

We consider unknowns, velocities and pressures, that satisfy (\ref{eq:strucvp}) and the system (\ref{eq:weak_mom}) for test functions such that
\[\p_{z} \varphi = 0\]
and
\begin{equation}
\label{eq:test_vec}
\vv(t,x,z) = \Big(\vv_{H}(t,x), \;(z-z_{b})\,V(t,x)\Big)^{'},\quad\quad \vv_{|_{\p I_{F}(t)}} = 0,
\end{equation}
where $\vv_{H}$ and $V(t,x)$ are smooth functions that do not depend on $z$.\\

We will now develop \eqref{eq:weak_mom} in order to obtain the mass and momentum conservation equations that satisfy the weak solution for this family of test functions for each layer.

\subsubsection*{Mass conservation}
Let $\varphi = \varphi(t, x)$ a scalar test function and $\vu_\a$ a weak solution of (\ref{eq:weak_mom}), from the mass conservation equation we get

$$
\begin{array}{rl}
0 = & \dint_{\O_{\a}(t)} (\g\cdot\vu_{\a})\,\varphi \,d\O  \, =\\
      &  \\
   = & \dint_{I_{F}(t)}\varphi(t, x)\Bigg(\dint_{z_{\a-\frac{1}{2}}}^{z_{\a+\frac{1}{2}}}\Big(\g\!_{x}\cdot\vu_{H,\a} +\p_{z}w_\a \Big)dz\Bigg)\,dx \, =  \\
      &  \\
      =& \dint_{I_{F}(t)}\varphi(t, x)\Bigg(\dint_{z_{\a-\frac{1}{2}}}^{z_{\a+\frac{1}{2}}}\g\!_{x}\cdot\vu_{H,\a}dz +w_{\a+\frac{1}{2}}^{-} - w_{\a-\frac{1}{2}}^{+}\Bigg)\,dx,\\
      \\
\end{array}
$$
for all $\a=1,...,N$. Using Leibnitz's rule
\[\dint_{z_{\a-\frac{1}{2}}}^{z_{\a+\frac{1}{2}}}\g\!_{x}\cdot\vu_{H,\a}dz = \g_{x}\cdot\left( \dint_{z_{\a-\frac{1}{2}}}^{z_{\a+\frac{1}{2}}}\vu_{H,\a}dz\right) - \vu_{H,\a+\frac{1}{2}}^{-}\cdot\g_{x}z_{\a+\frac{1}{2}} + \vu_{H,\a-\frac{1}{2}}^{+}\cdot\g_{x}z_{\a-\frac{1}{2}},\]
and this leads to
\[
\dint_{I_{F}(t)}\varphi(t, x)\,\Bigg(\g\!_{x}\cdot \left(h_{\a}\vu_{H,\a}\right) -\vu_{H,\a}\cdot\g\!_{x}z_{\a+\frac{1}{2}} + w_{\a+\frac{1}{2}}^- +\; \vu_{H,\a}\cdot\g\!_{x}z_{\a-\frac{1}{2}} \;-\; w_{\a-\frac{1}{2}}^+\Bigg)\,dx = 0.
\]
\\
\noindent Taking into account (\ref{eq:mass_interfCond}) and $\p_{t}h_{\a} = \p_{t}z_{\a+\frac{1}{2}} - \p_{t}z_{\a-\frac{1}{2}}$ we obtain the equation

\begin{equation*}
\label{eq:mass_flux_weak}
\begin{array}{rl}
\dint_{I_{F}(t)}\varphi(t, x)\,\Bigg(\p_{t}h_{\a} \;+\; \g\!_{x}\cdot \left (h_{\a}\vu_{H,\a}\,\right) -G_{\a+\frac{1}{2}}+G_{\a-\frac{1}{2}}\Bigg)\,\,dx & = 0,\\
\end{array}
\end{equation*}
for all $\varphi(t, .) \in L^2(I_{F}(t))$. Thus we get the mass conservation law for each layer

\begin{equation}
\label{eq:mass_flux}
\p_{t}h_{\a} + \g\!_{x}\cdot(h_{\a}\vu_{H,\a}) = G_{\a+\frac{1}{2}} - G_{\a-\frac{1}{2}},
\quad  \a=1,...,N
\end{equation}
 where $G_{N+1/2}$ and $G_{1/2}$ stand for the mass exchange with the free surface and the bottom respectively and both should be given data.
\color{red}

\bigskip
\color{black}
\noindent
\subsubsection*{Momentum conservation}
First we develop the variational formulation of momentum equation taking into account that\\
\\
\noindent $\bullet$ $
\dint_{\O_{\a}(t)}\g\cdot \left(p_{\a}\mathcal{E}\right)\cdot \vv\ d\O \ =\\$
\\
\[
\begin{array}{l}
= - \dint_{\O_{\a}(t)} \left( p_{\a}\mathcal{E} \right) : \g\vv d\O\ - \dint_{\G_{\a+\frac{1}{2}}(t)}\left( p_{\a+\frac{1}{2}}\mathcal{E}\ \vv \right)\cdot \vn_{\a+\frac{1}{2}} d\G\ + \dint_{\G_{\a-\frac{1}{2}}(t)} \left( p_{\a-\frac{1}{2}}\mathcal{E} \vv\right) \cdot \vn_{\a-\frac{1}{2}}\ d\G\ .
\end{array}\]

\noindent $\bullet$ $
\dint_{\O_{\a}(t)}\g\cdot \left(\bd{T}_{\a}\mathcal{E}\right)\cdot \vv\ d\O = - \dint_{\O_{\a}(t)}\left(\bd{T}_{\a}\cdot\mathcal{E}\right) : \g\vv\ d\O\  - $
\\
\[
\begin{array}{l}
\\
-\dint_{\G_{\a+\frac{1}{2}}(t)}\left(\left( \bd{T}_{\a+\frac{1}{2}}^{-}\mathcal{E}\right) \vv \right) \cdot \vn_{\a+\frac{1}{2}} d\G\ + \dint_{\G_{\a-\frac{1}{2}}(t)}\left(\left( \bd{T}_{\a-\frac{1}{2}}^{+}\mathcal{E}\right) \vv \right) \cdot \vn_{\a-\frac{1}{2}}\ d\G\ .
\end{array}\]

Now we can write the momentum equation as\\
\begin{equation}\label{eq:mom2}
\begin{array}{rl}
\displaystyle -\frac{1}{\varepsilon}\dint_{\O_{\a}(t)}\r\vec{\bd{f}}\cdot\vv\,d\O = & \dint_{\O_{\a}(t)}\r\p_{t}\vu_{\varepsilon,\a}\cdot\vv\,d\O + \dint_{\O_{\a}(t)}\r\Big(\vu_{\varepsilon,\a}\cdot\g\vu_{\a}\Big)\cdot\vv\ d\O\,- \\
& \\
& -\displaystyle \frac{1}{\varepsilon}\dint_{\O_{\a}(t)} \left( p_{\a}\mathcal{E}\right) : \g\vv\ d\O  + \displaystyle \frac{1}{\varepsilon}\dint_{\O_{\a}(t)}\left(\bd{T}_{\a}\mathcal{E}\right): \g\vv\ d\O\,+\\
& \\
& + \displaystyle \frac{1}{\varepsilon}\dint_{\G_{\a+\frac{1}{2}}(t)}\left(\left( -p_{\a+\frac{1}{2}}\mathcal{E}+\bd{T}_{\a+\frac{1}{2}}^{-}\mathcal{E}\right) \vn_{\a+\frac{1}{2}} \right) \cdot \vv\ d\G\,-\\
 &\\
& -\displaystyle \frac{1}{\varepsilon}\dint_{\G_{\a-\frac{1}{2}}(t)}\left(\left( -p_{\a-\frac{1}{2}}\mathcal{E}+\bd{T}_{\a-\frac{1}{2}}^{+}\mathcal{E}\right) \vn_{\a-\frac{1}{2}} \right) \cdot \vv\ d\G.
\end{array}
\end{equation}

Let $\vv \in H^1(\O_{\a})$ be a test function satisfying (\ref{eq:test_vec}). We develop the momentum equation in (\ref{eq:mom2}) by integrating with respect to the variable $z$ and by identifying the horizontal and vertical component of the vector test function $\vv$. In addition, taking into account the hydrostatic pressure framework, we can leave out the equation corresponding to the vertical component. This is equivalent to considering the vector test function where the vertical component vanishes, i.e. $\vv = \left(\vv_H(t,x),0\right)$. Therefore, the horizontal momentum equation reads, for a weak solution $\vu$ and for all $\a=1,...,N$:

\begin{equation} \label{eq:hmomentum_flux_weak}
\begin{array}{l}
\displaystyle -\frac{1}{\varepsilon}\dint_{\O_{\a}(t)}\r\vec{\bd{f}}\cdot (\vv_{H},0)\,d\O = \dint_{\O_{\a}(t)}\r\p_{t}(\vu_{H,\a},\varepsilon w_{\a})\cdot (\vv_{H},0)\,d\O \ + \\
\\
+\ \dint_{\O_{\a}(t)}\r\Big( (\vu_{H,\a},\varepsilon w_{\a})\cdot\g (\vu_{H},w_{\a})\Big)\cdot (\vv_{H},0)\,d\O \ -\\
\\
-\displaystyle \frac{1}{\varepsilon}\dint_{\O_{\a}(t)}
\left( p_{\a}\mathcal{E}\right) : \g (\vv_{H},0)\,d\O  + \displaystyle \frac{1}{\varepsilon}\dint_{\O_{\a}(t)}\left(\bd{T}_{\a}\mathcal{E}\right): \g (\vv_{H},0)\,d\O \ +\\
\\
+ \displaystyle \frac{1}{\varepsilon}\dint_{\G_{\a+\frac{1}{2}}(t)}\left(\left( -p_{\a+\frac{1}{2}}\mathcal{E}+\bd{T}_{\a+\frac{1}{2}}^{-}\mathcal{E}\right) \vn_{\a+\frac{1}{2}} \right) \cdot (\vv_{H},0)\,d\G\ -\\
\\
 -\displaystyle \frac{1}{\varepsilon}\dint_{\G_{\a-\frac{1}{2}}(t)}\left(\left( -p_{\a-\frac{1}{2}}\mathcal{E}+\bd{T}_{\a-\frac{1}{2}}^{+}\mathcal{E}\right) \vn_{\a-\frac{1}{2}} \right) \cdot (\vv_{H},0)\,d\G.
\end{array}
\end{equation}
\\

\noindent We develop each term of this equation, taking into account that \[\p_{z} \vu_{H,\a} = \p_{z} \vv_{H} = \vv_{|_{\p I_{F}(t)}} = 0.\]

\noindent $\bullet$ $\dint_{\O_{\a}(t)}\r\p_{t}\vu_{H,\a}\cdot\vv_{H}\,d\O = \dint_{I_{F}(t)} \inta \r\p_{t}\vu_{H,\a}\cdot \vv_{H}\,dzdx =  \dint_{I_{F}(t)}\r h_{\a}\p_{t}\vu_{H,\a}\cdot\vv_{H}\,dx$.\\
\\
\noindent $\bullet$ $\dint_{\O_{\a}(t)}\r\Big(\vu_{H,\a}\cdot\g_{x}\vu_{H,\a}\Big)\cdot\vv_{H}\,d\O =
\dint_{I_{F}(t)}\r \left( h_{\a}\vu_{H,\a}\cdot\g_{x}\vu_{H,\a}\right)\cdot \vv_{H}dx$.\\

\noindent $\bullet$ $
\dint_{\O_{\a}(t)}p_{\a}\,\g_{x}\cdot\vv_{H}\,d\O = \dint_{I_{F}(t)} \left(\inta p_{\a} \,dz\right)\g_{x}\cdot\vv_{H}\,dx = \\$
\[
\begin{array}{l}
=\dint_{I_{F}(t)}\g_{x}\left( \inta p_{\a}\ dz\right)\cdot \vv_{H}dx + \dint_{\p I_{F}(t)} \left(\inta p_{\a}\,dz\right)\, \vv_{H}\cdot \vn d\G = \\
\\
= -\dint_{I_{F}(t)}\left(\inta \g_{x}p_{\a}dz + \left.p_{\a}\frac{dz}{dx}\right]^{z_{\a+\frac{1}{2}}}_{z_{\a-\frac{1}{2}}} \right)\cdot \vv_{H}dx = (\ast).\\
\end{array}\]

\noindent Moreover,
\[
\begin{array}{l}
\displaystyle \inta \g_{x}p_{\a}dz = \inta \g_{x}\left(p_{S}+\dfrac{\r}{Fr^{2}}\left(z_{b}+h-z\right)\right)dz =\\
\\
= \displaystyle \inta \left(\g_{x}p_{S}+\dfrac{\r}{Fr^{2}}\g_{x}\left(z_{b}+h\right)\right)dz
=h_{\a} \g_{x}p_{S}+\dfrac{\r}{Fr^{2}}h_{\a} \g_{x}\left(z_{b}+h\right).
\end{array}
\]
\\
Then, we continue the computation of (3). We obtain
\[\begin{array}{l}
(\ast) = -\dint_{I_{F}(t)}\left( h_{\a}\g_{x}p_{S} +\dfrac{\r}{Fr^{2}}h_{\a}\g_{x}\left(z_{b}+h \right)
+p_{\a+\frac{1}{2}}\g_{x}z_{\a+\frac{1}{2}} -p_{\a-\frac{1}{2}}\g_{x}z_{\a-\frac{1}{2}} \right)\cdot \vv_{H}dx = \\
\\
= -\dint_{I_{F}(t)}\left(h_{\a}\g_{x}p_{S} +\dfrac{\r}{Fr^{2}}h_{\a}\g_{x}\left(z_{b}+h \right)\right)\cdot \vv_{H}dx\, - \\
\\
-\dint_{I_{F}(t)}p_{\a+\frac{1}{2}}\vn_{\a+\frac{1}{2}}\cdot (\vv_{H},0)\sqrt{1+\abs{\g_{x}z_{\a+\frac{1}{2}}}^{2}}\,dx +
\dint_{I_{F}(t)}p_{\a-\frac{1}{2}}\vn_{\a-\frac{1}{2}}\cdot (\vv_{H},0)\sqrt{1+\abs{\g_{x}z_{\a-\frac{1}{2}}}^{2}}\,dx.
\end{array}\]
\\
\noindent $\bullet$ $
\dint_{\O_{\a}(t)}\bd{T}_{H,\a}: \g_{x}\vv_{H}\ d\O = \dint_{I_{F}(t)} \left(\inta \bd{T}_{H,\a}: \g_{x}\vv_{H} \ dz\right)dx =\\$
\[
\begin{array}{l}
= \dint_{I_{F}(t)} \left(\inta \bd{T}_{H,\a}dz\right): \g_{x}\vv_{H}dx = - \dint_{I_{F}(t)}\g_{x}\cdot\left( \inta \bd{T}_{H,\a}\ dz\right)\cdot \vv_{H}dx \,+ \\
\\
+ \dint_{\p I_{F}(t)} \left(\inta \bd{T}_{H,\a}\ dz\right)\cdot \vv_{H}\cdot \vn d\G = - \dint_{I_{F}(t)}\g_{x}\cdot\left( \inta \bd{T}_{H,\a}\ dz\right)\cdot \vv_{H}dx.\\

\end{array}\]

Because $\p_{z}\vu_{H,\a} = 0$, we obtain that  $\|D_{\varepsilon}(\vu_{\a})\|$  is independent of z up to order $\varepsilon$, since
\[ D_{\varepsilon}(\vu_{\a})  = \left(\begin{matrix}
D_{H}(\vu_{H,\a}) & 0\\
&\\
0 & 2\p_{z}w
\end{matrix}\right) . \]

Then, from \eqref{eq:etaepsilon}, we get

\[\dint_{z_{\a-\frac{1}{2}}}^{z_{\a+\frac{1}{2}}} \bd{T}_{H,\a}\ dz = \dint_{z_{\a-\frac{1}{2}}}^{z_{\a+\frac{1}{2}}} \varepsilon\eta^{0}D_{H}(\vu_{H,\a})\ dz = O(\varepsilon).  \]
Therefore, we can neglect the term $\g_x\cdot\left(\dint_{z_{\a-\frac{1}{2}}}^{z_{\a+\frac{1}{2}}} \bd{T}_{H,\a}\ dz\right)$, which corresponds to the horizontal diffusion, since we are interested in the first order model.\\

\noindent $\bullet$ $
\dint_{\G_{\a+\frac{1}{2}}(t)}\left(\left( -p_{\a+\frac{1}{2}}\mathcal{E}+\bd{T}_{\a+\frac{1}{2}}^{-}\mathcal{E}\right) \vn_{\a+\frac{1}{2}} \right) \cdot (\vv_{H},0)\ d\G\ =\\
\\ = \dint_{I_{F}(t)}\left(\left( -p_{\a+\frac{1}{2}}\mathcal{E}+\bd{T}_{\a+\frac{1}{2}}^{-}\mathcal{E}\right) \vn_{\a+\frac{1}{2}} \right) \cdot (\vv_{H},0)\sqrt{1+\abs{\g_{x}z_{\a+\frac{1}{2}}}^{2}}\,dx$.\\

\bigskip
\noindent Introducing these calculations in (\ref{eq:hmomentum_flux_weak}) and taking into account that $\vec{\bd{f}} = \left(0,\frac{1}{Fr^2}\right)'$, we obtain
\[\begin{array}{l}
\dint_{I_{F}(t)}\Big(  \r h_{\a}\p_{t}\vu_{H,\a} + \r h_{\a}\vu_{H,\a}\cdot\g_{x}\vu_{H,\a}\ +h_{\a}\g_{x}p_{S} +\dfrac{\r}{Fr^{2}}h_{\a}\g_{x}\left(z_{b}+h \right)\Big)\cdot\vv_{H}\, dx\ +\\
\\
+ \dint_{I_{F}(t)}\left(p_{\a+\frac{1}{2}}\vn_{\a+\frac{1}{2}}\cdot (\vv_{H},0) \sqrt{1+\abs{\g_{x}z_{\a+\frac{1}{2}}}^{2}}- p_{\a-\frac{1}{2}}\vn_{\a-\frac{1}{2}}\cdot (\vv_{H},0) \sqrt{1+\abs{\g_{x}z_{\a-\frac{1}{2}}}^{2}}\right)\,dx\ +\\
\\
+\displaystyle \frac{1}{\varepsilon} \dint_{I_{F}(t)}\left(\left(\left( -p_{\a+\frac{1}{2}}\mathcal{E}+\bd{T}_{\a+\frac{1}{2}}^{-}\mathcal{E}\right) \vn_{\a+\frac{1}{2}} \right) \cdot (\vv_{H},0)\sqrt{1+\abs{\g_{x}z_{\a+\frac{1}{2}}}^{2}} \ -\right.\\
\\
-\left.\left(\left( -p_{\a-\frac{1}{2}}\mathcal{E}+\bd{T}_{\a-\frac{1}{2}}^{+}\mathcal{E}\right) \vn_{\a-\frac{1}{2}} \right) \cdot (\vv_{H},0)\sqrt{1+\abs{\g_{x}z_{\a-\frac{1}{2}}}^{2}}\right)\,dx =0.
\end{array}\]
Note that $\displaystyle \frac{1}{\varepsilon}p_{\a+\frac{1}{2}}\mathcal{E}\vn_{\a+\frac{1}{2}} \cdot(\vv_{H},0)= p_{\a+\frac{1}{2}}\vn_{\a+\frac{1}{2}}\cdot (\vv_{H},0)$, then
\[\begin{array}{l}
\dint_{I_{F}(t)}\Big( \r h_{\a}\p_{t}\vu_{H,\a} + \r h_{\a}\vu_{H,\a}\cdot\g_{x}\vu_{H,\a}
+ h_{\a}\g_{x}p_{S} +\dfrac{\r}{Fr^{2}}h_{\a}\g_{x}\left(z_{b}+h \right)\Big)\cdot \vv_{H}\,dx\ +\\
\\
+ \dint_{I_{F}(t)}\left(\left(\displaystyle \frac{1}{\varepsilon}\bd{T}_{\a+\frac{1}{2}}^{-}\mathcal{E} \vn_{\a+\frac{1}{2}} \right) \cdot (\vv_{H},0)\sqrt{1+\abs{\g_{x}z_{\a+\frac{1}{2}}}^{2}} \ -\right.\\
\\
-\left.\left(\displaystyle \frac{1}{\varepsilon}\bd{T}_{\a-\frac{1}{2}}^{+}\mathcal{E} \vn_{\a-\frac{1}{2}} \right) \cdot (\vv_{H},0)\sqrt{1+\abs{\g_{x}z_{\a-\frac{1}{2}}}^{2}}\right)dx=0.
\end{array}\]
Moreover,
\[\begin{array}{l}
\left(\bd{T}_{\a+\frac{1}{2}}^{-}\mathcal{E} \vn_{\a+\frac{1}{2}} \right) \cdot (\vv_{H},0)\sqrt{1+\abs{\g_{x}z_{\a+\frac{1}{2}}}^{2}} = \left(\begin{matrix}
\varepsilon \bd{T}_{H,\a+\frac{1}{2}}^{-} & T_{xz,\a+\frac{1}{2}}^{-}\\
\varepsilon T_{xz,\a+\frac{1}{2}}^{'^{-}} & T_{zz,\a+\frac{1}{2}}^{-}
\end{matrix}\right)\left(\begin{matrix}
\g_{x}z_{\a+\frac{1}{2}}\\
-1
\end{matrix}\right)\left(\begin{matrix}
\vv_{H}\\
0
\end{matrix}\right) \ =\\
\\
=\ \left(\varepsilon\bd{T}_{H,\a+\frac{1}{2}}^{-}\g_{x}z_{\a+\frac{1}{2}} - T_{xz,\a+\frac{1}{2}}^{-}\right)\cdot \vv_{H}.
\end{array}\]

\noindent Therefore,

\[\begin{array}{l}
\dint_{I_{F}(t)}\left[\r h_{\a}\p_{t}\vu_{H,\a} + \r h_{\a}\vu_{H,\a}\cdot\g_{x}\vu_{H,\a} +h_{\a}\g_{x}p_{S} +\dfrac{\r}{Fr^{2}}h_{\a}\g_{x}\left(z_{b}+h \right)\ +\right.\\
\\
\left.+\left(\bd{T}_{H,\a+\frac{1}{2}}^{-}\g_{x}z_{\a+\frac{1}{2}} - \displaystyle \frac{1}{\varepsilon}T_{xz,\a+\frac{1}{2}}^{-}\right) - \left(\bd{T}_{H,\a-\frac{1}{2}}^{+}\g_{x}z_{\a-\frac{1}{2}} - \displaystyle \frac{1}{\varepsilon}T_{xz,\a-\frac{1}{2}}^{+}\right)\right]\cdot \vv_{H}dx=0.
\end{array}\]
\\

\noindent And this yields, for each layer $\a = 1,\cdots,N$, the momentum equation

\[
\begin{array}{l}
\r h_{\a}\p_{t}\vu_{H,\a} + \r h_{\a}\vu_{H,\a}\cdot\g_{x}\vu_{H,\a} + h_{\a}\g_{x}p_{S} +\dfrac{\r}{Fr^{2}}h_{\a}\g_{x}\left(z_{b}+h \right)
 +\\
\\
+\left(\bd{T}_{H,\a+\frac{1}{2}}^{-}\g_{x}z_{\a+\frac{1}{2}} - \displaystyle \frac{1}{\varepsilon}T_{xz,\a+\frac{1}{2}}^{-}\right) - \left(\bd{T}_{H,\a-\frac{1}{2}}^{+}\g_{x}z_{\a-\frac{1}{2}} - \displaystyle \frac{1}{\varepsilon}T_{xz,\a-\frac{1}{2}}^{+}\right) = 0.
\end{array}
\]
\\
\begin{remark}\label{re:compH}
Observe that
\[\bd{T}_{H,\a+\frac{1}{2}}^{-}\g_{x}z_{\a+\frac{1}{2}} - \displaystyle \frac{1}{\varepsilon}T_{xz,\a+\frac{1}{2}}^{-} = \left[\displaystyle \frac{1}{\varepsilon}\bd{T}_{\a+\frac{1}{2}}^{-}\mathcal{E}\vn_{\a+\frac{1}{2}}\sqrt{1+\abs{\g_{x}z_{\a+\frac{1}{2}}}^{2}}\right]_{H},\]
where $[\ \cdot\ ]_{H}$ denotes the first component. Now by (\ref{rk_momemtum_prop}) we obtain
\[\begin{array}{l}
\left[\displaystyle \frac{1}{\varepsilon}\bd{T}_{\a+\frac{1}{2}}^{-}\mathcal{E}\vn_{\a+\frac{1}{2}}\sqrt{1+\abs{\g_{x}z_{\a+\frac{1}{2}}}^{2}}\right]_{H} =
\left[ \displaystyle \frac{1}{\varepsilon}\widetilde{\bd{T}}_{\a+\frac{1}{2}}\mathcal{E} \vn_{\a+\frac{1}{2}}\sqrt{1+\abs{\g_{x}z_{\a+\frac{1}{2}}}^{2}} - \frac{1}{2} \r G_{\a+\frac{1}{2}}\mathbb{B}\left[\vu\right]_{|_{\G_{\a+\frac{1}{2}}(t)}}\right]_{H} =\\
\\
= \displaystyle \widetilde{\bd{T}}_{H,\a+\frac{1}{2}}\g_{x}z_{\a+\frac{1}{2}} - \displaystyle \frac{1}{\varepsilon}\displaystyle \widetilde{T}_{xz,\a+\frac{1}{2}} - \frac{1}{2}\r G_{\a+\frac{1}{2}}\left(\vu_{H,\a+1}-\vu_{H,\a}\right),
\end{array}\]

\noindent and analogously

\[
\left[\displaystyle \frac{1}{\varepsilon}\bd{T}_{\a-\frac{1}{2}}^{+}\mathcal{E}\vn_{\a-\frac{1}{2}}\sqrt{1+\abs{\g_{x}z_{\a-\frac{1}{2}}}^{2}}\right]_{H} = \displaystyle \widetilde{\bd{T}}_{H,\a-\frac{1}{2}}\g_{x}z_{\a-\frac{1}{2}} - \displaystyle \frac{1}{\varepsilon}\displaystyle \widetilde{T}_{xz,\a-\frac{1}{2}} + \frac{1}{2}\r G_{\a-\frac{1}{2}}\left(\vu_{H,\a}-\vu_{H,\a-1}\right).
\]\\
\end{remark}

\noindent Remark \ref{re:compH} allows us to re-write the momentum equation
\[\begin{array}{l}
\r h_{\a}\p_{t}\vu_{H,\a} + \r h_{\a}\vu_{H,\a}\cdot\g_{x}\vu_{H,\a}
 +h_{\a}\g_{x}p_{S} +\dfrac{\r}{Fr^{2}}h_{\a}\g_{x}\left(z_{b}+h \right) +\\
\\
+\left(\displaystyle \widetilde{\bd{T}}_{H,\a+\frac{1}{2}}\g_{x}z_{\a+\frac{1}{2}} - \displaystyle \frac{1}{\varepsilon}\displaystyle \widetilde{T}_{xz,\a+\frac{1}{2}}\right) - \left(\displaystyle \widetilde{\bd{T}}_{H,\a-\frac{1}{2}}\g_{x}z_{\a-\frac{1}{2}} - \displaystyle \frac{1}{\varepsilon}\displaystyle \widetilde{T}_{xz,\a-\frac{1}{2}}\right) = \\
\\
= \displaystyle \frac{1}{2}\r G_{\a+\frac{1}{2}}\left(\vu_{H,\a+1}-\vu_{H,\a}\right) + \displaystyle \frac{1}{2}\r G_{\a-\frac{1}{2}}\left(\vu_{H,\a}-\vu_{H,\a-1}\right).
\end{array}\]\\

\noindent By combining the previous equation with (\ref{eq:mass_flux}) we get

\[\begin{array}{l}
\r \p_{t}\left(h_{\a}\vu_{H,\a}\right) + \r \g_{x}\cdot(h_{\a}\vu_{H,\a}\otimes\vu_{H,\a}) +h_{\a}\g_{x}p_{S} +\dfrac{\r}{Fr^{2}}h_{\a}\g_{x}\left(z_{b}+h \right)\ +\\
\\
+ \left(\displaystyle \widetilde{\bd{T}}_{H,\a+\frac{1}{2}}\g_{x}z_{\a+\frac{1}{2}} - \displaystyle \frac{1}{\varepsilon}\displaystyle \widetilde{T}_{xz,\a+\frac{1}{2}}\right) - \left(\displaystyle \widetilde{\bd{T}}_{H,\a-\frac{1}{2}}\g_{x}z_{\a-\frac{1}{2}} - \displaystyle \frac{1}{\varepsilon}\displaystyle \widetilde{T}_{xz,\a-\frac{1}{2}}\right) = \\
\\
= \displaystyle \frac{1}{2}\r G_{\a+\frac{1}{2}}\left(\vu_{H,\a+1}+\vu_{H,\a}\right) - \displaystyle \frac{1}{2}\r G_{\a-\frac{1}{2}}\left(\vu_{H,\a}+\vu_{H,\a-1}\right).
\end{array}\]
\\
Note that
\[ \begin{array}{l}
 \displaystyle \widetilde{\bd{T}}_{H,\a+\frac{1}{2}}\g_{x}z_{\a+\frac{1}{2}} - \displaystyle \frac{1}{\varepsilon}\displaystyle \widetilde{T}_{xz,\a+\frac{1}{2}}=
\left[\displaystyle \frac{1}{\varepsilon}\widetilde{\bd{T}}_{\a+\frac{1}{2}}\mathcal{E}\vn_{\a+\frac{1}{2}}\sqrt{1+\abs{\g_{x}z_{\a+\frac{1}{2}}}^{2}}\right]_{H} = \\
\\
 = \left[\eta^{0}\displaystyle \widetilde{D}_{\a+\frac{1}{2}}\mathcal{E}\vn_{\a+\frac{1}{2}}\right]_{H}\sqrt{1+\abs{\g_{x}z_{\a+\frac{1}{2}}}^{2}} = \varepsilon\eta^{0}_{\a+\frac{1}{2}}\displaystyle D_{H} \left( \frac{\vu_{H,\a+\frac{1}{2}}^+ + \vu_{H,\a+\frac{1}{2}}^-}{2} \right) \cdot \g\!_{x} z_{ \a+\frac{1}{2}}\,  -\\
 \\
-\ \varepsilon\eta^{0}_{\a+\frac{1}{2}}\left( \g\!_{x} \left( \frac{w_{\a+\frac{1}{2}}^+ + w_{\a+\frac{1}{2}}^-}{2} \right)  \right)' - \displaystyle \frac{1}{\varepsilon}\eta^{0}_{\a+\frac{1}{2}}\Q_{H,\a+\frac{1}{2}} = - \dfrac{1}{\varepsilon}\eta^{0}_{\a+\frac{1}{2}}\Q_{H,\a+\frac{1}{2}} + O(\varepsilon).
 \end{array}\]\\

\noindent We define\\
\begin{equation}
\label{eq.fric_term}
\vec{K}_{\a+\frac{1}{2}}= - \dfrac{1}{\varepsilon}\eta^{0}_{\a+\frac{1}{2}}\Q_{H,\a+\frac{1}{2}},
\end{equation}

\noindent where $\eta^{0}_{\a+\frac{1}{2}}$ is a first order approximation of $\eta$ at $z=z_{\a+\frac{1}{2}}$. We obtain

\begin{equation} \label{eq:approx_eta_interfaz}
\eta^{0}_{\a+\frac{1}{2}} = \dfrac{\mu(I_{\a+\frac{1}{2}}) p_{\a+\frac{1}{2}}}{\mbox{max}\left(\| Q_{H,\a+\frac{1}{2}} \|,\frac{\mu(I_{{\a+\frac{1}{2}}})p_{\a+\frac{1}{2}}}{\eta_{M}}\right)}\, ,
\end{equation}
with

 \begin{equation} \label{eq:def_I_interfaz}
 p_{\a+\frac{1}{2}} = p_{S}+ \dfrac{\r}{Fr^{2}}\sum_{\beta=\a+1}^N h_{\beta}\, , \quad\quad
 I_{\a+\frac{1}{2}} = \frac{d_s \| Q_{H,\a+\frac{1}{2}} \|}{\sqrt{p_{\a+\frac{1}{2}}/ \r_{s}}}\, .
 \end{equation}

 These expressions of $\eta_{\a+\frac{1}{2}}$ and $I_{\a+\frac{1}{2}}$ are obtained from definitions \eqref{eq:regularisation} and (\ref{eq:def_I}), respectively, by considering the hydrostatic pressure approximation (\ref{eq:pressure}) with the definition of $\rho$ (\ref{eq:def_rho}) and with the following first order approximation of $\|D(\vu)\|$ at $z=z_{\a+\frac{1}{2}}$,

\begin{equation} \label{eq:approx_norm_Du_o1}
\|D(\vu)\|_{|_{z=z_{\a+\frac{1}{2}}}} \approx \| Q_{H,\a+\frac{1}{2}} \|.
\end{equation}

\bigskip

\noindent By re-writing the momentum equation again, we obtain up to order $\varepsilon$,
\begin{equation}\label{eq:momentum_flux}
\begin{array}{l}
\r \p_{t}\left(h_{\a}\vu_{H,\a}\right) + \r \g_{x}\cdot(h_{\a}\vu_{H,\a}\otimes\vu_{H,\a})
+h_{\a}\g_{x}p_{S} +\dfrac{\r}{Fr^{2}}h_{\a}\g_{x}\left(z_{b}+h \right) =\\
\\
=  \vec{K}_{\a-\frac{1}{2}} - \vec{K}_{\a+\frac{1}{2}} + \displaystyle \frac{1}{2}\r G_{\a+\frac{1}{2}}\left(\vu_{H,\a+1}+\vu_{H,\a}\right) - \displaystyle \frac{1}{2}\r G_{\a-\frac{1}{2}}\left(\vu_{H,\a}+\vu_{H,\a-1}\right),
\end{array}
\end{equation}

\noindent the horizontal momentum conservation laws, for $\smash\a=1,...,N$.

%

\begin{remark}
We must impose friction at the bottom (\,$\G_{\a-\frac{1}{2}}$ with $\a = 1$). We can translate (\ref{eq:boundCond_bot}) into the notation  of the multilayer approach, giving
\begin{equation*}
\label{eq:boundCond_bot_mult}
\begin{array}{l}
w_{|_{\G_{\frac{1}{2}}}} = 0,\\
\\
\displaystyle \frac{1}{\varepsilon} \eta_{\frac{1}{2}} Q_{H,\frac{1}{2}} = \mu(I) p_{\frac{1}{2}}\dfrac{\vu_{H,\frac{1}{2}}^{+}}{\abs{\vu_{H,\frac{1}{2}}^{+}}}.
\end{array}
\end{equation*}
Therefore to impose the friction condition, we should change definition (\ref{eq.fric_term}) of $\vec{K}_{\frac{1}{2}}$, taking into account that
\[\vu_{H,\frac{1}{2}}^{+} = \vu_{H,1}.\]
Then we obtain
\begin{equation}
\label{eq:friction_term}
\vec{K}_{\frac{1}{2}} =  - \dfrac{1}{\varepsilon}\eta^{0}_{\frac{1}{2}}\Q_{H,\frac{1}{2}} = -\mu(I)^{0}p_{\frac{1}{2}}\dfrac{\vu_{H,1}}{\abs{\vu_{H,1}}}.
\end{equation}
\end{remark}


\par

\vskip 0.7 cm

\par

\subsection{Final Model}\label{se_sub:orig_var}
We have obtained the dimensionless final system given by (\ref{eq:defG}), (\ref{eq:mass_interfCond}), (\ref{eq:mass_flux}) and (\ref{eq.fric_term})-(\ref{eq:friction_term}). The last step is to return to the original variables taking into account subsection (\ref{nondim_var}). We obtain the final multilayer  system, for $\smash\a = 1,...,N$,

\begin{align}
\label{eq:FinalModel}
\left\{
\begin{array}{l}
\p_{t}h_{\a} + \g\!_{x}\cdot(h_{\a}\vu_{H,\a}) = G_{\a+\frac{1}{2}} - G_{\a-\frac{1}{2}}, \\
\\
\r\p_{t}\left(h_{\a}\vu_{H,\a}\right) \;+\; \g\!_{x}\cdot\left(\r h_{\a}\vu_{H,\a}\otimes\vu_{H,\a}\right) \, + \\[4mm]
\quad+\;h_{\a}\g_{x}p_{S} + \r gh_{\a}\g_{x} \left(z_{b}+h\right) \, =\, \vec{K}_{\a-\frac{1}{2}} - \vec{K}_{\a+\frac{1}{2}}\; +\\[4mm]
\quad+\;\dfrac{1}{2} \r G_{\a+\frac{1}{2}}\left(\vu_{H,\a+1} + \vu_{H,\a}\right) \;-\; \dfrac{1}{2} \r G_{\a-\frac{1}{2}}\left(\vu_{H,\a} + \vu_{H,\a-1}\right),
\\
\end{array}
\right.
\end{align}
\\
where\\
\begin{equation}\label{eq:FinalModel2}
\begin{array}{l}
G_{\a+\frac{1}{2}} = \p_{t}z_{\a+\frac{1}{2}} + \vu_{H,\a+1}\cdot\g\!_{x}z_{\a+\frac{1}{2}} - w_{\a+\frac{1}{2}}^{+} = \p_{t}z_{\a+\frac{1}{2}} + \vu_{H,\a}\cdot\g\!_{x}z_{\a+\frac{1}{2}} - w_{\a+\frac{1}{2}}^{-},\\[4mm]
\vec{K}_{\a+\frac{1}{2}} = - \eta_{\a+\frac{1}{2}} \Q_{H,\a+\frac{1}{2}} = \left[\eta_{\a+\frac{1}{2}}\widetilde{D}_{\a+\frac{1}{2}} \cdot \vn_{\a+\frac{1}{2}}\right]_H \left( \sqrt{1+\abs{\g\!_{x}z_{ \a+\frac{1}{2}}}^2} \right),\\[4mm]
\hspace{-1.5cm}\mbox{and}\\[4mm]
\vec{K}_{\frac{1}{2}} = -\mu(I)\r gh\dfrac{\vu_{H,1}}{\abs{\vu_{H,1}}}.
\end{array}
\end{equation}
\\

System \eqref{eq:FinalModel} must be closed by setting the vertical partition of the domain.  For this, we can write the thickness of the pre-set layer based on the total height. That is,  we set $h_{\a} = l_{\a}\,h \,$ where $l_{\a}\ $ is a positive constant, for $\a = 1,\cdots,N$, and
\begin{equation*} \label{eq:layersum}
\dsum_{\a=1}^N l_{\a} = 1.
\end{equation*}

\noindent Note that $G_{\a+\frac{1}{2}}$ can be written, by summing the mass equations from 1 to $\a$, as
\begin{align}
\label{eq:cont_mass_al}
G_{\a+\frac{1}{2}} = G_{\frac{1}{2}} + \dsum_{\b=1}^\a \left(\p_{t}h_{\b} + \g_{x}\cdot( h_{\b}\vu_{H,\b})\right).
\end{align}

\noindent Moreover, for the special case $\a=N$ and assuming no mass transfer with the atmosphere, i.e. $G_{N+\frac{1}{2}}=0$, the above equation leads to
\begin{align*}
\label{eq:cont_mass}
\p_{t}h  + \g_{x}\cdot\Biggl(  h\dsum_{\b=1}^N l_{\b}\vu_{H,\b}\Biggr) =  -\; G_{\frac{1}{2}}.
\end{align*}

\noindent By introducing this in \eqref{eq:cont_mass_al} we obtain
\begin{equation}\label{eq:cont_mass_G}
 G_{\a+\frac{1}{2}} = \;G_{\frac{1}{2}} +  \dsum_{\b=1}^\a l_{\b}\Biggl( \g_{x}\cdot\left(h\vu_{H,\b}\right) -  \dsum_{\gamma=1}^N\g_{x}\cdot\left(l_{\gamma}h\vu_{H,\gamma}\right) -\; G_{\frac{1}{2}}\Biggr).
\end{equation}

\noindent Let us define $\smash L_{\a} := l_{1} + \dots + l_{\alpha}$ and $\xi_{\a,\gamma} = \dsum_{\b=1}^\a (\delta_{\b \gamma} - l_{\b})l_{\gamma}$, where  $\delta_{\b \gamma}$ is the standard  Kronecker symbol. That is,
\begin{align*}
\xi_{\a,\gamma} =
\begin{cases}
\bigl(  1 - (l_{1} + \dots + l_{\alpha}) \bigr)l_{\gamma}, & \text{if $\gamma \leq \a$} ,\\
\\
- (l_{1} + \dots + l_{\alpha}) l_{\gamma}, &  \text{otherwise},
\end{cases}
\end{align*}
for $\alpha,\;\gamma \in \{1,\dots,N\}$.  Then, we can write the mass transfer \eqref{eq:cont_mass_G} in the interface $\G_{\a+1/2}$ as
\begin{equation}
\label{eq:trans_masa_vel}
G_{\a+\frac{1}{2}} =  \left(1 - L_{\a}\right)\,G_{\frac{1}{2}} + \dsum_{\gamma=1}^N \xi_{\a,\gamma}\g_{x}\cdot(h\vu_{H,\gamma}),
\end{equation}
for $\a = 1,\dots,N$.\\

Next, considering $\vec{\bd{q}}_{H,\a} = h\vu_{H,\a}$ and after some straightforward  calculations, the final system \eqref{eq:FinalModel}-\eqref{eq:FinalModel2} is re-written, for $\a=1,\cdots,N$, as
\begin{equation*}
\label{eq:swHydro_syst_num}
\left\{
\begin{array}{l}
\p_{t}h  + \g_{x}\cdot\Biggl( \dsum_{\b=1}^N l_{\b}\vec{\bd{q}}_{H,\b}\Biggr) =  -\; G_{\frac{1}{2}}, \\
\\
\p_{t}\vec{\bd{q}}_{H,\a} + \g_{x}\cdot\left(\dfrac{\vec{\bd{q}}_{H,\a}\otimes\vec{\bd{q}}_{H,\a}}{h}\right) + \g_x\left(g\dfrac{h^2}{2} + \dfrac{p_{S}h}{\r} \right)
\;-\; \dfrac{p_{S}}{\r}\p_{x}h\ +\\[4mm]
\quad+ \dsum_{\gamma=1}^N\dfrac{1}{2hl_{\a}}\bigg(  \left(\vec{\bd{q}}_{H,\a} + \vec{\bd{q}}_{H,\a-1}\right)\xi_{\a-1,\gamma} - \left(\vec{\bd{q}}_{H,\a+1} + \vec{\bd{q}}_{H,\a}\right)\xi_{\a,\gamma}\bigg)\,\g_{x}\cdot\left(\vec{\bd{q}}_{H,\gamma}\right) \ = \\[4mm]
\quad = \dfrac{1}{2hl_{\a}}  \bigg( \left(\vec{\bd{q}}_{H,\a+1} + \vec{\bd{q}}_{H,\a}\right)\left(1 - L_{\a}\right)
\ -\left(\vec{\bd{q}}_{H,\a} + \vec{\bd{q}}_{H,\a-1}\right)\left(1 - L_{\a-1}\right)\bigg)\,G_{\frac{1}{2}}\ -\\[4mm]
\quad-gh\g_{x}z_{b} + \dfrac{1}{\r l_{\a}} \Big(\vec{K}_{\a-\frac{1}{2}} - \vec{K}_{\a+\frac{1}{2}}\bigg).
\end{array}
\right.
\end{equation*}
\\
\subsection{Energy associated with the final model}
In this section we study the energy balance of the obtained system \eqref{eq:FinalModel}-\eqref{eq:FinalModel2}.

\begin{theorem}
Denoting the energy of the layer $\a=1,\cdots,N$ for the system \eqref{eq:FinalModel}-\eqref{eq:FinalModel2} by
\[E_{\a} = h_{\a}\left( \dfrac{\abs{\vu_{H,\a}}^{2}}{2} + \dfrac{p_{S}}{\r} + g\left(z_{b}+\dfrac{h}{2}\right)\right),\]
the following dissipative energy inequality is satisfied:
\[\begin{array}{l}
\r \p_{t}\left(\dsum_{\a=1}^{N}E_{\a}\right) + \r \g_{x}\cdot\left[\dsum_{\a=1}^{N}\vu_{H,\a}\left(E_{\a} + \r gh_{\a}\dfrac{h}{2}\right)\right] \leq  h_{\a}\p_{t}\left(p_{S}+\r gz_b\right) - \\
\\
-\r gh\abs{\vu_{H,1}}\mu(I) - \dfrac{\abs{\vu_{H,N}}^{2}}{h_{N}}\eta_{N+\frac{1}{2}} - \dsum_{\a=1}^{N-1}\dfrac{\left(\vu_{H,\a+1}-\vu_{\a}\right)^{2}}{h_{\a+\frac{1}{2}}}\eta_{\a+\frac{1}{2}} -G_{\frac{1}{2}}\left(p_{S} + \r g(z_{b}+h)\right).
\end{array}\]
\end{theorem}
\noindent {\sc Proof.-} \\
Firstly we multiply the momentum equation for each layer by $\vu_{H,\a}$ and use the mass equation to simplify the convective terms. Secondly we sum up the obtained equation for layers $\a=1$ to $\a=N$ and then we obtain that the global system has a dissipative energy balance.\\


Now, we write the momentum conservation equation in terms of the velocity using the continuity equation

\begin{equation}
\label{eq:energ_vel}
\begin{array}{l}
\r h_{\a}\p_{t}\vu_{H,\a} + \r \left(\g_{x}\vu_{H,\a}\right)h_{\a}\vu_{H,\a} + h_{\a}\g_{x}p_{S} + \r gh_{\a}\g_{x} \left(z_{b}+h\right)= \\
\\
=\vec{K}_{\a-\frac{1}{2}} - \vec{K}_{\a+\frac{1}{2}}+ \dfrac{1}{2} \r G_{\a+\frac{1}{2}}\left(\vu_{H,\a+1} - \vu_{H,\a}\right) + \dfrac{1}{2} \r G_{\a-\frac{1}{2}}\left(\vu_{H,\a} - \vu_{H,\a-1}\right).
\end{array}
\end{equation}
\\
Multiplying (\ref{eq:energ_vel}) by $\vu_{H,\a}$ we obtain
\\
\\
\[\begin{array}{l}
\r h_{\a}\vu_{H,\a}\cdot\p_{t}\vu_{H,\a} + \r
\left( \left(\g_{x}\vu_{H,\a}\right)\, h_{\a}\vu_{H,\a}\right)\cdot\vu_{H,\a}
 + h_{\a}\vu_{H,\a}\cdot\g_{x}\left(p_{S} + \r g \left(z_{b}+h\right)\right) =\\
\\
=\vu_{H,\a}\cdot\left(\vec{K}_{\a-\frac{1}{2}} - \vec{K}_{\a+\frac{1}{2}}\right) + \dfrac{1}{2} \r G_{\a+\frac{1}{2}}\left(\vu_{H,\a+1}\cdot\vu_{H,\a} - \abs{\vu_{H,\a}}^{2}\right)+ \\
\\
\quad + \dfrac{1}{2} \r G_{\a-\frac{1}{2}}\left(\abs{\vu_{H,\a}}^{2} - \vu_{H,\a}\cdot\vu_{H,\a-1}\right).
\end{array}\]
\\
This gives
\[\begin{array}{l}
\left( \left(\g_{x}\vu_{H,\a}\right)\,h_{\a}\vu_{H,\a}\right)\cdot\vu_{H,\a} = \g_{x}\left(\dfrac{\abs{\vu_{H,\a}}^{2}}{2}\right)\cdot  h_{\a}\vu_{H,\a} =\\
\\
= \g_{x}\cdot\left(\dfrac{\abs{\vu_{H,\a}}^{2}}{2} h_{\a}\vu_{H,\a}\right) - \dfrac{\abs{\vu_{H,\a}}^{2}}{2}\g_{x}\cdot\left(h_{\a}\vu_{H,\a}\right),
\end{array}
\]
which can be re-written as
\\
\begin{equation}
\label{eq:energ_vel_u}
\begin{array}{l}
\r h_{\a}\vu_{H,\a}\cdot\p_{t}\vu_{H,\a} + \r
\g_{x}\cdot\left(\dfrac{\abs{\vu_{H,\a}}^{2}}{2} h_{\a}\vu_{H,\a}\right) - \r\dfrac{\abs{\vu_{H,\a}}^{2}}{2}\g_{x}\cdot\left(h_{\a}\vu_{H,\a}\right)
 +\\
 \\
 + h_{\a}\vu_{H,\a}\cdot\g_{x}\left(p_{S} + \r g \left(z_{b}+h\right)\right)  = \vu_{H,\a}\cdot\left(\vec{K}_{\a-\frac{1}{2}} - \vec{K}_{\a+\frac{1}{2}}\right) + \\
 \\
 +\dfrac{1}{2} \r G_{\a+\frac{1}{2}}\left(\vu_{H,\a+1}\cdot\vu_{H,\a} - \abs{\vu_{H,\a}}^{2}\right) + \dfrac{1}{2} \r G_{\a-\frac{1}{2}}\left(\abs{\vu_{H,\a}}^{2} - \vu_{H,\a}\cdot\vu_{H,\a-1}\right).
\end{array}
\end{equation}

\noindent Let us consider the mass conservation equation multiplied by $\left(\r\dfrac{\abs{\vu_{H,\a}}^{2}}{2} + p_{S}+\r g(z_{b}+h)\right)$,
\begin{equation}
\label{eq:energ_mass}
\begin{array}{l}
\r\dfrac{\abs{\vu_{H,\a}}^{2}}{2}\p_{t}h_{\a} + \r\dfrac{\abs{\vu_{H,\a}}^{2}}{2}\g_{x}\cdot\left(h_{\a}\vu_{H,\a}\right) + \left( p_{S}+\r g(z_{b}+h)\right)\p_{t}h_{\a} + \left( p_{S}+\r g(z_{b}+h)\right)\g_{x}\cdot\left(h_{\a}\vu_{H,\a}\right)=\\
\\
= \r\dfrac{\abs{\vu_{H,\a}}^{2}}{2}\left(G_{\a+\frac{1}{2}}-G_{\a-\frac{1}{2}}\right) + (p_{S}+\r g(z_{b}+h))\left(G_{\a+\frac{1}{2}}-G_{\a-\frac{1}{2}}\right).
\end{array}
\end{equation}

\noindent Summing (\ref{eq:energ_vel_u}) and (\ref{eq:energ_mass}), we obtain the equation
\[
\begin{array}{l}
\r \p_{t}\left(\dfrac{\abs{\vu_{H,\a}}^{2}}{2}h_{\a}\right) + \r
\g_{x}\cdot\left(\dfrac{\abs{\vu_{H,\a}}^{2}}{2} h_{\a}\vu_{H,\a}\right)  + (p_{S}+\r g(z_{b}+h))\p_{t}h_{\a}\,+\\
 \\
 + \, \g_{x}\cdot\left(h_{\a}\left( p_{S} + \r g \left(z_{b}+h\right)\right) \vu_{H,\a}\right)  = \vu_{H,\a}\cdot\left(\vec{K}_{\a-\frac{1}{2}} - \vec{K}_{\a+\frac{1}{2}}\right)+\\
 \\
 + \, G_{\a+\frac{1}{2}}\left(\r\dfrac{\vu_{H,\a+1}\cdot\vu_{H,\a}}{2} + p_{S} + \r g(z_{b}+h)\right) - G_{\a-\frac{1}{2}}\left(\r\dfrac{\vu_{H,\a}\cdot\vu_{H,\a-1}}{2} + p_{S} + \r g(z_{b}+h)\right).
\end{array}
\]
\\
We can reformulate this as\\
\[
\begin{array}{l}
\r \p_{t}\left[ h_{\a}\left(\dfrac{\abs{\vu_{H,\a}}^{2}}{2}+\dfrac{p_{S}}{\r} + g(z_{b}+h)\right)\right] -h_{\a}\p_{t}p_{S}-\r gh_{\a}\p_{t}z_{b} - \r gh_{\a}\p_{t}h \,+ \\
\\
+ \,\r \g_{x}\cdot\left[ h_{\a}\left(\dfrac{\abs{\vu_{H,\a}}^{2}}{2}+\dfrac{p_{S}}{\r} + g(z_{b}+h)\right)\vu_{H,\a}\right]  = \vu_{\a}\left(\vec{K}_{\a-\frac{1}{2}} - \vec{K}_{\a+\frac{1}{2}}\right) + \\
 \\
 + \, G_{\a+\frac{1}{2}}\left(\r\dfrac{\vu_{H,\a+1}\cdot\vu_{H,\a}}{2} + p_{S} + \r g(z_{b}+h)\right) - G_{\a-\frac{1}{2}}\left(\r\dfrac{\vu_{H,\a}\cdot\vu_{H,\a-1}}{2} + p_{S} + \r g(z_{b}+h)\right).
\end{array}
\]

\noindent Note that
\[-\r gh_{\a}\p_{t}h = -\r gh_{\a}\p_{t}\dfrac{h}{2}-\r gh_{\a}\p_{t}\dfrac{h}{2} = -\r g\p_{t}\left(h_{\a}\dfrac{h}{2}\right) + \r gh\p_{t}\dfrac{h_{\a}}{2} -\r gh_{\a}\p_{t}\dfrac{h}{2}\, ,\]
\\
then,
\[
\begin{array}{l}
\r \p_{t}\left[ h_{\a}\left(\dfrac{\abs{\vu_{H,\a}}^{2}}{2}+\dfrac{p_{S}}{\r} + g\left(z_{b}+\dfrac{h}{2}\right)\right)\right] + \dfrac{\r g}{2}\left( h\p_{t}h_{\a} - h_{\a}\p_{t}h\right) - h_{\a}\p_{t}p_{S}-\r gh_{\a}\p_{t}z_{b} \, + \\
\\
+ \,\r \g_{x}\cdot\left[ h_{\a}\left(\dfrac{\abs{\vu_{H,\a}}^{2}}{2}+\dfrac{p_{S}}{\r} + g\left(z_{b}+\dfrac{h}{2}\right)\right)\vu_{H,\a}+ gh_{\a}\dfrac{h}{2}\vu_{H,\a}\right]  = \vu_{\a}\left(\vec{K}_{\a-\frac{1}{2}} - \vec{K}_{\a+\frac{1}{2}}\right) + \\
 \\
 + \, G_{\a+\frac{1}{2}}\left(\r\dfrac{\vu_{H,\a+1}\cdot\vu_{H,\a}}{2} + p_{S} + \r g(z_{b}+h)\right) - G_{\a-\frac{1}{2}}\left(\r\dfrac{\vu_{H,\a}\cdot\vu_{H,\a-1}}{2} + p_{S} + \r g(z_{b}+h)\right).
\end{array}
\]
\\
Denoting \[E_{\a} = h_{\a}\left( \dfrac{\abs{\vu_{H,\a}}^{2}}{2} + \dfrac{p_{S}}{\r} + g\left(z_{b}+\dfrac{h}{2}\right)\right),\]
we have for $\a = 1,...,N$ the following energy equality
\begin{equation}
\label{eq:energy_layer}
\begin{array}{l}
\underbrace{\r \p_{t}E_{\a}}_{(1)} +
\dfrac{\r g}{2}\underbrace{\left( h\p_{t}h_{\a} - h_{\a}\p_{t}h\right)}_{(2)}  + \\
\\
+ \underbrace{\r \g_{x}\cdot\left[\left(E_{\a} + \r gh_{\a}\dfrac{h}{2}\right)\vu_{H,\a}\right]}_{(3)}  = \underbrace{\vu_{\a}\left(\vec{K}_{\a-\frac{1}{2}} - \vec{K}_{\a+\frac{1}{2}}\right)}_{(4)} + h_{\a}\p_{t}\left(p_{S}+\r gz_{b}\right)+ \\
 \\
 +\underbrace{G_{\a+\frac{1}{2}}\left(\r\dfrac{\vu_{H,\a+1}\cdot\vu_{H,\a}}{2} + p_{S} + \r g(z_{b}+h)\right) - G_{\a-\frac{1}{2}}\left(\r\dfrac{\vu_{H,\a}\cdot\vu_{H,\a-1}}{2} + p_{S} + \r g(z_{b}+h)\right).}_{(5)}
\end{array}
\end{equation}

\noindent Now we sum up \eqref{eq:energy_layer} from $\smash\a = 1$ to $\smash\a = N$. We take into account that $\sum_{\a=1}^N h_{\a}=h$, $G_{N+\frac{1}{2}} = 0$ (there is no transfer with the atmosphere) and $\vu_{H,0} = \vu_{H,N+1} = 0$ (velocity of the bottom and atmosphere respectively). This gives, term by term:

\begin{enumerate}[(1)]
\item $\r \p_{t}\left(\dsum_{\a=1}^{N}E_{\a}\right).$\\
\item $h\p_{t}\left(\dsum_{\a=1}^{N}h_{\a}\right) - \dsum_{\a=1}^{N}h_{\a}\p_{t}h = h\p_{t}h - h\p_{t}h = 0.$\\
\item $\r \g_{x}\cdot\left[\dsum_{\a=1}^{N}\vu_{H,\a}\left(E_{\a} + \r gh_{\a}\dfrac{h}{2}\right)\right].$
\item Taking into account $K_{\a+\frac{1}{2}} = - \eta_{\a+\frac{1}{2}} \Q_{H,\a+\frac{1}{2}}$ and $K_{\frac{1}{2}} = -\r gh\dfrac{\vu_{H,1}}{\abs{\vu_{H,1}}}\mu(I)$ when we sum all the layers, we obtain 

\[-\r gh\abs{\vu_{H,1}}\mu(I) + \vu_{H,N}\eta_{N+\frac{1}{2}}\Q_{H,N+\frac{1}{2}} - \dsum_{\a=1}^{N-1}\left(\vu_{H,\a+1}-\vu_{\a}\right)\eta_{\a+\frac{1}{2}}\Q_{H,\a+\frac{1}{2}} \, ,\]
where $\Q_{H,\a+\frac{1}{2}}$ is an approximation of $\p_{z}\vu_{H}$ in $\G_{\a+\frac{1}{2}}$. We consider

\[\Q_{H,\a+\frac{1}{2}} = \dfrac{\vu_{H,\a+1}-\vu_{H,\a}}{h_{\a+\frac{1}{2}}}\ ;\quad \quad \Q_{H,N+\frac{1}{2}} = \dfrac{\vu_{H,N+1}-\vu_{H,N}}{h_{N}},\]

with $h_{\a+\frac{1}{2}}$ being the distance between the midpoints of layers $\a$ and $\a+1$. This gives
\[-\r gh\abs{\vu_{H,1}}\mu(I) - \dfrac{\abs{\vu_{H,N}}^{2}}{h_{N}}\eta_{N+\frac{1}{2}} - \dsum_{\a=1}^{N-1}\dfrac{\left(\vu_{H,\a+1}-\vu_{\a}\right)^{2}}{h_{\a+\frac{1}{2}}}\eta_{\a+\frac{1}{2}},\]
which is a dissipative term.\\

\item Considering $G_{N+\frac{1}{2}} = 0$, we have
\[ - G_{\frac{1}{2}}\left(p_{S} + \r g(z_{b}+h)\right).\]
\end{enumerate}
Finally, by summarising (1)-(5), the proof is completed.

\hfill $\square$

\section{Numerical tests}\label{se:numericalTest}
The numerical approximation is performed in 2D (downslope and normal directions). We re-write the model as a nonconservative hyperbolic system with source terms as in \cite{EnriqueMultilayer}. Then a splitting procedure is considered. First, we set aside the term that appears in the internal interfaces and a standard path-conservative finite volume method is applied. These path-conservative methods were introduced in \cite{ParesPathCons}. To deal with the Coulomb friction term, we use the hydrostatic reconstruction introduced in \cite{AudusseBouchutReconst}, which is applied in \cite{BouchutReconst} to solve the Saint-Venant system with Coulomb friction. The main advantage of this reconstruction is its great stability.\\

The second step is to solve the contribution of the term in the internal interfaces, which represents the mass and momentum exchange between layers. In this step, a semi-implicit scheme is employed, taking into account the regularisation of $\|D(\vu_\a)\|$ mentioned in Section \ref{se:muI} in order to avoid the singularity when $\|D(\vu_\a)\|$ vanishes.\\

In order to validate the Multilayer Shallow Model (denoted MSM hereafter) with the $\mu(I)$ rheology, we compare it to (1) a 2D analytical solution for steady uniform flows over an inclined bed and (2) laboratory experiments of granular collapses over an inclined plane covered by an erodible bed made of the same material.

\subsection{Analytical solution}
\label{se_sub:solAnal}


Let us first compare the model to the 2D analytical solution deduced in \cite{LagreeStaronPopinet} for a uniform flow over an inclined plane of slope $\theta$ and thickness $H>0$. This solution is obtained by imposing zero pressure and zero shear stress at the free surface and a no-slip condition at the bottom.\\

 \begin{figure}[!h]
 \begin{center}
 \includegraphics[width=0.6\textwidth]{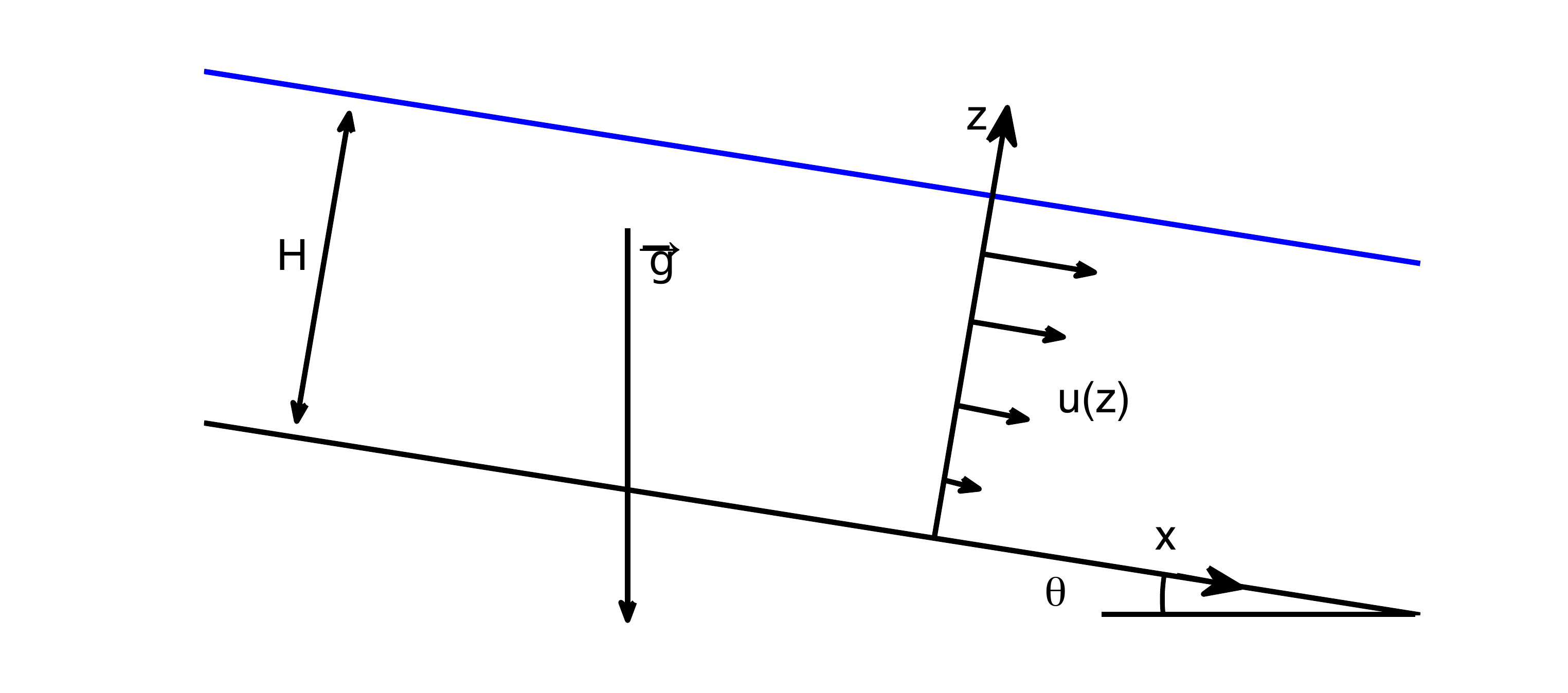}
 \caption{\label{fig:solAnal}\footnotesize \textit{Sketch of the analytical solution.}}
 \end{center}
 \end{figure}

By denoting $u$ and $v$ the downslope and normal velocities, $p$ the pressure and $\tau$ the shear stress and by taking the rheological parameters defined in Section \ref{se:muI}, the analytical solution reads

\begin{equation}
\label{eq:solAnal}
\left\{\begin{array}{l}
u(z) = \dfrac{2}{3d}I_{0}\left(\dfrac{tan\theta - \mu_{s} }{\Delta \mu - tan\theta+\mu_{s}}\right)\sqrt{\varphi_s gcos\theta}\left(H^{3/2}-\left(H-z\right)^{3/2}\right),\\
\\
u(z=0) = 0,\ \ \  v = 0,\\
\\
 p(z) = \r gcos\theta\left(H-z\right),\\
\\
\tau(z) = \mu(I)p = \r gsin\theta\left(H-z\right),\\
\\
p(z=H) = 0, \ \ \ \tau(z=H)=0,\\
\\
\mu(I) = tan(\theta),\hspace{6cm}\mbox{for }z\in(0,H).
\end{array}\right.
\end{equation}


For the numerical simulation, as in the analytical solution, we consider a uniform flow with constant thickness $H = 1$ m and velocity $u = v = 0$ m$\cdot$ s$^{-1}$  at the initial time $t= 0$ s. The boundary condition at the free surface and at the bottom have been set as in (\ref{eq:solAnal}). At the right and left boundary, we use open boundary conditions.\\

\noindent Note that at the free surface we have
\[\begin{array}{l}
p = 0 \quad\quad\mbox{and}\quad\quad\|D(u)\| = \p_z u = \dfrac{I_0}{d}\left(\dfrac{tan\theta - \mu_{s} }{\Delta \mu - tan\theta+\mu_{s}}\right)\sqrt{\varphi_s gcos\theta\left(H-z\right)} = 0.
\end{array}\]
As a result, we cannot use the regularisation \eqref{eq:regularisation} since its denominator$$\text{max}\left(\|D(\vu)\|,\frac{\mu(I)p}{\eta_{M}}\right)$$vanishes at the free surface. In this case we use the regularisation
\[\eta = \frac{\mu(I)p}{\sqrt{\|D(\vu)\|^2 + \delta^2}},\]
where $\delta>0$ is a small parameter (see \cite{EngelmanReg}).\\

\begin{figure}[!h]
 \begin{center}
 \includegraphics[scale = 0.41]{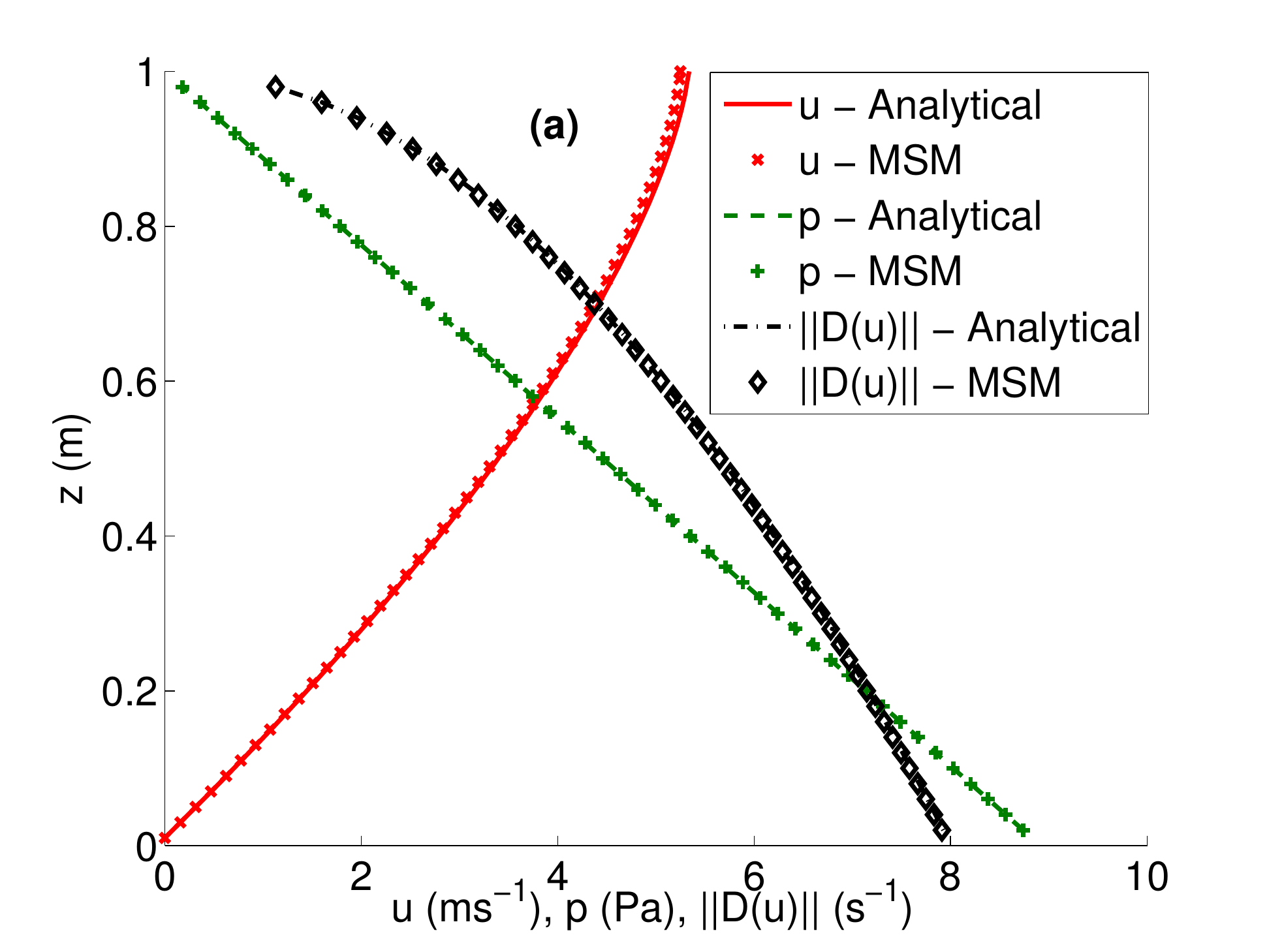}
 \includegraphics[scale = 0.41]{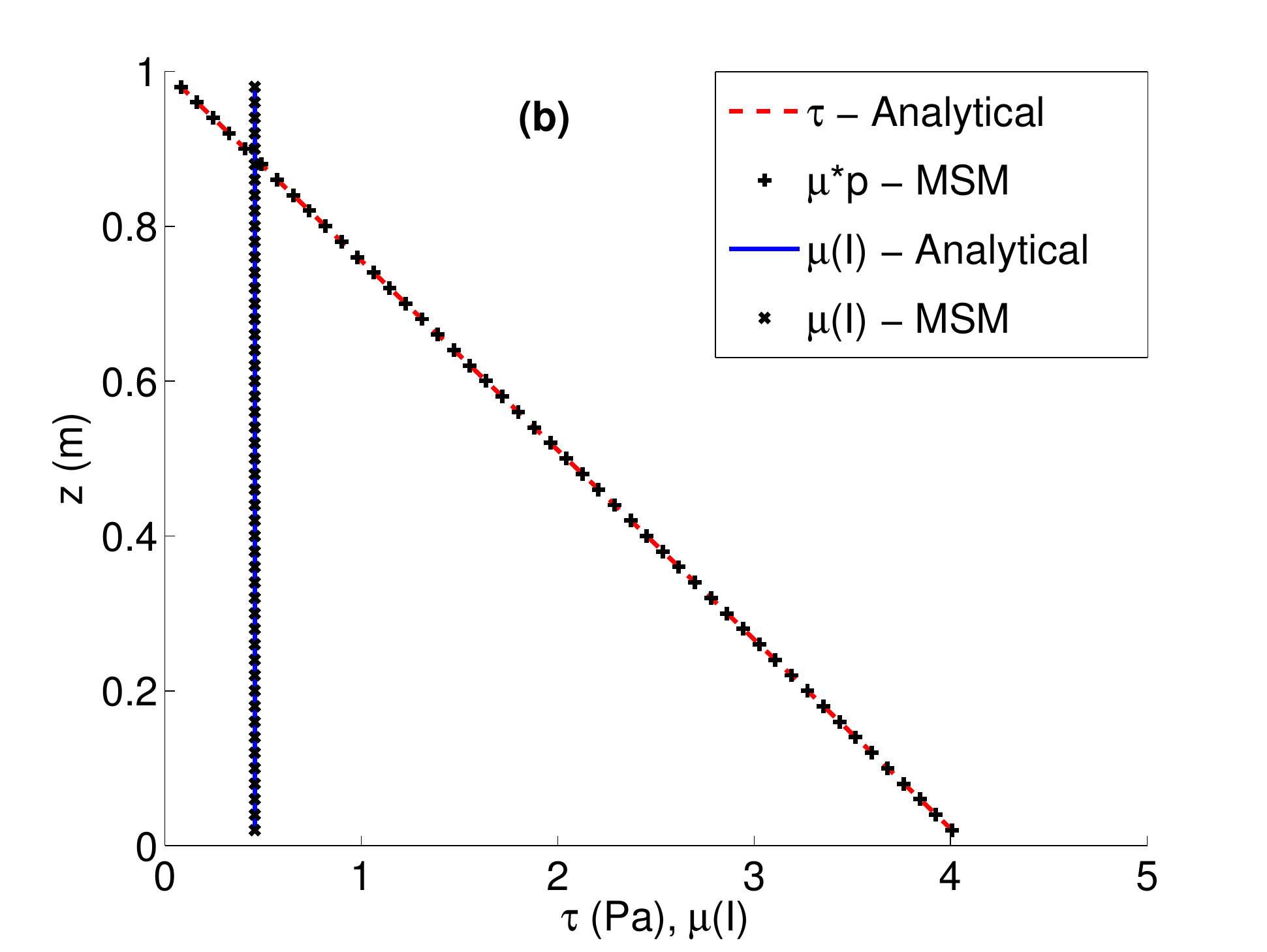}
\\
\includegraphics[scale = 0.41]{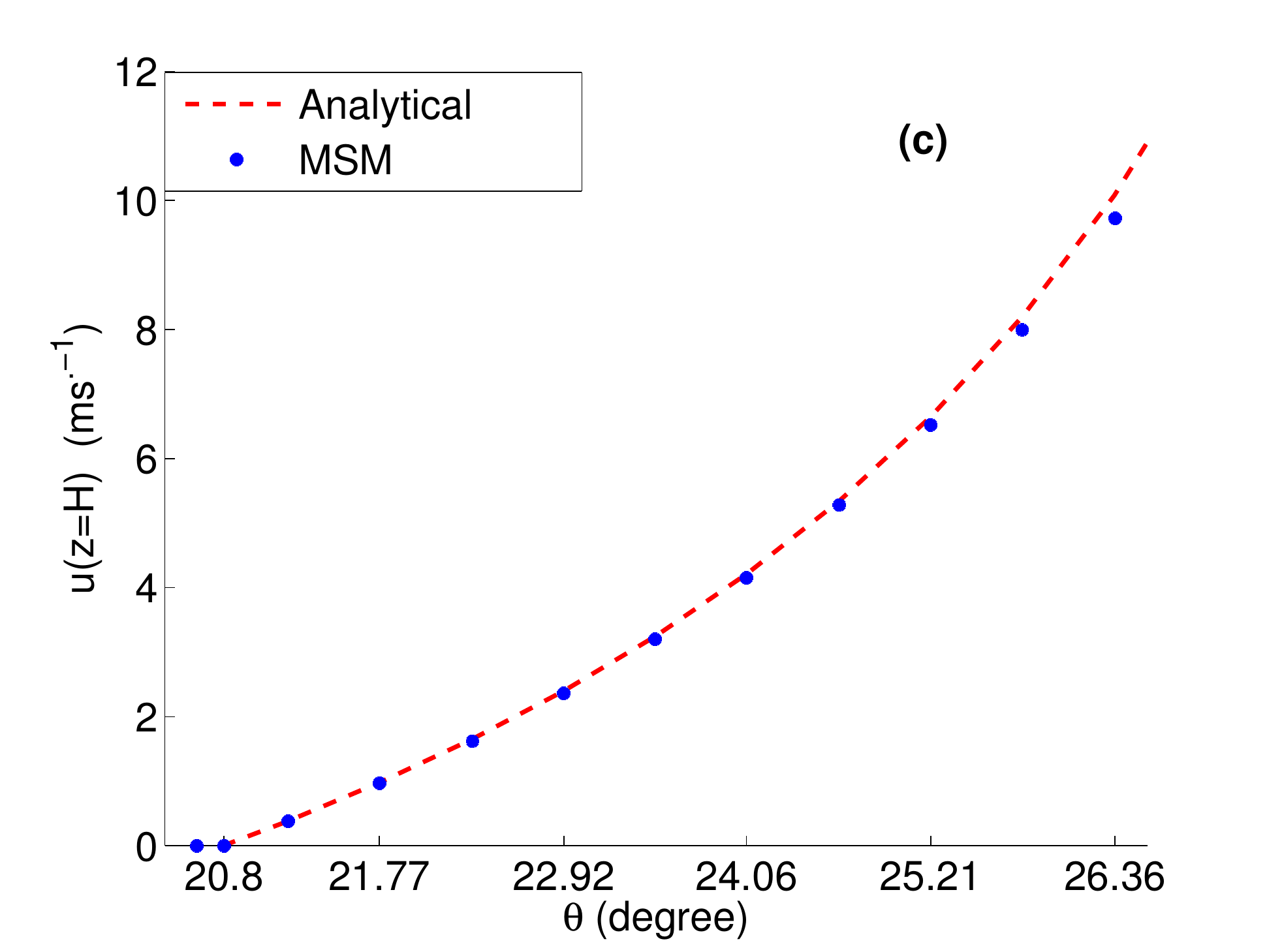}
 \caption{\label{fig:solAnal1} \footnotesize \textit{Comparison between the analytical solution (dashed and solid lines) and the simulations obtained using the MSM with the $\mu(I)$ rheology (symbols). (a) Analytical and simulated downslope horizontal velocity $u$, pressure $p$ and strain rate $\|D(\vu)\|$; (b) Analytical and simulated shear stress and friction coefficient $\mu(I)$; (c) Comparison between the simulated (symbols) and the exact (dashed line) horizontal velocity at the free surface as a function of the slope angle.}}
 \end{center}
 \end{figure}

\noindent We choose the rheological parameters $I_0 = 0.279$ and $\mu_s = 0.363 \approx \tan(19.95^{\circ})$, $\mu_2 = 0.74 \approx \tan(31.8^{\circ})$ and the particle diameter $d_s = 4$ cm with solid volume fraction $\varphi_s = 0.62$. The slope angle is taken as $\theta = 0.43 \mbox{ rad} \approx 24.64^\circ$. Figure \ref{fig:solAnal1} shows the good agreement between the simulated and exact solutions for the profiles of the velocity, pressure, shear stress, $\mu(I)$ and $\|D(u)\|$. It also shows the downslope velocity at the free surface as a function of the slope angle. These results are computed using 50 layers in the MSM. \\

\begin{figure}[!h]
 \begin{center}
 \includegraphics[scale = 0.41]{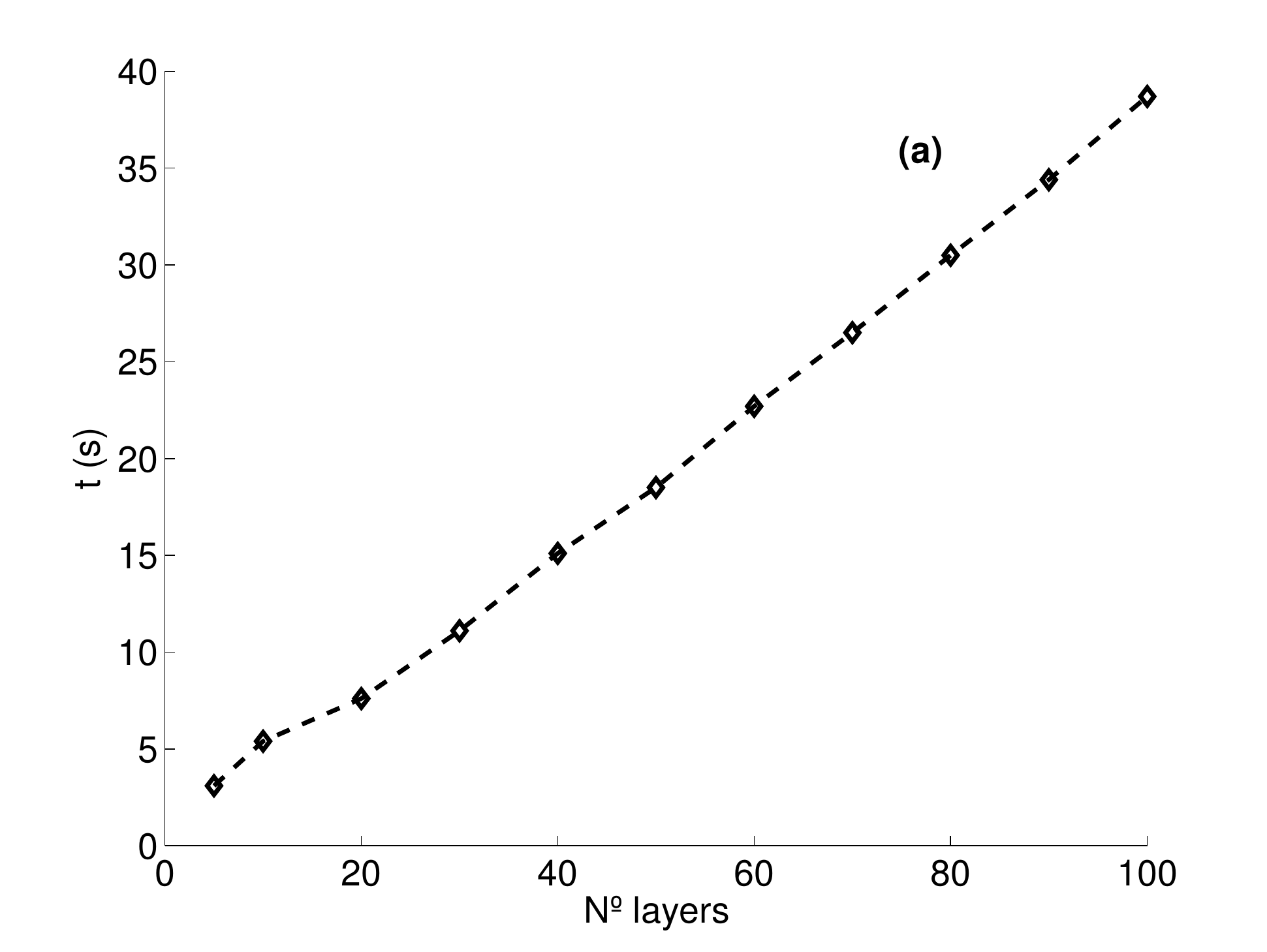}
 \includegraphics[scale = 0.41]{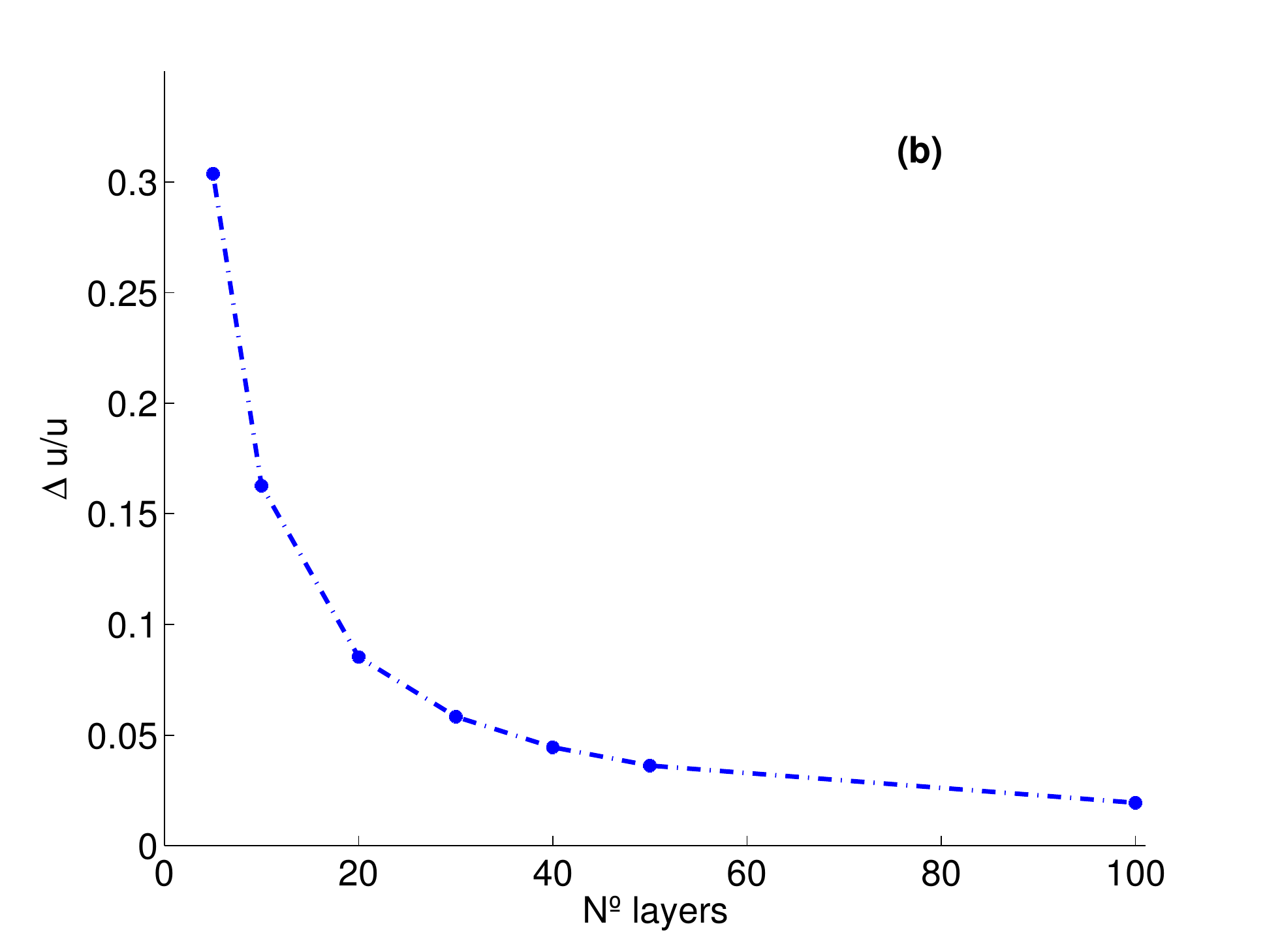}
 \caption{\label{fig:solAnal2} \footnotesize \textit{(a) Computing time as a function of the number of layers in the MSM and (b) relative error between the computed and exact velocity for simulations over a slope of $\theta=24.63^\circ$.}}
 \end{center}
 \end{figure}

Figure \ref{fig:solAnal2} shows the computing time required to simulate 50 seconds and the relative error between the computed velocity and the exact solution using a different number of layers. The error is computed by
\begin{equation}\label{eq:error}
\dfrac{\Delta u}{u} = \sqrt{\dfrac{\sum_{i=0}^{M}(u_{ex,i}-u_{sim,i})^2}{\sum_{i=0}^{M}u_{ex,i}^2}},
\end{equation}
where $u_{ex}$ (respectively $u_{sim}$) is the analytical (respectively computed) velocity and $M$ is the number of partitions of the mesh in the horizontal direction (in this case $M=20$). Note that for slopes smaller than arctan$(\mu_s)$, the surface velocity is zero because the mass does not flow. The error decreases as the number of layers increases and is less than 10$\%$ for 20 layers. The main error occurs near the free surface where the gradient of the horizontal velocity is large.

\subsection{Comparison with laboratory experiments}
We will now use the multilayer shallow model to simulate the laboratory experiments performed in \cite{MangeneyErosion}. The objectives are threefold: (1) to evaluate if the model with the $\mu(I)$ rheology  gives a reasonable approximation of the flow dynamics and deposits of real granular flows, (2) to observe if it reproduces the increase in runout distance observed for increasing thickness of the erodible bed above a critical slope angle $\theta_c \in [12^o, 16^o]$ and (3) to show how the multilayer approach improves the results compared to the classical depth-averaged Saint-Venant model (i.e. monolayer model).\\


\begin{figure}[!h]

 \begin{center}
 \includegraphics[scale = 0.3]{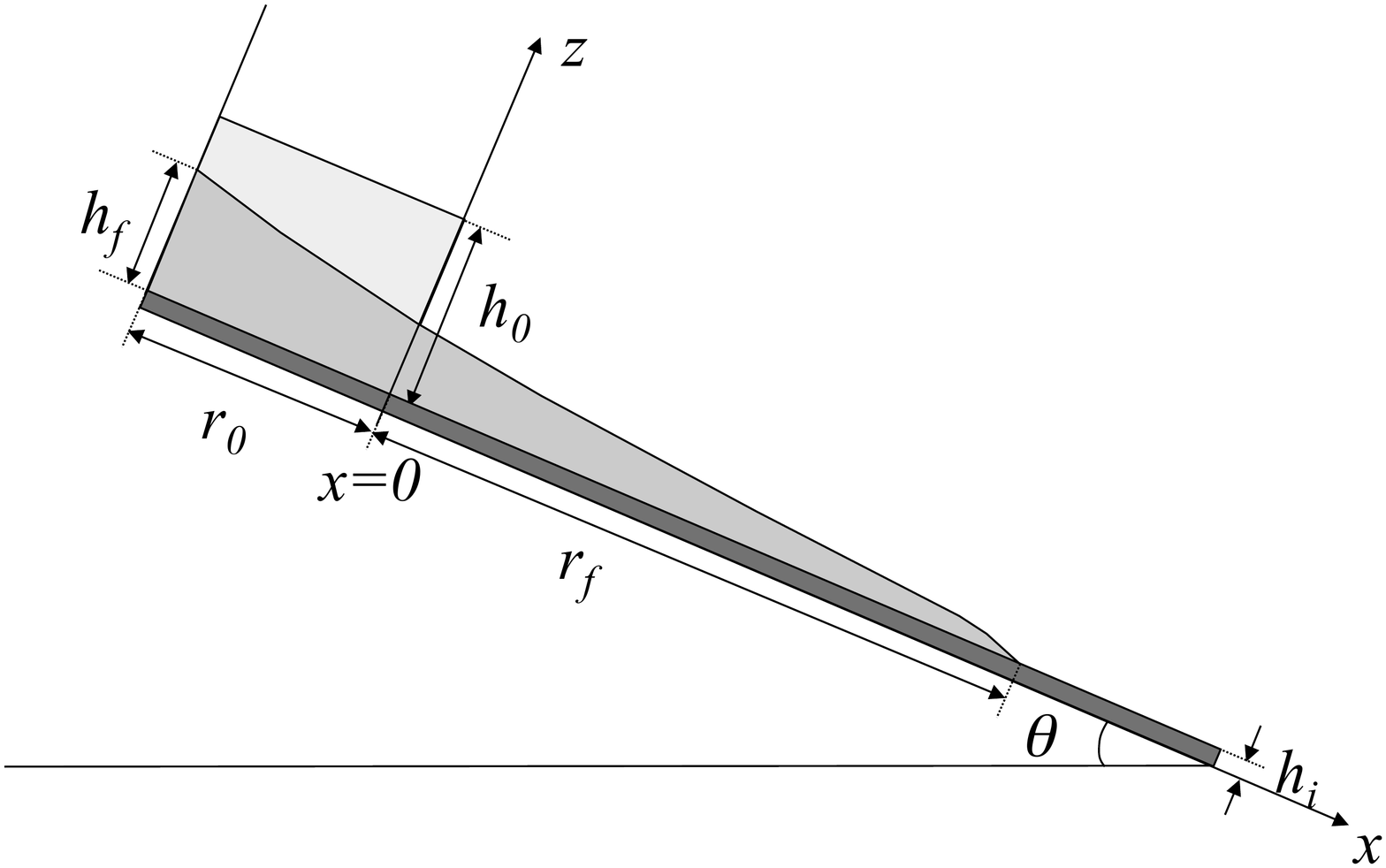}
 \caption{\footnotesize \label{fig:granularColapse}\textit{Sketch of the initial and final state of the granular collapse. A granular column with a thickness $h_0 = 14$ cm and a length $r_0=20$ cm is released on an inclined plane of slope $\theta$. The plane is covered by an erodible bed of thickness $h_i$ made of the same material. When the flow stops, the maximum final thickness is $h_f$ and its final extent $r_f$.}}
 \end{center}
 \end{figure}

The variable $r_f$ denotes the runout distance, i.e. the length of the deposit measured from the position of the front of the released material at the initial time located at $x=0$, $t_f$ denotes the flow time from $t=0$ s to the time when the material stops and $h_f$ denotes the maximum final thickness of the deposit (see Figure \ref{fig:granularColapse}).\\

In the laboratory experiments performed in \cite{MangeneyErosion}, subspherical glass beads of diameter $d_s=0.7$ mm were used. They were cohesionless and highly rigid. The particle density $\r_s = 2500\ kg m^{-3}$ and volume fraction $\varphi_s = 0.62$ were estimated, leading to an apparent flow density $\r = \varphi_s \r_s = 1550\  kg m^{-3}$.\\

In order to use the $\mu(I)$ rheology, the rheological parameters ($\mu_s,\ \mu_2$ and $I_0$) are taken as in \cite{IonescuMangeney}, according to the measurements made in the experiments of \cite{MangeneyErosion} and \cite{PouliquenForterre}, where the effect of lateral wall friction is taken into account empirically. These parameters can be obtained by fitting the curve $h_{stop}(\theta)$, where $h_{stop}$ is the thickness of the deposit lying on the slope when the supply is stopped after steady uniform flow (see \cite{Pouliquen1999a} for more details). As a result, we take here $I_0 = 0.279$, as in \cite{JopForterrePouliquen}, $\mu_s =$ tan$(25.5^{\circ}) \approx 0.48$ and $\mu_s = 0.74 \approx$ tan$(36.5^{\circ})$.\\

This experiment has been simulated for different slopes $\theta$ and thicknesses $h_i$ of the erodible bed: $\theta = 16^{\circ}$ and $h_i = 1.4,\ 2.5,\ 5$ mm, $\theta = 19^{\circ}$ and $h_i = 1.5,\ 2.7,\ 5.3$ mm, $\theta = 22^{\circ}$ and $h_i = 1.82,\ 3.38,\ 4.6$ mm, $\theta = 23.7^{\circ}$ and $h_i = 1.5,\ 2.5,\ 5$ mm. Note that the model does not take into account the effect of removing the gate during the initial instants even though it has a non-negligible impact on the flow dynamics as shown in \cite{IonescuMangeney}. For instance, when the gate is taken into account, even with no friction along it, the flow is substantially slowed down however the deposit is almost unchanged. All the simulations are performed using 20 layers.\\

We compare hereafter (i) the constant and variable friction rheologies and (ii) the monolayer and multilayer approaches.

\subsubsection{Deposit profiles}
\begin{figure}[!h]
\begin{center}
\includegraphics[width=1\textwidth]{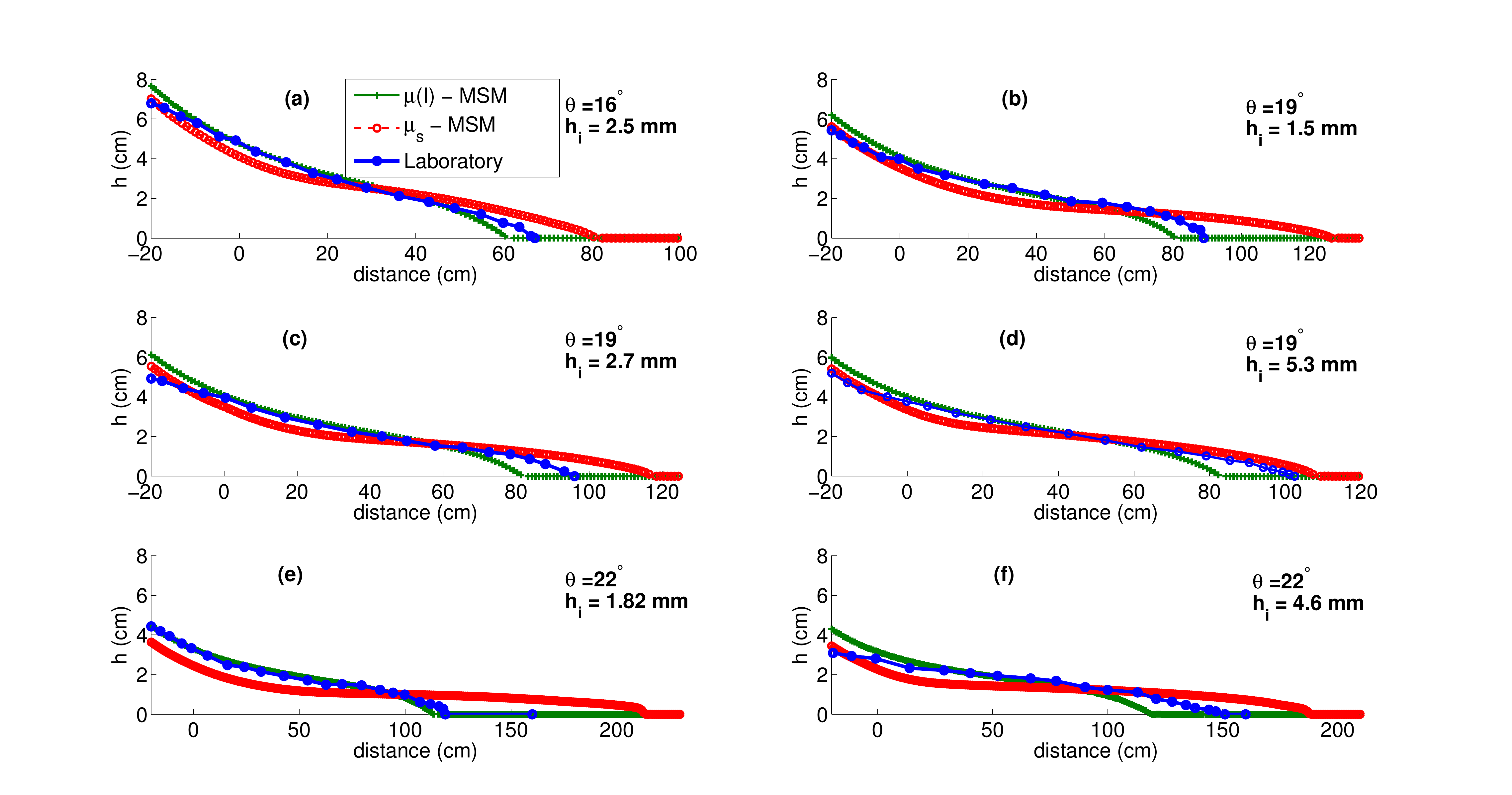}
\end{center}
\caption{\footnotesize \label{fig:muI_mus} \it{Deposit obtained in the experiments (solid-circle blue line) and with the Multilayer Shallow Model using a constant friction coefficient $\mu_s$ (dotted-circle red line) and a variable friction coefficient $\mu(I)$ (solid-cross green line), for different slopes $\theta$ and erodible bed thicknesses $h_i$.}}
 \end{figure}

Let us compare the deposits simulated with the $\mu(I)$ rheology and with a constant friction coefficient $\mu_s$ for different slopes $\theta$ and erodible bed thicknesses $h_i$. Figure \ref{fig:muI_mus} shows that the deposit calculated with the variable friction coefficient $\mu(I)$ is closer to the experimental deposit than the one calculated with a constant friction coefficient $\mu_s$. The runout distance with the constant coefficient $\mu_s$ is always too long except at $\theta=19^o$ and $h_i=5.3$ mm (see Figure \ref{fig:muI_mus}d). To properly reproduce the runout distance with a constant friction coefficient, we need to increase its value. For example, with a slope $\theta = 16^\circ$ and an erodible bed thickness $h_i = 2.5$ mm (Figure \ref{fig:muI_mus}a), we need to use the value $\mu_s =\mbox{tan}(27.3^\circ)$ to produce the runout observed in the laboratory experiments.\\

\begin{figure}[!h]
\begin{center}
\includegraphics[width=0.7\textwidth]{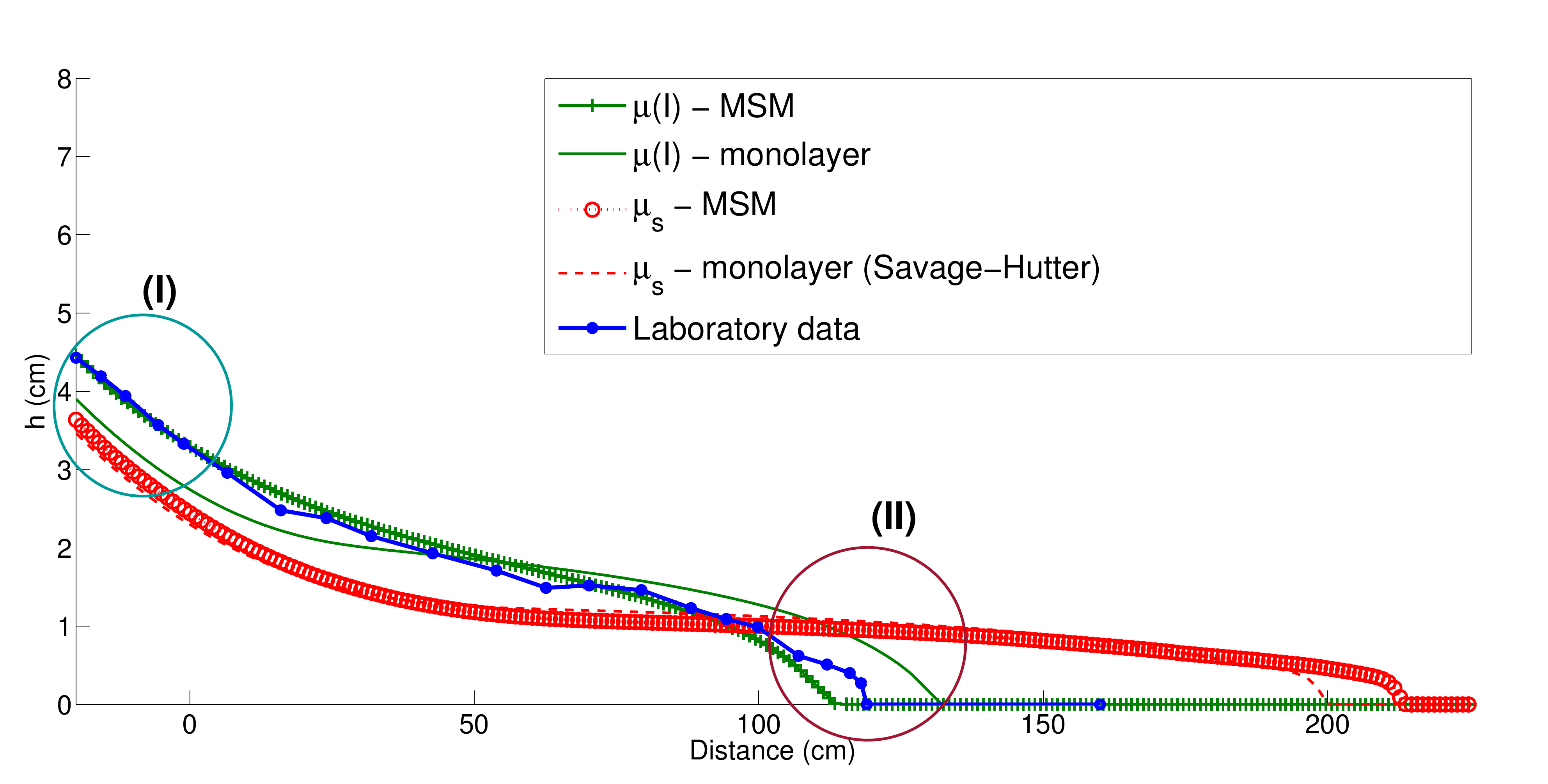}
\end{center}
 \caption{\footnotesize \label{fig:muI_mus_22grad} \it{Deposit obtained in the experiments (solid-circle blue line), with the Multilayer Shallow Model using a constant friction coefficient $\mu_s$ (dotted-circle red line) and a variable friction coefficient $\mu(I)$ (solid-cross green line) and with the monolayer model with $\mu_s$ (dashed red line) and $\mu(I)$ (solid green line) for a slope $\theta = 22^\circ$ and an erodible bed thickness $h_i=1.82$ mm.}}
 \end{figure}

Figure \ref{fig:muI_mus_22grad} shows, for a slope $\theta = 22^\circ$ and $h_i = 1.82$ mm, the final deposit obtained using the constant or variable friction coefficients for both the multilayer and monolayer models. The difference between the multilayer and monolayer models is stronger when using the $\mu(I)$ rheology. For instance, the monolayer approach changes the full deposit profiles for the $\mu(I)$ rheology, while it only changes the front position for $\mu_s$. The multilayer approach makes it possible to obtain a deposit shape which is very close to the experiments with the $\mu(I)$ rheology. More generally, the shape of the deposit is closer to the observations with $\mu(I)$ than with $\mu_s$ in the Multilayer Shallow Model.

\subsubsection{Effect of the erodible bed}
\begin{figure}[!h]
\begin{center}
\includegraphics[width=1\textwidth]{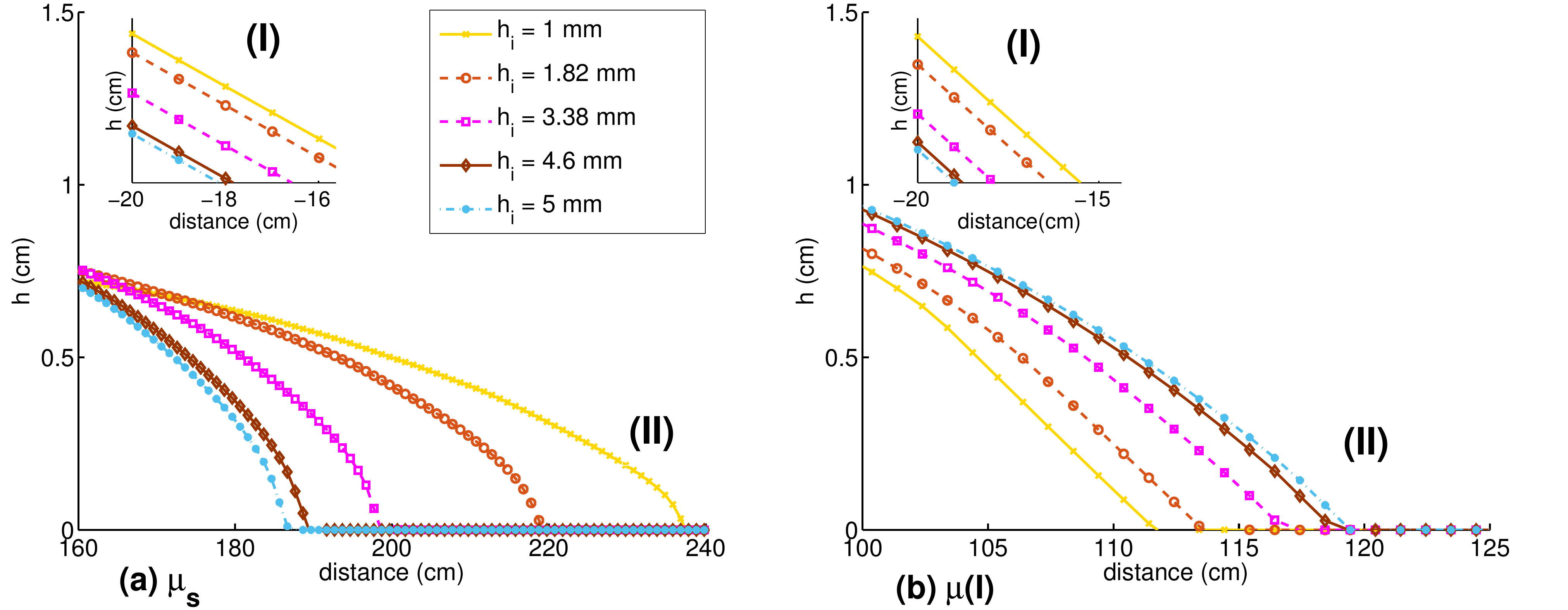}
\end{center}
 \caption{\footnotesize \label{fig:hi_22grad_tf} \it{Influence of the thickness of the erodible bed on the runout distance $r_f$ and on the maximum final thickness $h_f$ (smaller graphs) with the Multilayer Shallow Model using a constant friction coefficient $\mu_s$ (left hand side) and with a variable friction coefficient $\mu(I)$ (right hand side), for a slope $\theta = 22^\circ$.}}
 \end{figure}

Figure \ref{fig:hi_22grad_tf} shows two zooms, one near the front (I) and one near the maximum thickness of the deposit (II), for $\theta = 22^\circ$ and different values of $h_i$ (see Figure \ref{fig:muI_mus_22grad} for the approximate location of these zooms). With the variable coefficient $\mu(I)$, the runout distance $r_f$ increases as the thickness of the erodible bed $h_i$ increases (see Figure \ref{fig:hi_22grad_tf}b(II)) as observed in laboratory experiments. On the other hand, with a constant friction coefficient $\mu_s$ (Figure \ref{fig:hi_22grad_tf}a(II)), the runout distance $r_f$ decreases with increasing $h_i$. Note that in both cases the maximum final thickness $h_f$ decreases with increasing $h_i$ as in the experiments (Figure \ref{fig:hi_22grad_tf}a(I),b(I)).

\begin{figure}[!h]
\begin{center}
\includegraphics[width=1\textwidth]{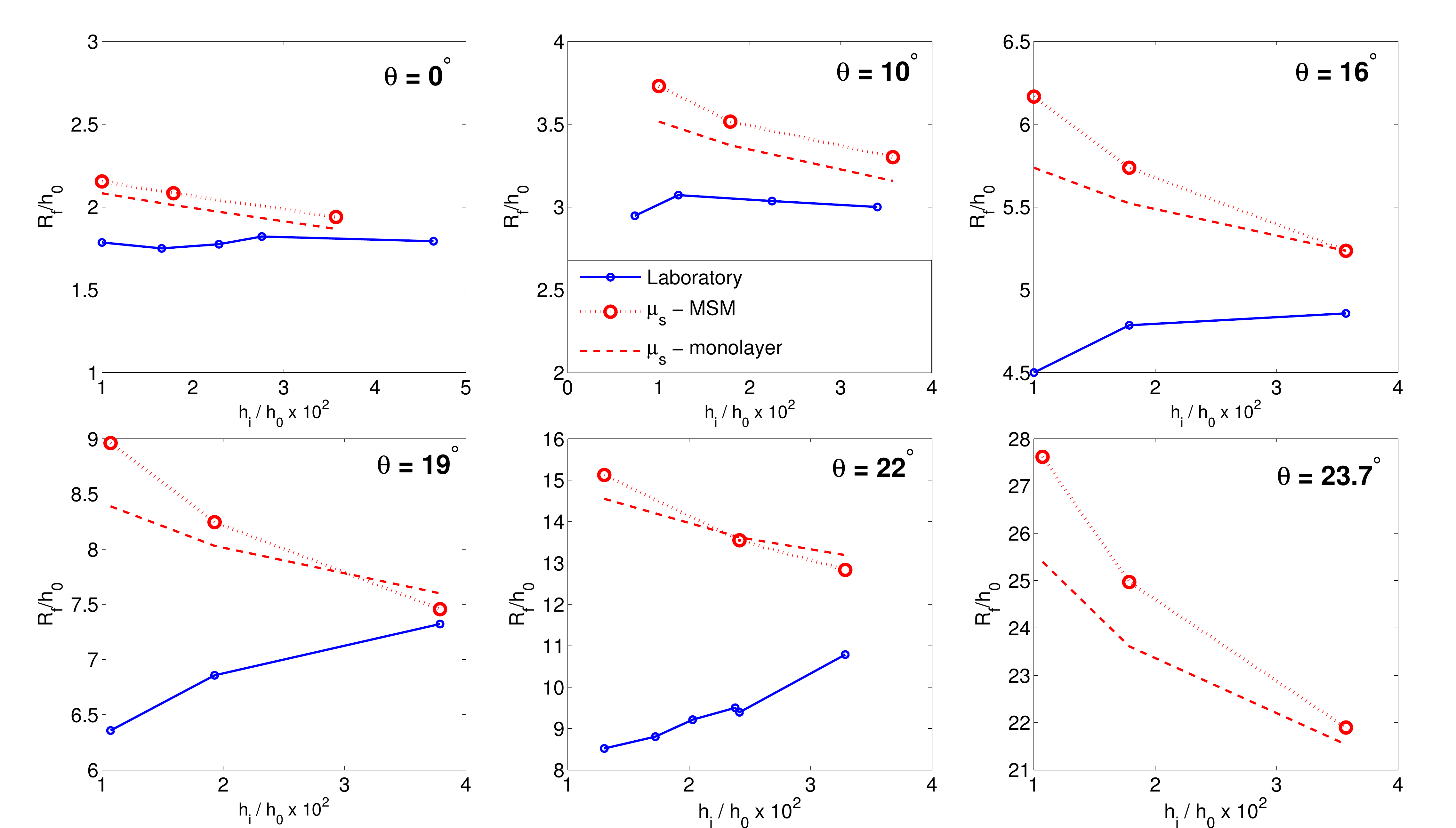}
\end{center}
 \caption{\footnotesize \label{fig:Runout_mus} \it{Influence of the thickness $h_i$ of the erodible bed on the final runout $r_f$ for slopes $\theta =\ 0^{\circ},\ 10^{\circ},\ 16^{\circ},\ 19^{\circ},\ 22^{\circ},\ 23.7^{\circ}$ observed in the experiments of \cite{MangeneyErosion} (solid-circle blue line) and obtained with different simulations using the Multilayer Shallow Model with a constant friction coefficient $\mu_s$, with 20 layers (dotted-circle red line) and with one layer, i.e. the Savage-Hutter model \cite{SavageHutter} (dashed red line). There is no laboratory data for $\theta = 23.7^\circ$. Normalisation using $h_0 = 14$ cm.}}
  \end{figure}

Figure \ref{fig:Runout_mus} shows that the decrease in runout distance with increasing $h_i$ for constant friction $\mu_s$ is observed for all slopes, e.g. $\theta =\ 0^{\circ},\ 10^{\circ},\ 16^{\circ},\ 19^{\circ},\ 22^\circ,\ 23.7^\circ$. For the $\mu_s$-model, the multi- and monolayer models follow the same trend. Note that this nonphysical decrease in runout distance with increasing $h_i$ has been demonstrated analytically in \cite{FaccanoniMangeney} for the monolayer model. Moreover, laboratory experiments show that when the thickness of the erodible bed increases, for slopes $\theta\geq \theta_c$, where $\theta_c\in [12^\circ,16^\circ]$ is a critical slope, the runout distance $r_f$ and the stopping time $t_f$ both increase while the maximum final thickness $h_f$ decreases. Note that there is no trend concerning the runout when the thickness $h_i$ is increased for slopes $\theta < \theta_c$ ($\theta =\ 0^{\circ},\ 10^{\circ}$) in the laboratory experiments.

\begin{figure}[!h]
\begin{center}
\includegraphics[width=1\textwidth]{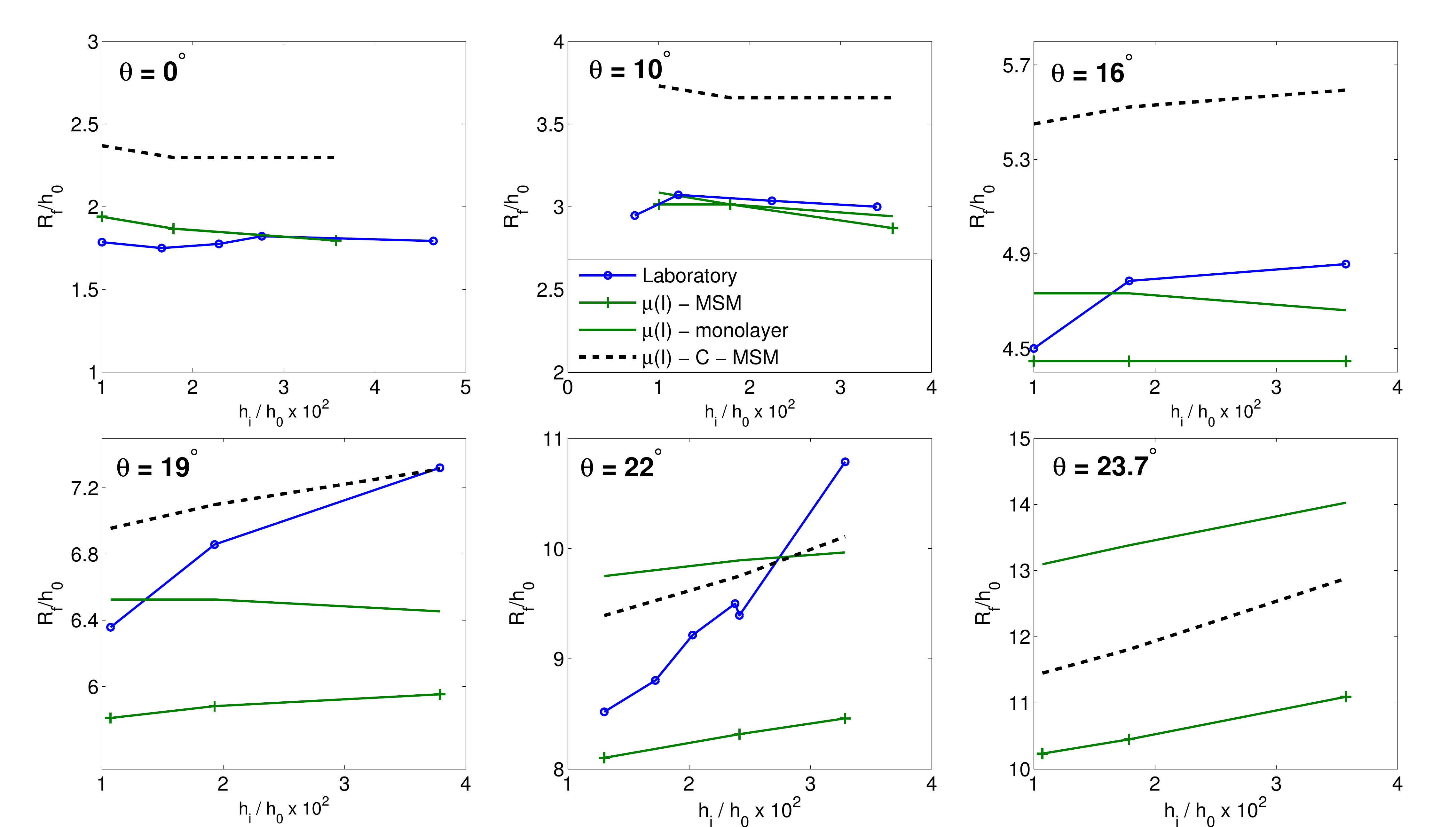}
\end{center}
 \caption{\footnotesize \label{fig:Runout_muI} \it{Influence of the thickness $h_i$ of the erodible bed on the final runout $r_f$ for slopes $\theta = 16^{\circ},\ 19^{\circ},\ 22^{\circ},\ 23.7^{\circ}$ observed in the experiments of \cite{MangeneyErosion} (solid-circle blue line) and obtained with different simulations using the Multilayer Shallow Model with the $\mu(I)$ rheology, with 20 layers (solid-cross green line) and with one layer (solid green line) and with the Multilayer Shallow Model with the correction of $\|D(\vu_a)\|$ (dashed black line). Normalisation using $h_0 = 14$ cm.}}
 \end{figure}

Figure \ref{fig:Runout_muI} shows that the increase of runout distance observed in the experiments for increasing $h_i$ is qualitatively well reproduced with the $\mu(I)$ Multilayer Shallow Model. With the $\mu(I)$ Multilayer Shallow Model, the runout increase with $h_i$ is actually larger for higher slopes, as observed experimentally: at $\theta=16^\circ$, the runout distance is almost unaffected by the thickness of the erodible bed while it increases by $26.9\%$ at $\theta=22^\circ$ when the thickness of the erodible bed increase from $1.82$ mm to $4.6$ mm.  Note that in the $\mu(I)$ MSM, the increase of the runout distance appears on slopes $\theta > 16^\circ$, higher than $\theta_c$ in the experiments. Actually, it appears starting with the slope $\theta = 18^\circ$. When using the $\mu(I)$ monolayer model, the runout distance is higher than for the Multilayer Shallow Model whatever the slope and thickness of the erodible bed. Based on the values of the runout distance in these cases, it is hard to discriminate which of the monolayer or multilayer models is closer to the experiments. However, in the $\mu(I)$ monolayer model, the runout distance at $\theta = 16^{\circ},\ 19^{\circ}$ decreases when $h_i$ increases, contrary to the experimental data. For $\theta=22^\circ$ and $\theta=23.7^\circ$, the monolayer and multilayer $\mu(I)$ models reproduce qualitatively the increase in runout with $h_i$. Note that for $\theta = 0^\circ, 10^\circ$ ($\theta < \theta_c$), the $\mu(I)$ models predict a very slight decrease in the runout distance.

As a result, the Multilayer Shallow Model with the $\mu(I)$ rheology provides the results that are the closest to observations even though the effect of erosion is still much smaller than in the experiments (the runout distance increases by $4.4\%$ for a slope $\theta = 22^\circ$ and from $1.82$ mm to $4.6$ mm of thickness of the erodible bed, while it increases by $26.9\%$ in the experiments).\\

\begin{figure}[!h]
\begin{center}
\includegraphics[width=0.8\textwidth]{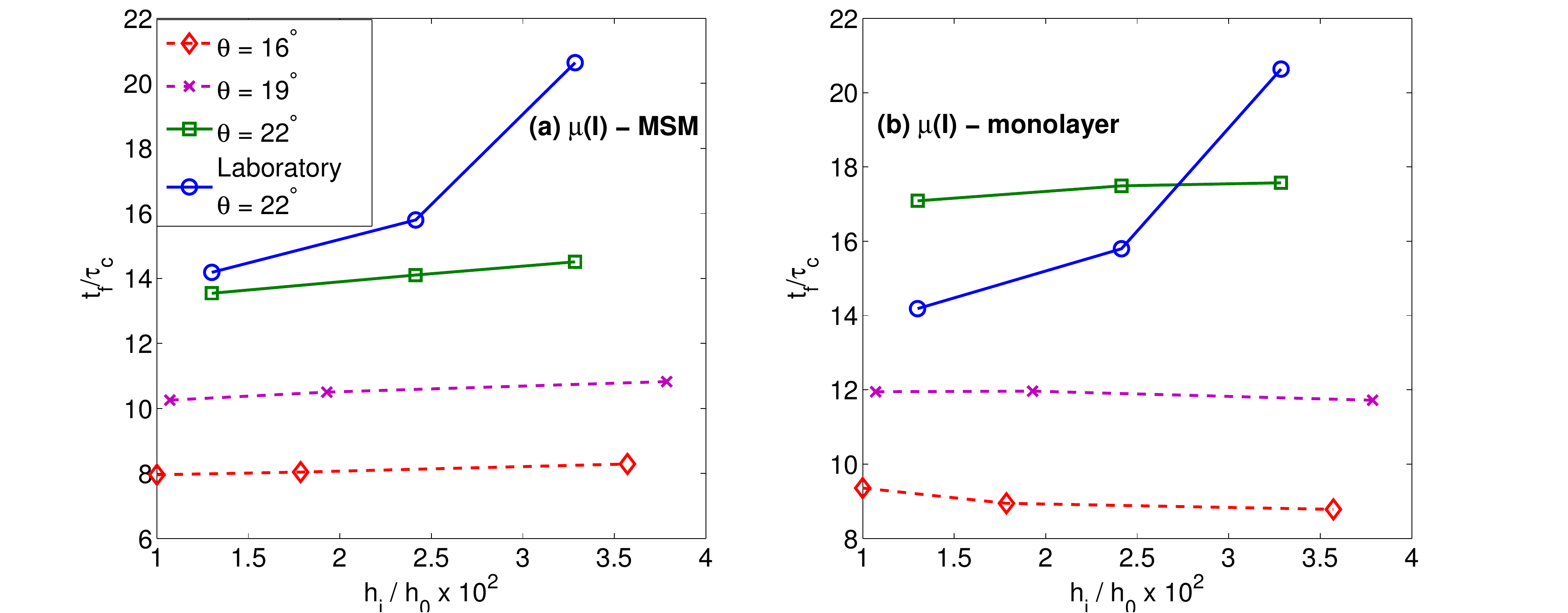}
\end{center}
 \caption{\footnotesize \label{fig:tiempos} \it{Influence of the thickness $h_i$ of the erodible bed on the final time $t_f$, for slopes $\theta = 16^{\circ}$ (dashed-diamond red line), $\theta= 19^{\circ}$ (dashed-cross magenta line), $\ 22^{\circ}$ (solid-square green line) and the values observed in the experiments of \cite{MangeneyErosion} for slope $\theta = 22^{\circ}$ (solid-circle blue line), obtained using (a) the Multilayer Shallow Model and (b) the monolayer model, with the $\mu(I)$ rheology with 20 layers. Normalisation using $\tau_c = \sqrt{h_0/(g\,cos\theta)}$ and $h_0 = 14$ cm.}}
 \end{figure}

In Figure  \ref{fig:tiempos}, the final time (time at which the front stops) is plotted as a function of the thickness of the erodible bed for $\theta=16^o$, $\theta=19^o$ and $\theta=22^o$. Moreover, for $\theta=22^o$,  we also plot the experimental data. Experimental data show that the final time increases when the thickness of the erodible bed increases. In Figure  \ref{fig:tiempos}a, we can see that this is true for all the values of $\theta$ for the multilayer method. However, in Figure  \ref{fig:tiempos}b, for the monolayer model, we observe that it is only true for the highest value, $\theta=22^o$. At the same time, the final time decreases when the erodible bed increases for $\theta=16^o$ and $\theta=19^o$.\\

The advantage of the multilayer models is that we obtain a variable profile of the downslope velocity, in contrast with the constant profile of the monolayer model. It makes it possible to obtain a better approximation of $\|D(\vu)\|$ (see equation (\ref{eq:approx_norm_Du_o1})). As a consequence, this improves the approximation of the inertial number $I$ (see equations (\ref{eq:def_I}) and (\ref{eq:def_I_interfaz})), which is a key number in the variable friction coefficient with $\mu(I)$.\\

As the main advantage of the multilayer model is the improvement of the approximation of $\|D(\vu)\|$, we present two approximations that can be made with the multilayer model. First, let us recall that a first order approximation corresponds to the definition (\ref{eq:approx_norm_Du_o1}). This approximation considers only the leading order term, i.e. $\|D(\vu)\|_{z=z_{\a+\frac{1}{2}}}  \approx \|\p_{z}\vu_{H} |_{z=z_{\a+\frac{1}{2}}} \|=\| Q_{H,\a+\frac{1}{2}} \|$. Note that in dimensionless form, we have
\begin{equation}\label{eq:normAdim}
\|D(\vu)\| = \sqrt{\frac{1}{\varepsilon^2}\left(\p_{z}\vu_{H}\right)^2 + 4\left(\p_{x}\vu_{H}\right)^2 + 2\p_x w \p_{z}\vu_{H} + \varepsilon^2 \left(\p_{x}w \right)^2}.
\end{equation}

We can improve the approximation of $\|D(\vu)\|$ at the interfaces $z=z_{\a+\frac{1}{2}}$ by considering the approximation taking into account second order terms in the previous equation. For the numerical tests, we consider the following approximation $\|D(\vu)\|$ at the interfaces,

\begin{equation}
\label{eq:norm_amp}
\|D(\vu)\|_{z=z_{\a+\frac{1}{2}}}   \approx \sqrt{\| Q_{H,\a+\frac{1}{2}}\|^2  +      \left(\p_{x}  ( \vu_{H,\a+1}+ \vu_{H,\a} )   \right)^2}.
\end{equation}
Note that this definition corresponds to an approximation of
$$
\|D(\vu)\| \approx \sqrt{\left(\p_{z}\vu_{H}\right)^2 + 4\left(\p_{x}\vu_{H}\right)^2}.
$$
at $z=z_{\a+\frac{1}{2}}$. Nevertheless, in \eqref{eq:normAdim}, the term $2\p_z\vu_{H} \p_x w $ is not taken into account although it is of the same order as $4(\p_x \vu_{H})^2$. This is because when an approximation of this term is added, we obtain results that are very similar to those obtained when considering \eqref{eq:norm_amp}. Furthermore adding this term implies an additional computational cost since pre-calculated vertical velocities are required. Note that \eqref{eq:norm_amp} is a second order correction while we have developed a first order model that neglects other second order terms. This correction however highlights the importance of second order terms in granular collapses over erodible beds. \\

The model corresponding to the multilayer approximation with the $\mu(I)$ rheology will hereafter be denoted $\mu(I)$-MSM when $\|D(\vu)\|$ is approximated by  (\ref{eq:approx_norm_Du_o1}). When $\|D(\vu )\|$ is approximated by the correction (\ref{eq:norm_amp}), we denote the model $\mu(I)$-C-MSM. Figure \ref{fig:Runout_muI} shows that the correction of $\|D(\vu)\|$ corresponding to $\mu(I)$-C-MSM improves the simulation of both the runout extent and the influence of the erodible bed. They both increase, leading to a better agreement with laboratory experiments. \\

\subsubsection{Flow dynamics and velocity profiles}

\begin{figure}[!h]
\begin{center}
\includegraphics[width=1\textwidth]{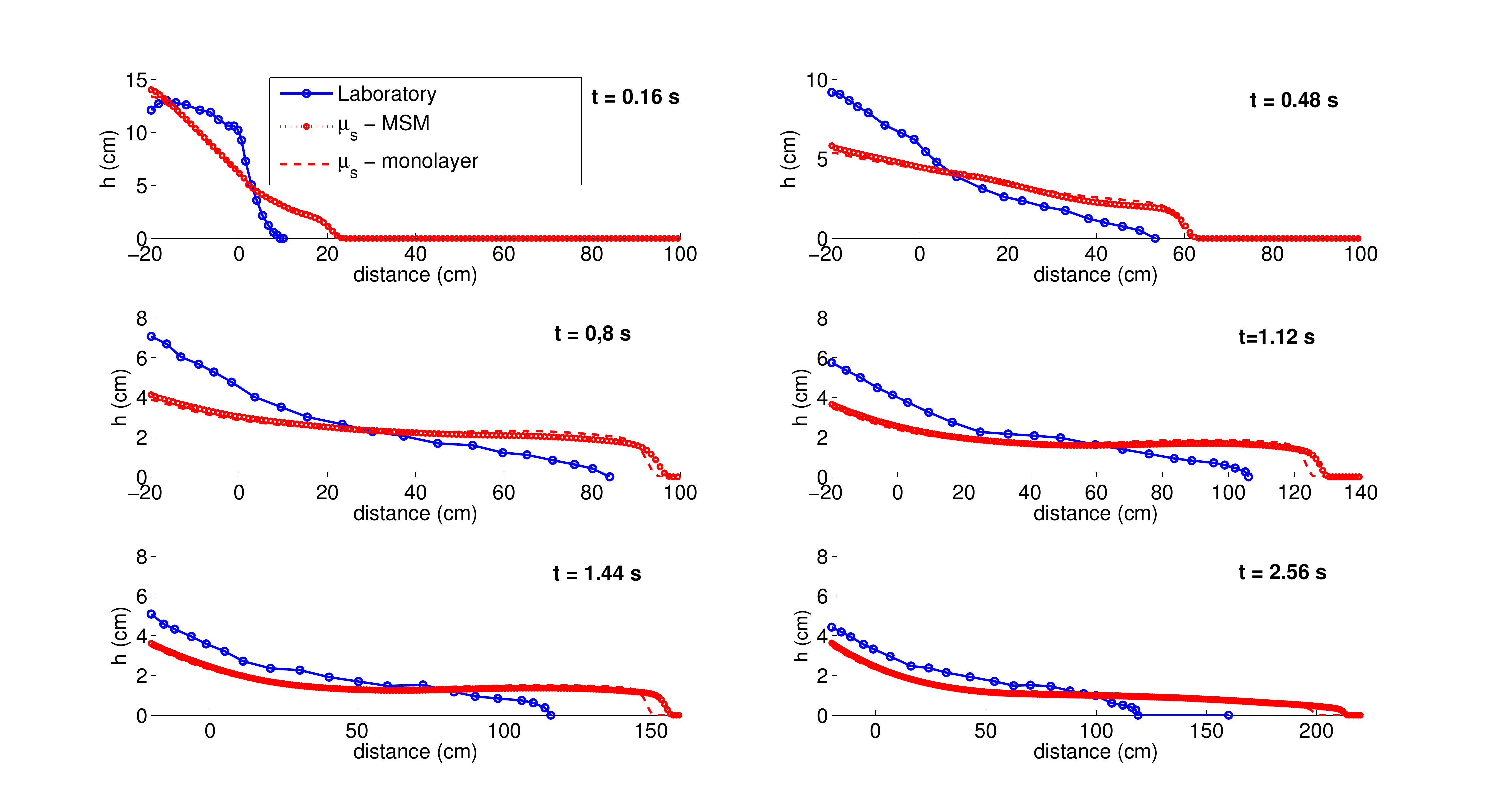}
\end{center}
  \caption{\footnotesize \label{fig:mus_1c_20c} \it{Thickness of the granular mass at different times in the experiments (solid-circle blue line) and with the Multilayer Shallow Model using a constant friction coefficient $\mu_s$ and either 20 layers (dotted-circle red line) or one layer (Savage-Hutter model, dashed red line), for the slope $\theta = 22^\circ$ and erodible bed thickness $h_i = 1.82$ mm.}}
 \end{figure}

\begin{figure}[!h]
\begin{center}
\includegraphics[width=1\textwidth]{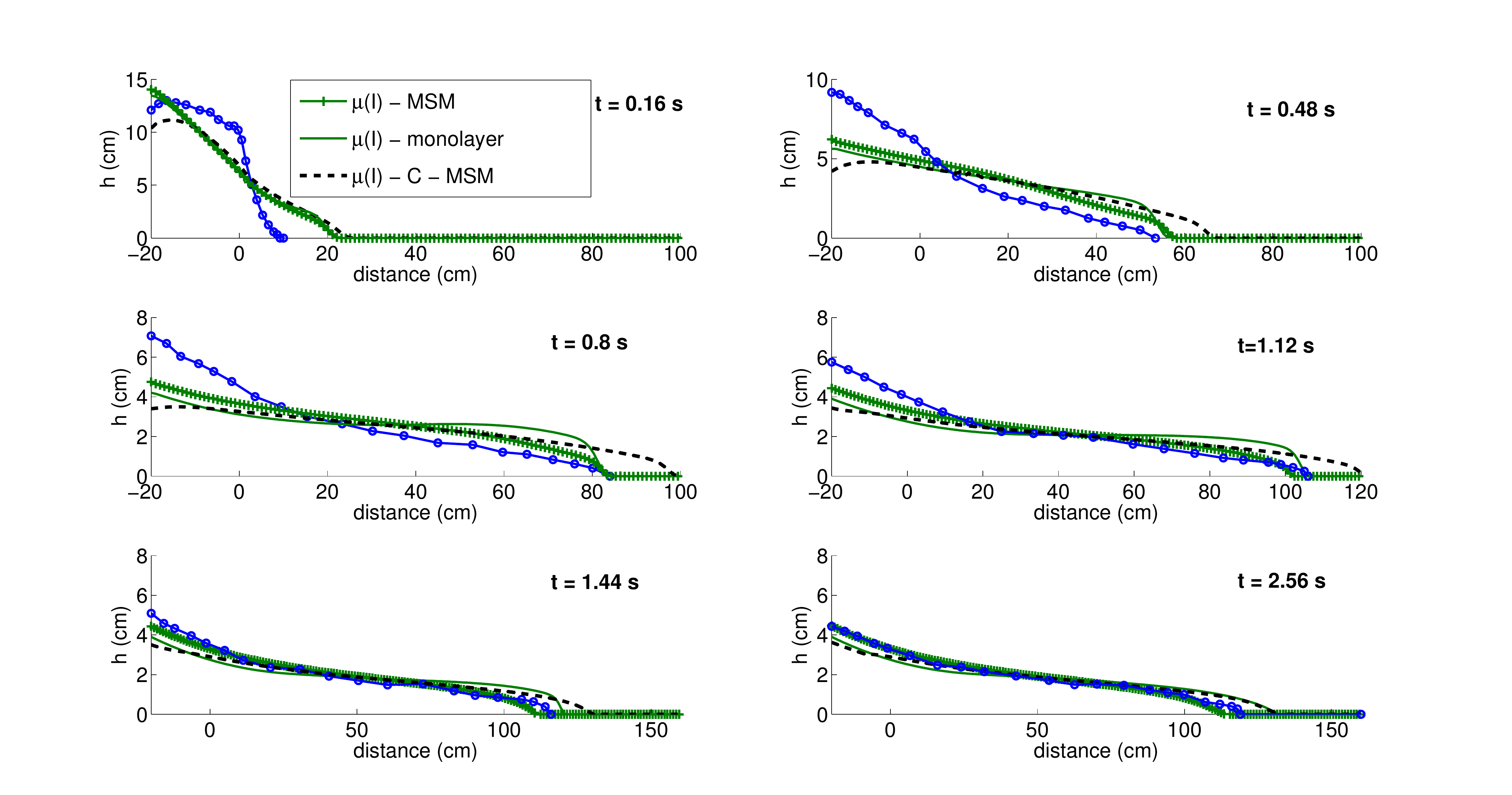}
\end{center}
 \caption{\footnotesize \label{fig:muI_1c_20c} \it{Thickness of the granular mass at different times in the experiments (solid-circle blue line), with the $\mu(I)$ Multilayer Shallow Model using either 20 layers (solid-cross green line) or one layer (solid green line) and with the correction of $\|D(\vu_a)\|$ (dashed black line) for the slope $\theta = 22^\circ$ and erodible bed thickness $h_i = 1.82$ mm. }}
 \end{figure}

Figures \ref{fig:mus_1c_20c} and \ref{fig:muI_1c_20c} show the time change of the granular column thickness for a slope $\theta = 22^\circ$ and an erodible bed of thickness $h_i=1.82$ mm for $\mu_s$ and $\mu(I)$, respectively, for both the monolayer and multilayer models. As observed for the deposit, the difference between the thickness profiles simulated with the multilayer and the monolayer model is stronger for $\mu(I)$ than for $\mu_s$. The $\mu(I)$-MSM makes it possible to increase the maximum thickness of the flow and decrease the thickness of the front. This is an important result as the shape of the front may be an indicator of the flow rheology \cite{Pouliquen1999b}, \cite{Jessop2012}. When a constant coefficient $\mu_s$ is used, very similar profiles are obtained with the Multilayer Shallow Model and monolayer model (Savage-Hutter model). As a result, the multilayer approach does not significantly improve the results when a constant friction coefficient is used. Note that during the initial instants, the simulated mass spread faster than in the experiments. This is partly due to the role of initial gate removal that is not taken into account here. However, this effect could not explain the strong difference between the simulation and experiments (see \cite{IonescuMangeney} for more details). The hydrostatic assumption may also be responsible for this overestimation of the spreading velocity (see e.g. \cite{MangeneySpreading}).\\

\begin{figure}[!h]
\begin{center}
\includegraphics[width=1\textwidth]{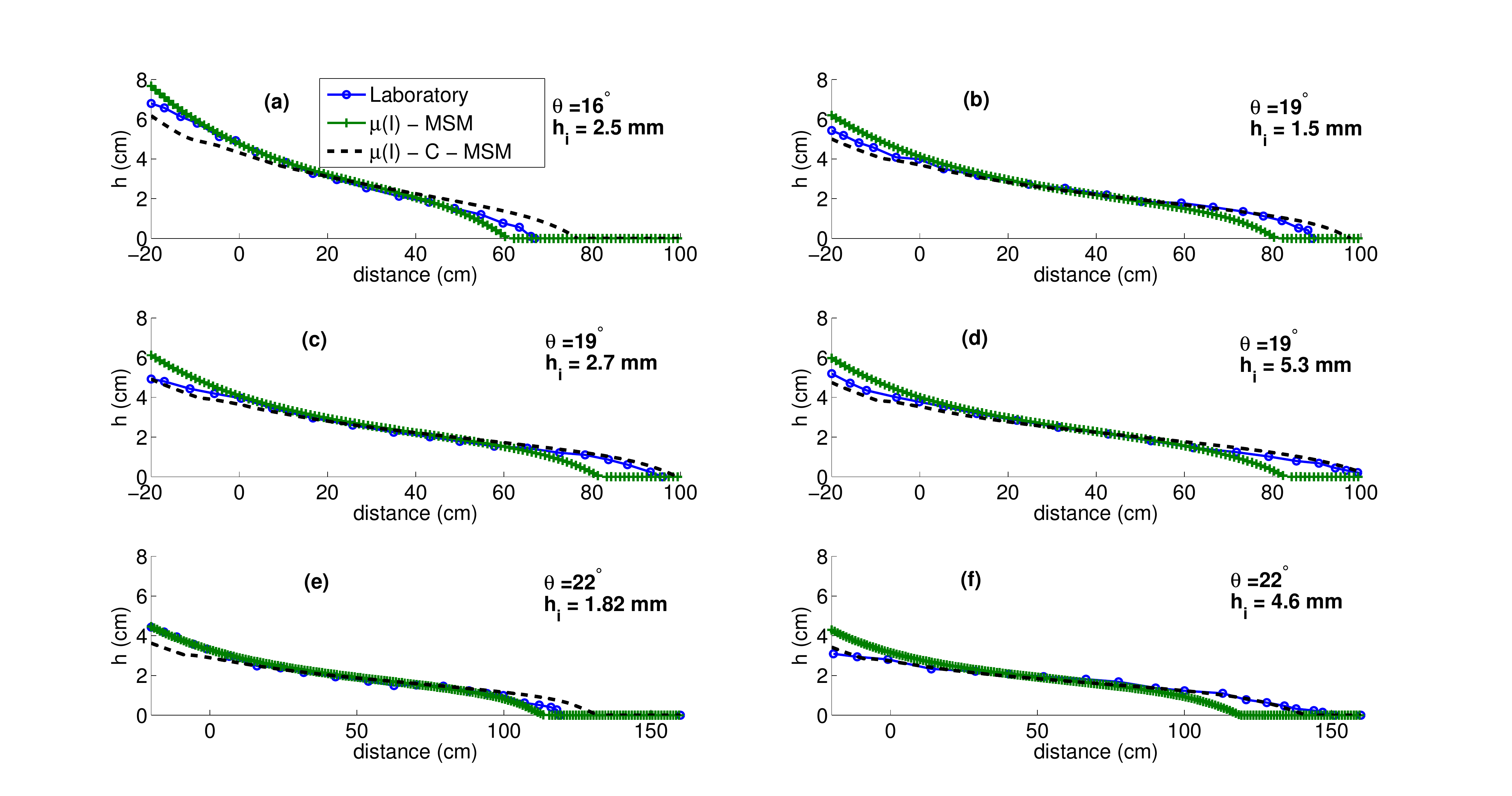}
\end{center}
 \caption{\footnotesize \label{fig:muI_muImod} \it{Simulated deposits at different slopes $\theta$ and erodible bed thicknesses $h_i$ with the Multilayer Shallow Model using the higher order correction of $\|D(\vu_\a)\|$ (\ref{eq:norm_amp}) (dashed black line) and without this correction (solid-cross green line). The deposits observed in the experiments is represented by solid-circle blue lines.}}
 \end{figure}

Figure \ref{fig:muI_muImod} shows that the second order correction in $\mu(I)$-C-MSM leads to simulated deposits that are generally closer to the experimental observations than those calculated with $\mu(I)$-MSM. In particular the deposits at $\theta=19^\circ$ and $\theta=22^\circ$ with $h_i =4.6$ mm are very well reproduced (Figure \ref{fig:muI_muImod}b,c,d,f). However, in some cases,  $\mu(I)$-MSM gives better results than $\mu(I)$-C-MSM, for example for $\theta=22^\circ$ with $h_i =1.82$ mm. This is true for the overall dynamics as illustrated in Figure \ref{fig:muI_1c_20c} that shows the time change of the granular column thickness. We can see that with $\mu(I)$-C-MSM, the avalanche is faster and the runout is overestimated and very similar to the runout obtained with the $\mu(I)$ monolayer model. As other second order terms than those included in the $\mu(I)$-C-MSM model are neglected, it is not easy to draw a firm conclusion on the improvement of results when using second order terms.

\begin{figure}[!h]
\begin{center}
\includegraphics[width=0.8\textwidth]{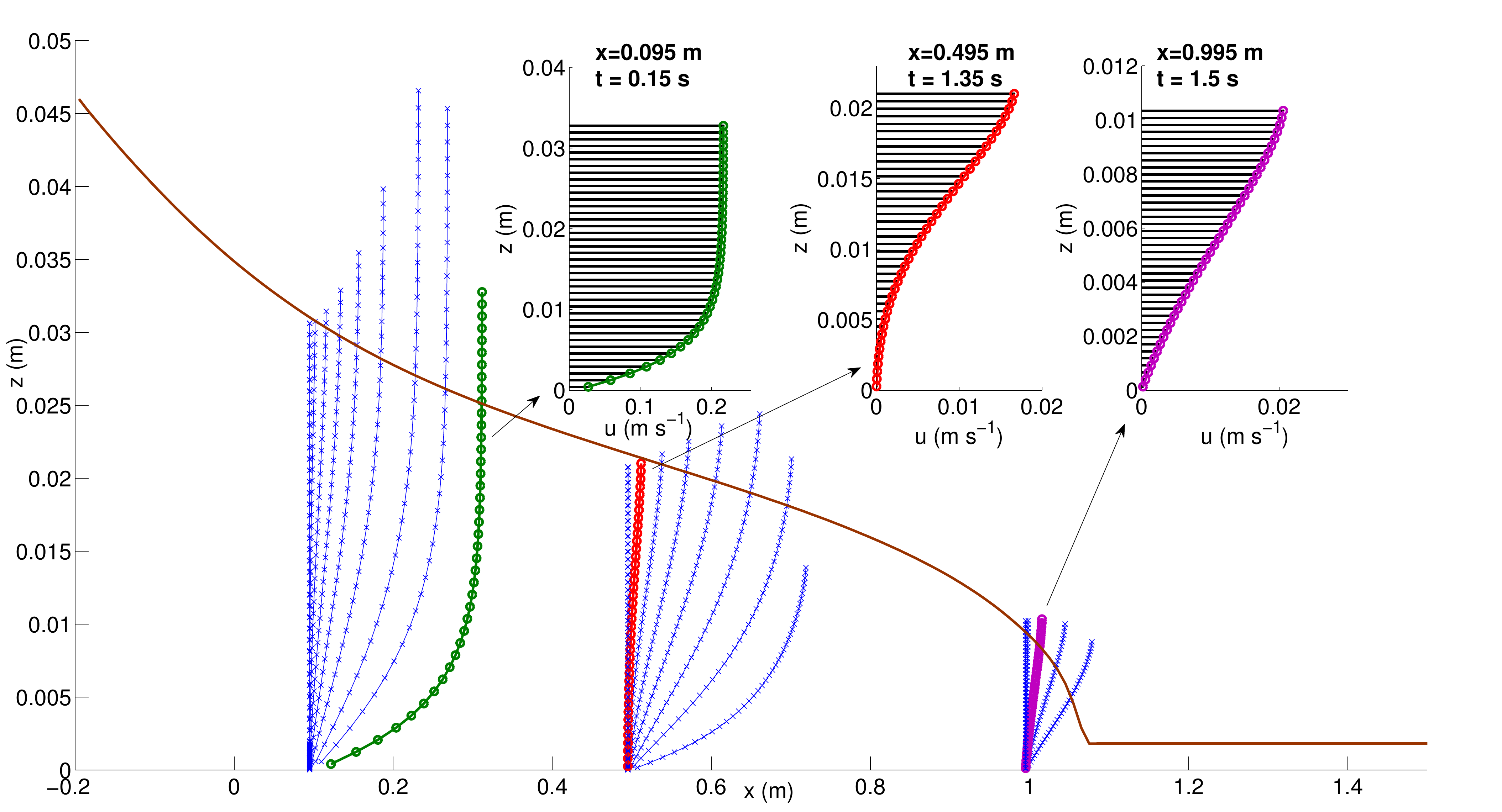}\\
\end{center}
 \caption{\footnotesize \label{fig:perfil_3puntos} \it{Normal profiles of the downslope velocity obtained with the Multilayer Shallow Model (40 layers) for $\theta = 22^{\circ}$ and $h_i = 1.82$ mm during granular collapse at different positions ($x=0.095,\,  0.495,\, 0.995$ m). For these positions, we represent the velocity profiles for different times, taken every 0.15 s (blue lines). The first selected profile (green) shows a profile at the beginning of the flow and the second (red) and third (magenta) profiles were measured during the stopping stage. The final deposit is represented by the solid brown line.}}
 \end{figure}

\begin{figure}[!h]
\begin{center}
\includegraphics[width=0.8\textwidth]{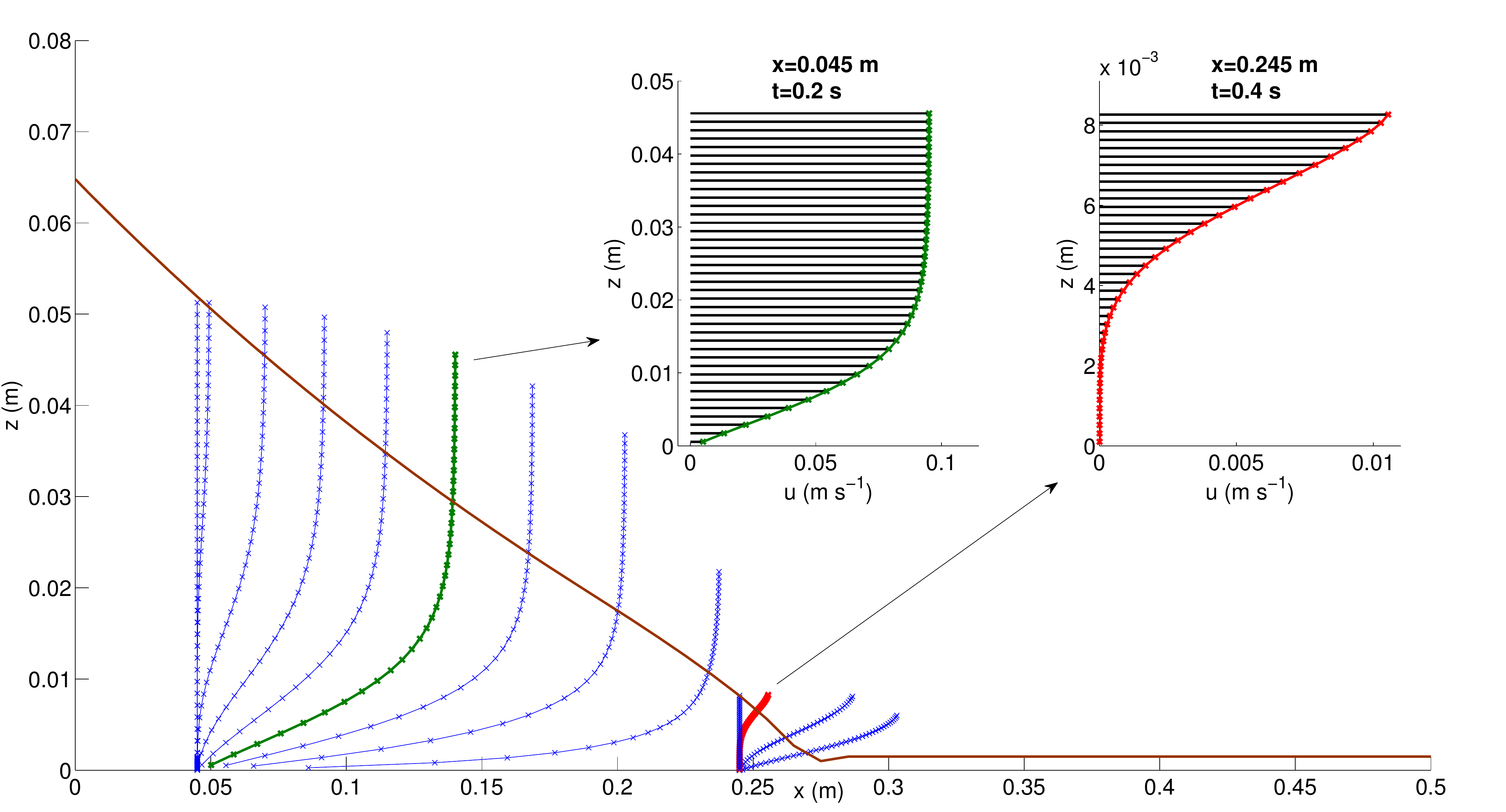}
\end{center}
 \caption{\footnotesize \label{fig:perfil_3puntos_hor} \it{Normal profiles of the downslope velocity obtained with the Multilayer Shallow Model (40 layers) for $\theta = 0^{\circ}$ and $h_i = 1.5$ mm during granular collapse at different positions ($x=0.045, 0.245$ m) and times taken every 0.05 s.}}
 \end{figure}

The Multilayer approach make it possible to obtain a normal profile of the downslope velocity. Figures \ref{fig:perfil_3puntos} and \ref{fig:perfil_3puntos_hor} show the normal profiles of the downslope velocity obtained at different times until the mass stops, for two different configurations of slopes and erodible beds. In order to obtain a more accurate profile, 40 layers are used in the Multilayer Shallow Model.

\begin{figure}[!h]
\begin{center}
\includegraphics[width=0.8\textwidth]{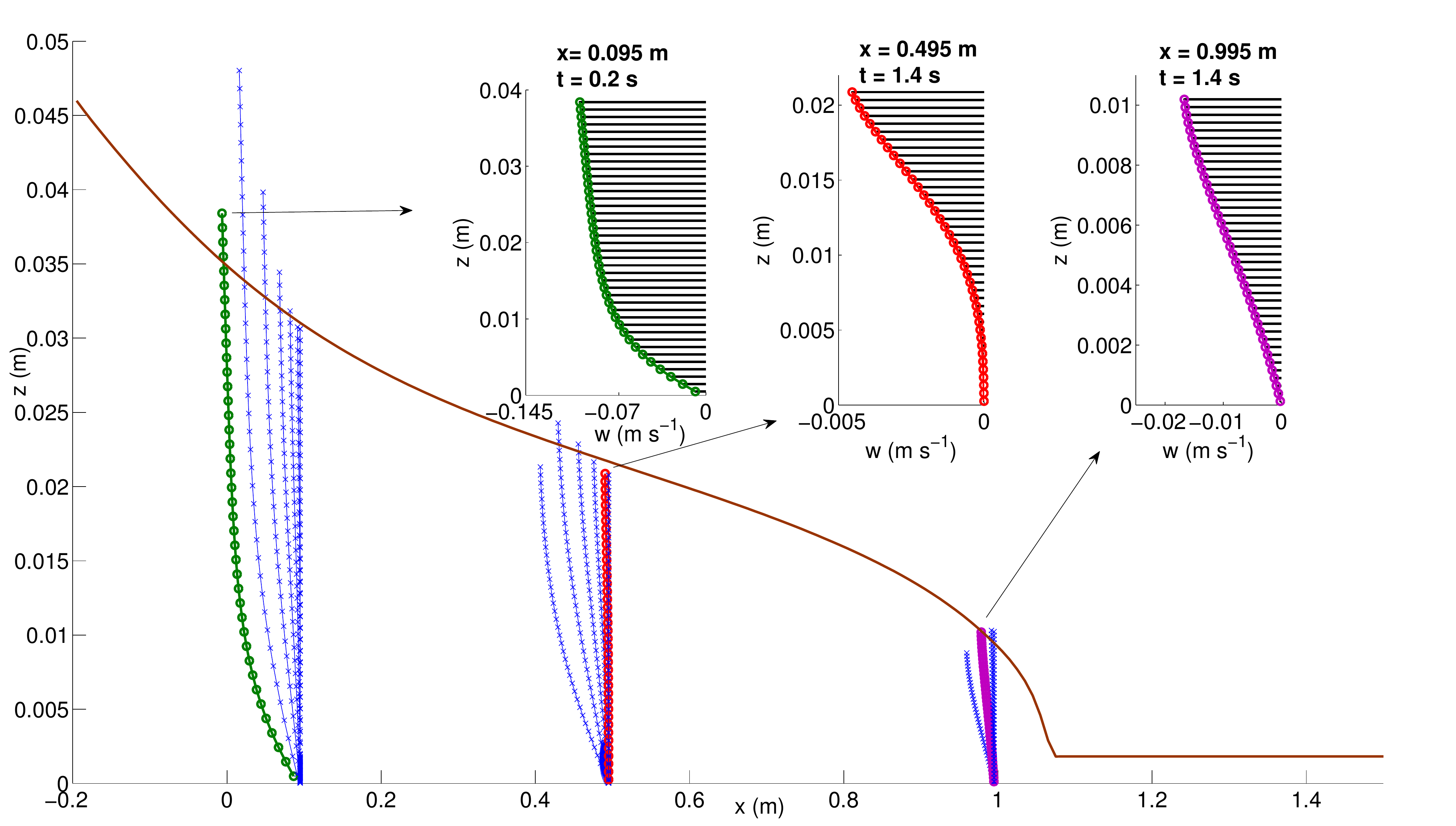}
\end{center}
 \caption{\footnotesize \label{fig:perfil_3puntos_vert} \it{Normal profiles of the normal velocity obtained with the Multilayer Shallow Model (40 layers) for $\theta = 22^{\circ}$ and for $h_i = 1.82$ mm during granular collapse at different positions ($x=0.095,\,  0.495,\, 0.995$ m) and times taken every 0.2 s.}}
 \end{figure}

The different kind of profiles observed in Figures \ref{fig:perfil_3puntos} and \ref{fig:perfil_3puntos_hor} are in good qualitative agreement with typical velocity profiles of granular flows \cite{GdRMidi} (see also \cite{LussoBouchutSimplified} and \cite{LussoBouchut}). The model predicts some sliding at the base of the flow as shown at $x=0.095$ m in Figure \ref{fig:perfil_3puntos} and at $x = 0.045$ m in Figure \ref{fig:perfil_3puntos_hor} (green profiles), in agreement with \cite{IonescuMangeney}. This suggests that a friction condition at the base could be more appropriate than the no-slip boundary condition suggested in some studies (see \cite{ChauchatMedale2} and \cite{LagreeStaronPopinet}). Note that the lower layers stop before the upper layers as observed experimentally.\\
Let us compare the averaged velocity obtained with the monolayer model to the average of the velocities over all the layers in the Multilayer Shallow Model.  In Figure \ref{fig:perfil_3puntos}, for the green profile (respectively red and magenta profiles), the velocity in the monolayer model is $1.01$ m/s (respectively $0.02$ and $0.14$ m/s) and $0.95$ m/s (respectively $0.03$ and $0.05$ m/s) for the averaged velocity in the Multilayer Shallow Model. Note that we obtain similar values for the first and second profiles. For the third profile, the averaged velocities strongly differ. Actually, at this position and time, the velocity profile corresponds to the stopping phase for the Multilayer model but not for the monolayer model. As a result, the velocity obtained in the Multilayer model is smaller than that obtained in the monolayer model.
Figure \ref{fig:perfil_3puntos_vert} shows the normal profile of normal velocity for the same configuration as Figure \ref{fig:perfil_3puntos}. Note that the normal velocities are always negative and that their absolute values are greater in the upper layers.\\

\section{Conclusion}\label{se:conclusions}

In this work, we have proposed a Multilayer Shallow Model for dry granular flows that considers a $\mu(I)$ rheology. The Multilayer approach has been applied as in \cite{EnriqueMultilayer}, thus leading to a solution of the resulting model that is a particular weak solution of the full Navier-Stokes equations. A regularisation method has been used to avoid the singularity occurring when $\|D(\vu)\|$ vanishes. A dissipative energy inequality has been proved for this model, which is an essential feature to guarantee that the calculated solution is physically meaningful.\\

The numerical solutions of this model have been compared to the 2D analytical solutions of 2D infinite granular layer flowing over an inclined plane proposed by \cite{LagreeStaronPopinet}. The Multilayer Shallow Model gives an accurate approximation of this 2D analytical solution.\\

By comparing the numerical results obtained with this new model to laboratory experiments, we have shown that the model qualitatively and sometimes quantitatively reproduces the granular column collapses over inclined erodible beds performed in \cite{MangeneyErosion}. The increase of the runout distance with increasing thickness of the erodible bed is only reproduced when using the Multilayer Shallow Model with the $\mu(I)$ rheology, although this increase is significantly underestimated. To our knowledge, this is the first time that a model has been able to reproduce this effect. The increase in runout distance appears for slopes $\theta \geq 18^\circ$ whereas it is observed for slopes $\theta \geq 16^\circ$ in the laboratory experiments. On the other hand, when using the monolayer $\mu(I)$ rheology, the increase of runout distance with the thickness of the erodible bed only occurs for slopes $\theta \ge 21^\circ$ . Moreover, in the monolayer model for $\theta = 19^\circ$, the runout distance decreases as the thickness of the erodible bed increases, contrary to observations. As a result, when using the $\mu(I)$ rheology, the multilayer model significantly improves the simulated deposits at different slopes over different thicknesses of the erodible bed compared to the monolayer model. In particular it changes the shape of the front. This is an important result as the shape of the front may be an indicator of the flow rheology \cite{Pouliquen1999b}, \cite{Jessop2012}. \\

When considering a constant friction coefficient, the multilayer approach only slightly changes the results compared to the monolayer model. Even with the Multilayer model, the use of a constant friction coefficient does not make it possible to reproduce the increase in runout distance with increasing thickness of the erodible bed. The opposite effect is observed. This confirms the analytical results of \cite{FaccanoniMangeney} obtained for the monolayer Savage-Hutter equations. \\

An important result is that this multilayer approach allows us to obtain the normal profiles of the downslope and normal velocities. These profiles qualitatively agree with the typical granular flow profiles during the developed flow and during the stopping phase \cite{GdRMidi}.\\

One of the differences between the multilayer and monolayer approaches is the accuracy of the approximation of the strain rate and consequently of the inertial number and the $\mu(I)$ friction coefficient. We have seen that the $\mu(I)$-C-MSM model, which introduces a second order correction to improve the approximation of the strain rate, generally improves the results. The increase in runout distance when the thickness of the erodible bed is increased is larger and therefore closer to the laboratory experiments. In addition, the critical slope above which the runout increases with the thickness of the bed erodible is $\theta \geq 16^\circ$, which is closer to the value observed in the experiments than the critical slope predicted by the model without the second order correction. This suggests that the extension of this shallow model up to the second order could be an important contribution. \\

\section*{Acknowledgements} 
This research has been partially supported by the Spanish Government and FEDER through the Research project MTM2012-38383-C02-02, by the Andalusian Government through the project P11-RNM7069, by the ANR contract ANR-11-BS01-0016 LANDQUAKES, the USPC PEGES project and the ERC contract ERC-CG-2013-PE10-617472 SLIDEQUAKES.
\color{black}
\bibliographystyle{plain}

\bibliography{Bibliograf} 
\end{document}